\documentclass[review,hidelinks,onefignum,onetabnum]{article}

\usepackage{lipsum}
\usepackage{amsfonts}
\usepackage{graphicx}
\usepackage{epstopdf}
\usepackage{algorithmic}
\usepackage{amssymb,amsmath,bm}
\usepackage{xcolor,chemarrow}
\usepackage{tabularx}
\usepackage{booktabs}
\usepackage{enumerate}
\usepackage{amsthm}
\usepackage{caption}

\newcommand\dbar{\rd\mkern-7mu\mathchar'26\mkern-2mu}

\newtheorem{theorem}{Theorem}
\newtheorem{remark}{Remark}

\title{Kinetic and Thermodynamic Descriptions of Open Systems of Complex Chemical Reactions with Multiple Scales}

\author{Liu Hong\thanks{School of Mathematics, Sun Yat-sen University, Guangzhou, Guangdong 510275, China ({hongliu@sysu.edu.cn}).}
\and Hong Qian\thanks{Department of Applied Mathematics, University of Washington, Seattle, Washington 98195-3925, U.S.A.  Current address: Center for Interdisciplinary Studies, Westlake University, Hangzhou, Zhejiang, 310030, China ({qianhong@westlake.edu.cn}).}}

\usepackage{amsopn}

\def\rd{{\rm d}}
\def\vn{{\bf n}}
\def\vx{{\bf x}}

\begin{document}

\maketitle

\begin{abstract}
The general theory of a complex system of nonlinear chemical reactions is a primary language of chemistry that includes chemical engineering and cellular biochemistry.  Its significance as an analytical framework, however, has not been fully appreciated outside the community of physical chemists.  In this review, we discuss the latest advances in the kinetics and Gibbsian thermodynamics of chemical reactions in a spatially homogeneous aqueous solution with a multiscale perspective on  complex systems. From the microscopic level of single reaction events which are purely stochastic in continuous time, one at a time among a set of molecules, to the macroscopic chemical reaction systems in bulk in terms of deterministic rate equations,  the mathematical descriptions of kinetic models for chemical reactions at different levels are presented in detail, with rigorous mathematical justifications presented.  In parallel with the kinetics of chemical reactions,  the irreversible thermodynamics of open systems and the stochastic thermodynamics along reactions trajectories are reviewed thoroughly.  As a novel feature, the mathematical theory of large deviations is shown to play a pivotal role in the thermodynamics of chemical reactions in equilibrium and in irreversible processes.  This review is expected to stimulate interests in and help defining multiscale phenomena and  nonequilibrium thermodynamics in many research fields on population dynamics of interacting species using chemical reactions as an analytic paradigm.
\end{abstract}


\newpage
\tableofcontents

\section{Introduction}
Multiscale modeling as a key concept in applied mathematics has been rapidly growing in the past two decades.\cite{eweinan}  One of the major driving forces of the field has been {\em computational mathematics} which begins with a set of mathematical equations | fundamental law(s) |  that describe a system rigorously and in the greatest detail. With this perspective, science and engineering become a matter of computational power, strategies (algorithms), and choice of mathematical approximations; Comparisons with Reality then follow.  From a scientific perspective, however, at the core of the notion of ``multiscale'' is the choices of mathematical representations, or idealizations, of Reality at different levels of descriptions.\cite{Chibbaro-book}  Here we emphasize {\em models}, or even {\em laws}, at different scales; they are not merely mathematical results via brute-force computations from the finest to the complex.  

The inter-connection between mathematical models at different levels of descriptions is one of the most intriguing topics both in science and in practice.  Statistical thermodynamics, as a theory in classical physics, deals precisely such issues.\cite{keizer1987statistical}  On the other hand, chemical reaction kinetics has always been a significant subject in applied mathematical studies,\cite{aris1991,epstein_book, erdi1989mathematical,seller_1984, shapiro_1969} and it is still a fertile area for quantitative modeling work with sufficient rigor, plenty of novelty,  as well as serving paradigms for other scientific research.\cite{feinberg, qian2021stochastic}

It is not an overstatement that, 
via chemical and mechanical engineerings respectively, the significance of the mathematical theory of chemical reaction kinetics \cite{erdi1989mathematical,epstein_book,aris1991,feinberg,qian2021stochastic} is on a par with analytical mechanics \cite{jeans1907,goldstein1951}.  One further notices an interesting one-to-one correspondence between the subject matters of the theory of chemical reactions, which includes both kinetics and Gibbs' chemical thermodynamics and its generalization,\cite{guggenheim1933} and that of Newtonian mechanics, which consists of kinema\-tics and dynamics. To be precise,

(1) The basic elements in chemical reactions are individual particles (atoms and molecules). The mathematical variables that represent them can be either discrete (as particle numbers) or continuous (as concentrations). This corresponds to the point masses
represented by positions and velocities in Newtonian mechanics. 

(2) In both chemical reaction theory and mechanics, interactions are the driving force for the changes in particle numbers (concentrations) or the movements of  point masses.

(3) There are well-established governing equations for both deterministic changes.  Macroscopic chemical reactions are characterized by first-order ordinary differential equations (in time, we neglect  spatial dependence following the assumption on homogeneous aqueous solution), while the Newtonian dynamics is governed by second-order differential equations as a manifestation of Newton's laws of motion. 

(4) In both cases, to be useful in engineering the fundamental laws have to be supplemented with specific {\em constitutive relations}.  These are the force field ${\bf F}({\bf x})$, where ${\bf F}$, ${\bf x}$ $\in\mathbb{R}^{3N}$ for $N$ point masses in mechanics and rate laws in chemistry $R_m({\bf c})$, $1\le m\le M$, ${\bf c}=(c_1,\cdots,c_N)$ for $N$ chemical species and $M$ reactions.

(5) Both dynamic equations are equipped with several conservation laws, like mass and energy.  In addition, there are conservations of momentum and/or angular momentum in mechanics, and there are conservations of atoms, chemical functional groups, and/or moieties in chemical reactions in a closed system.

(6) Isothermal chemical reactions in aqueous solutions occurring in an closed system is time irreversible and thus dissipative until they reach an chemical equilibrium.  The final state can be described by the celebrated Gibbsian equilibrium chemical thermodynamics. Similarly, the Newtonian mechanics can be either conservative or dissipative, depending on the nature of forces. The dissipative Newtonian mechanics behaves like the chemical reactions in many ways, while the conserved Newtonian mechanics can be transformed into equivalent Hamiltonian dynamics or Langrangian description. The latter is governed by the lease action principle, which in some way is quite similar to the minimum free energy principle in chemical reactions.

\begin{figure}[ht]
	\label{figure_1}
    \includegraphics[width=1.0\linewidth]{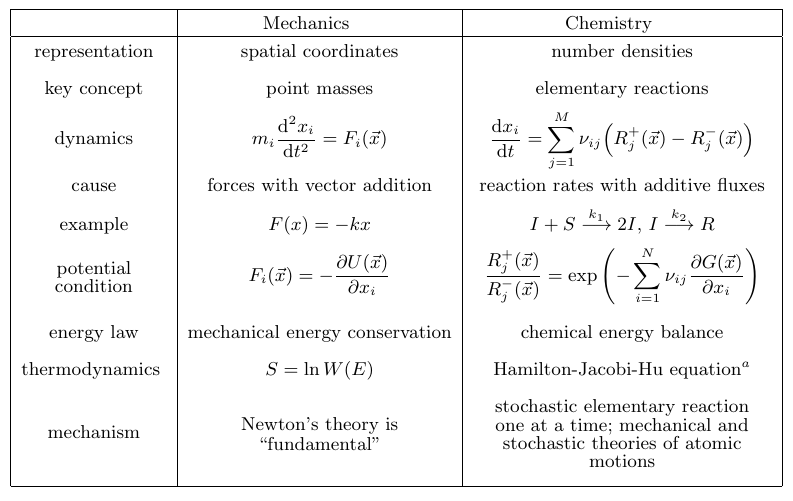}
	\caption{Macroscopic deterministic chemical reaction kinetics vs. Newtonian mechanics with contradistinctions: While most comparisons are pedantic, we call attention to that the force with a potential $F(\vec{x})=-\nabla U(\vec{x})$ gives rise to Hamiltonian conservative dynamics; and the existence of a Gibbs potential $G(\vec{x})$ guarantees detailed balance, thus chemical equilibrium, among reactions.  The pair of chemical reactions, $I+S\longrightarrow 2I$, $I\longrightarrow R$, represent the synthesis of $I$ from its precursor $S$ and the degradation of $I$.  As for Hamilton-Jacobi-Hu equation$^a$, see Eq. \eqref{HJE-Hu} and Ref. \cite{gang1986lyapounov}. 
}
\end{figure}

All in all, one clearly sees that \textbf{\textit{the theory of chemical reaction kinetics is to chemistry what theoretical mechanics is to classical physics}}.  Both are primary languages that provide an analytic description of the respective subjects with wide applicability. However, in contrast to the supreme status of Newtonian dynamics in Science with a long history, it is a pity that the merit of chemical reaction theory has not gained a full respect in theoretical science and applied mathematics outside chemistry and biochemistry.  This is one of the chief motivations for the present review.

In this work, we aim at establishing a comprehensive and consistent picture of a system of chemical reactions that occur at different levels of description with both kinetic and thermodynamic aspects. This becomes possible only recently due to the latest advances in the field. To be exact, our review starts with the mathematical representation/description of single stochastic elementary reaction occurring among several molecules by Poissonian random events.  Each such event leads to changes in molecular numbers of the reactants and products, whose evolution is characterized by a multivatiate integer-valued continuous-time Markov process; and whose probability evolution follows a chemical master equation.\cite{beard_qian_book}  It is well-known that, according to the Kurtz limit theorem a system of nonlinear chemical rate equations emerges naturally as a consequence of the law of large numbers.  Furthermore, the central limit theorem gives a system of asymptotic chemical Langevin equations with a corresponding chemical Fokker-Planck equation with small diffusivity \cite{gillespie2000chemical}.  Along this line, we establish microscopic, mesoscopic and macroscopic kinetic models for general chemical reactions step by step. The existence, uniqueness and asymptotic stability of the stationary solutions to steady-state equations of the respective models are addressed.

In parallel with kinetics of chemical reactions, we also discuss in details the stochastic thermodynamics of random chemical reaction events along each stochastic trajectory,  and nonequilibrium steady-state (NESS) thermodynamics based on the molecular number distribution, {\em e.g.} concentration fluctuations, of chemical reactions occurring in open systems.  These two formulations, as modern generalizations of Gibbs' thermodynamic theory for macroscopic chemical reactions in closed equilibrium systems, clearly reveal the intrinsic connections among kinetic theories of chemical reaction at different levels. Particularly, the applications of the mathematical theory of large deviations to chemical kinetics, such as the emergent chemical Hamilton-Jacobi equations for macroscopic chemical thermodynamics and Kramers' formula for the transition rate theory, are highlighted.

As a footnote, one also notices that there is a change of representations from so-called Lagrangian coordinates for point masses to Eulerian coordinates for counting in fluid mechanics.  In the latter $\vn(t+\rd t)=\vn(t)+\rd\vn(t)$ should be understood as elementary ``reactions'' that have exponential laws for sojourn times $\Delta t$.  The contradistinctions between ``individuals'' and ``population'' of individuals need to be appreciated:  In order to obtain meaningful statistical descriptions, the latter has to sacrifice great details contained in the former; individual stories give rise to a coherent statistical narrative of {\em l'homme moyen}.  Ultimately of course it is the entire ``life history'' of an individual that makes each one unique \cite{hopfield_jtb,pence2011,alekseev_symb_dyn}.

\begin{figure}[h]
	\centering	\includegraphics[width=1\linewidth]{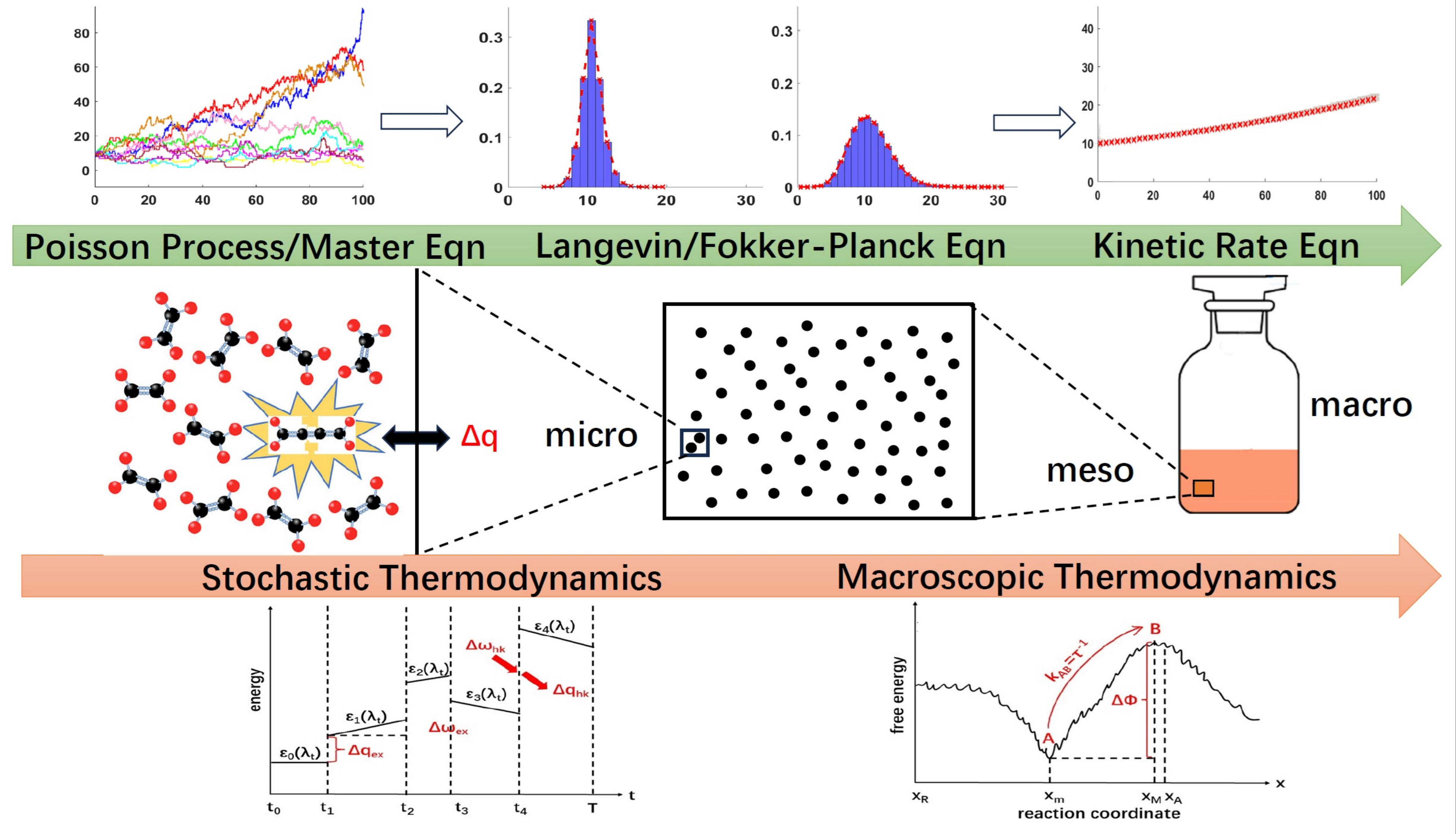}
	\caption{Kinetic and thermodynamic descriptions of chemical reactions at multiple scales. Left: microscopic counting of each and every reaction and individual molecules, one at a time \cite{xie}.  Middle: mesoscopic concentrations with stochastic fluctuations \cite{elson}.  Right: macroscopic deterministic kinetics and Gibbs' thermodynamics.}
	\label{multiscale modeling}
\end{figure}

\section{Chemical Reaction Systems at Different Scales}
{The chemical reactions can be viewed from different scales. From a macroscopic view, they occur in a deterministic way, characterized by a group of coupled ordinary differential equations, or partial differential equations when spatial heterogeneity is taken into consideration. In contrast, at the very fundamental molecular level, elementary chemical reactions are merely a series of Poisson random events involving several reactive molecules. Between the two, one can adopt a probabilistic description for reactions among a few molecules. These diverse descriptions and models are closely related in mathematics, and provide very detailed and richful kinetic information about the system under study. The various kinetic models developed from the chemical reactions in the last century will be reviewed in this section, while the corresponding thermodynamics at different levels will be addressed later.}

\subsection{Elementary Reactions as Poisson Random Events at Individual Molecule Level}

Without loss of generality, let us consider a rapidly stirred isothermal chemical reaction system that consists of $N$ components $\{X_1,X_2,...,X_N\}$, or chemical species, and $M$ possible reversible reactions in a given aqueous solution with volume $V$,
\begin{eqnarray}
	\label{chemical-reaction}
	\nu^+_{i1}X_1 + \nu^+_{i2}X_2 + \cdots + \nu^+_{iN}X_N \ \underset{r^-_i}{\overset{r^+_i}{\rightleftharpoonsfill{26pt}}} \  
	\nu^-_{i1}X_1 + \nu^-_{i2}X_2 + \cdots + \nu^-_{iN}X_N,
\end{eqnarray}
where $i=1,2,\cdots,M$.  In the equation, all stoichiometric coefficients are non-negative; most of them are in fact zero. We denote $\nu_{ij}=(\nu^-_{ij}-\nu^+_{ij})$ which gives the change in the number of species $j$ in the $i$th reaction; with respective forward and backward reaction rate functions $r_i^+(\vec{n};V)$ and $r_i^-(\vec{n};V)$, and $\vec{n}(t)=(n_1,\cdots,n_N)(t)$ being the numbers of molecules in the system at time $t$.  The functional forms of non-negative $r^{\pm}_i$ can be quite arbitrary; one concrete example is the {\em law of mass action} which will be specified later. If one of the forward and backward rates of a reaction is zero, we say the reaction is irreversible.  If a particular $\nu_{ij}^+=\nu_{ij}^-\neq 0$ but
$\nu_{ij}=\nu_{ij}^- - \nu_{ij}^+=0$, we say the species $j$ is a catalyst of the $i$th reaction; and if $\nu_{ij}^->\nu_{ij}^+\neq 0$, then the $i$th reaction in the forward direction is auto-catalytic. DNA replication reaction that requires a DNA template is a good example, which corresponds nicely a birth process in population dynamics.  

The central issue for chemical kinetic theory is to characterize how the number of each species changes with time due to reactions.  In the stochastic theory it is represented by each and every individual reaction that occurs one at a time.  According to the detailed molecular theory about collisions and reactions, for a reaction to happen it is necessary for the reacting particles (atoms or molecules) to first come together and then collide with one another. However, not all collisions bring about a reaction. In the classical description a collision will be effective in producing chemical change only when the colliding species possess an internal energy greater than the activation energy of the reaction and a suitable orientation favourable to the rearrangement of atoms and electrons. Mathematically, we can treat the appearance of effective collisions among reactants as a random process -- a series of random events with probability distributions for inter-event times.  This is precisely captured by the rate functions.  

Each {\em stochastic elementary reaction} has a rate function $r$ which is a function of $\vec{n}$ and $V$.  The defining assumption is that the probability distribution in continuous time is exponentially distributed with the $r(\vec{n},V)$ as the rate.  Given the number of reactants and/or possible reactions are sufficiently large, there are two important mathematical limit theorems that secure the exponential distribution.  First, Cox \cite{cox1954superposition} and Khinchin \cite{khinchin1960mathematical} independently proved that the superposition of $m$ independent and identically distributed (i.i.d.) renewal processes has an asymptotic exponential waiting time for the next event in the limit of $m\rightarrow\infty$.

\begin{theorem}[Superposition of Renewal Processes \cite{cox1954superposition, khinchin1960mathematical}]
	A renewal process is a sequence of time points $T_1,\cdots,T_n,\cdots$, where successive waiting times $X\equiv T_{i+1}- T_i\ge 0$ are independent and follow a common probability density function $f_X(x)$.  Let $\big\{T_i^{(k)}; i\ge 1,\, 1\le k\le m\big\}$ be $m$ i.i.d.  renewal processes all having the distribution $f_X(x)$.  Then the superposition of the $m$ renewal processes, a complete ordering of all the $T_i^{(k)}$ with increasing time,  and scaled by $m$:
	\[
	\Big\{ \tilde{T}_j: j\ge 1 \,\Big|\, \tilde{T}_{j+1}-\tilde{T}_j\ge 0 \Big\} =
	\bigcup_{k=1}^m \Big\{ mT_i^{(k)} \,\Big|\, 
	i\ge 1 \Big\}, 
	\]
	is a Poisson process in the limit of $m\rightarrow\infty$, with rate parameter $\mathbb{E}^{-1}[X]$.
\end{theorem}

This theorem can be fittingly applied to $m$ same enzyme molecules, each has a complicated distribution for the time between consecutive products, nevertheless when $m\ge 15$ the waiting time for product next arrival irrespective the enzyme is exponential.\cite{qian_bj_08} It points to the essential role of a Poisson process in stochastic chemical kinetics at the individual molecule level, one reaction at a time. 

In the stochastic setting, \textbf{\textit{the number of occurrences of a reaction per unit time is replaced by the inverse of the waiting time for the next reaction event}}.  We call a reaction with an exponential waiting time \textbf{\em{stochastic elementary}} \cite{epstein_book, keizer1987statistical,qian2021stochastic}.  In general non-elementary reactions can be modelled through a set of elementary reactions by introducing multiple intermediate species.  Therefore one can safely postulate all reactions in Eq. \eqref{chemical-reaction} as stochastic elementary, with given rate functions, as long as a chemical reaction can be conceptualized as an instantaneous ``rare'' event.

The waiting time of a single elementary reaction is exponentially distributed. What happens if we have two independent elementary reactions that occur simul\-taneously?  A simple calculation shows that for two independent, exponentially distributed $T_1$ and $T_2$
with rates $\lambda_1$ and $\lambda_2$, the random time of the next reaction to occurring, $T_{min}=\min\{T_1,T_2\}$, is again exponential with rate $\lambda_1+\lambda_2$, and the probability $\Pr\{T_{min}=T_i\}$ is $\lambda_i/(\lambda_1+\lambda_2)$, $i=1,2$.  Generalize this result to $n$ independent elementary reactions {\em in parallel} with rate $\lambda_1, \lambda_2, \cdots, \lambda_n$, we have $\Pr\{T_{min}=T_k\} = \lambda_k/(\lambda_1+\cdots+\lambda_n)$, $1\le k\le n$.  More importantly,
\begin{eqnarray*}
	&&\Pr\{T_{min}\geq t,T_{min}=T_k\}\\
	&=&\Pr\{T_1\geq T_k,\cdots,T_{k-1}\geq T_k,T_k\geq t,T_{k+1}\geq T_k,\cdots,T_n\geq T_k\}\\
	&=&\int_t^{\infty}\lambda_k e^{-\lambda_kt_k}\bigg(\prod_{l=1,l\neq k}^n\int_{t_k}^{\infty}\lambda_l e^{-\lambda_lt_l}\rd t_l\bigg)\rd  t_k=\frac{\lambda_k}{\sum_{i=1}^n\lambda_i}\exp\left( 
	-t\sum_{i=1}^n\lambda_i\right)\\
	&=&\Pr\{T_{min}\geq t\}\Pr\{T_{min}=T_k\}.
\end{eqnarray*} 
This means \textbf{\textit{for independently occurring elementary reactions, the random time of the first one and the probability of which one are independent.}} 

Above result naturally provides the theoretical underpinning of the
continuous-time Markov formulation and the Gillespie (also Boltz-Kalos-Lebowitz, Doob \cite{qian2021stochastic}) algorithm \cite{gillespie1976general} for stochastic simulation of chemical reactions.  The algorithm is a Monte Carlo method,  which consists of two basic steps --  (1) generate a random time at which the next event occurs and randomly select one among all possibilities as the next event (the probabilities of these two random variables are given by the formula above); (2) move the model time forward to the next event time and update the state of the system accordingly. These steps will be repeated again and again until certain stopping criteria is met.

The stochastic trajectories generated through the Gillespie algorithm, or more precisely the integer-valued continuous-time Markov process, can be expressed in terms of the Poisson processes with change of ``times'':
\begin{eqnarray}
	\label{chemical-poisson-pro}
	n_j(t)= n_j(0) &+&\sum_{i=1}^M \nu_{ij}\left\{Y_i^+\bigg[\int_{0}^tr_i^+(\vec{n}(s);V)\rd s\bigg] \right. 
	\\
	&-& \left.  Y_i^-\bigg[\int_{0}^tr_i^-(\vec{n}(s);V)\rd s\bigg]\right\},\quad j=1,\cdots, N,
	\nonumber
\end{eqnarray}
where $\vec{n}=(n_1,\cdots,n_N)$ is the numbers of chemical species $X_1,\cdots, X_N$ respectively. $Y_i^+(t)$ and $Y_i^-(t)$ stand for $2M$ independent standard Poisson processes all with $\mathbb{E}[Y(t)]=t$.

\subsection{Chemical Master Equation for Molecular Number Distribution}

We have shown, at the molecular level, how to use the molecule number counts of $N$ species in a rapidly stirred reaction vessel with volume $V$, $\vec{n} = \{n_1,n_2\cdots,n_N\}$ which are integer-valued random variables, to characterize the state of a chemical reaction system. In mathematics, it is well-known that this stochastic formulation can be transformed into an equivalent probability flow description. Denote the probability of a system in state $\vec{n}$ at time $t$ as $P(\vec{n},t)$, with the transition rates $r^+_i(\vec{n}; V)$ for the forward and $r^-_i(\vec{n};V)$ for the backward directions of the $i$th reaction
respectively, it obeys the so-called chemical master equation (CME),
\begin{eqnarray}
	\label{chemical-master-eq}
	\frac{\rd P(\vec{n},t)}{\rd t}=&&\sum_{i=1}^M\bigg[r^+_i(\vec{n}-\vec{\nu}_i; V)P(\vec{n}-\vec{\nu}_i,t)-r^+_i(\vec{n}; V)P(\vec{n},t)\\
	&&\qquad-r^-_i(\vec{n}; V)P(\vec{n},t)+r^-_i(\vec{n}+\vec{\nu}_i;V)P(\vec{n}+\vec{\nu}_i,t)\bigg],\nonumber
\end{eqnarray}
where $\vec{\nu}_i$ is the vector that contains the stoichiometric coefficients of the $i$th reaction. The chemical master equation actually is a manifestation of the law of total probability by considering all possible changes in probability due to reactions, one $\rd t$ step at a time.

It is noted that the CME above is a special case of general master equations for continuous-time Markov chain,
\begin{eqnarray}
	\label{master-equation}
	\frac{\rd P_n(t)}{\rd t}=\sum_{n'}\Big(W_{n,n'}P_{n'}-W_{n',n}P_n\Big),
\end{eqnarray}
where $P_n(t)\geq0$ is the probability for finding the system in state $n$ at time $t$, and $W_{n,n'}\geq0$ $(n\neq n')$ denotes the transition rate from state $n'$ to state $n$, with $W_{nn}=-\sum_{n'\neq n}W_{n',n}$. By definition each column sum of a transition rate matrix is zero, i.e. $\sum_{n}W_{n,n'}=0$.

Without loss of generality and for finite state space, the transition rate matrix $(W_{n,n'})$ is assumed to satisfy the \textbf{\textit{irreducibility condition}}, which claims that, for any states $n_0\neq n_1$, there is a sequence of states $n'_1,n'_2,\cdots,n'_l$, such that $n_0=n'_1,n_1=n'_l$ and $W_{n'_{i+1},n'_i}>0$ for $1\leq i\leq l-1$. In other words, between each pair of states $n_0$ and $n_1$, there exists a pathway with all positive transition rates. Otherwise, the whole system could be split into multiple isolated sub-systems. Under the irreducibility condition, it is straightforward to show $0$ is a unique eigenvalue for a matrix with zero column sum, with $(1,\cdots,1)^T$ as its  left eigenvector. More importantly, according to the Perron-Frobenius Theorem, $(W_{n,n'})$ is a stable matrix -- all other eigenvalues have negative real parts.\cite{schnakenberg1976, peng2018generalized}

If one further supposes that the initial $P_n(0)=P_n^0$ with $0\leq P_n^0\leq 1$ and $\sum_i P_n^0=1$ for all states $n$, it can be shown that there exists a unique \textbf{\textit{nonequilibrium steady state (NESS)}} with a stationary probability distribution $\{P_n^{ss}\}$ as the long-time solution to the master equation:\cite{schnakenberg1976}
\begin{equation} 
	\lim_{t\to\infty} P_n(t) = P_n^{ss}, 
	\text{ and } \, 
	\sum_{n'} W_{n,n'}P_{n'}^{ss} = 
	\sum_{n'} W_{n',n}P_n^{ss}.
\end{equation} 
One class of master equations is of particular importance, that is, its stationary probability distribution $\{P_n^{eq}\}$ further satisfies the \textbf{\textit{condition of detailed balance}}, $W_{n,n'}P_{n'}^{eq}=W_{n',n}P_n^{eq}$ for each pair of $n$ and $n'$. In this case, $\{P_n^{eq}\}$ is called an equilibrium distribution.

\begin{theorem}[Uniqueness of NESS for Finite-State Markov Process \cite{schnakenberg1976}]
	If a finite Markov transition rate matrix $(W_{n,n'})$ satisfies the irreducibility condition, and if the initial distribution satisfies $P_n(0)=P_n^0$ with $0\leq P_n^0\leq 1$ and $\sum_n P_n^0=1$ for all states $n$, then one has
	\begin{enumerate}[(i)]
		\item the transition rate matrix $(W_{n,n'})$ has exactly one non-degenerate eigenvalue $\lambda_0=0$;
		\item all other eigenvalues of $(W_{n,n'})$ have negative real parts;
		\item the NESS $\{P_n^{ss}\}$ is unique in the full space of probabilities, a simplex in $\mathbb{R}^L$, and every solution starting at arbitrary points in the probability simplex is asympto\-tically stable with respect to the $\{P_n^{ss}\}$.
	\end{enumerate}
\end{theorem}

The physical meanings of equilibrium state vs. NESS become clear by introducing the concept of ``probability fluxes'': Denote $J_{n,n'}=W_{n,n'}P_{n'}$ as the flux from state $n'$ to state $n$. Then $J_n^{in}=\sum_{n'\neq n}W_{n,n'}P_{n'}$ and $J_n^{out}=\sum_{n'\neq n}W_{n',n}P_n$ are interpreted respectively as the total influx from all other states to state $n$ and the total efflux leaving state $n$. The NESS, thus, is understood as a state that satisfies an overall balance between inflows and outflows $J_n^{in}=J_n^{out}$ for each and every $n$.  An equilibrium state requires a much stronger balance -- between each pair of fluxes $J_{n,n'}=J_{n',n}$.  The former implies the system has at least one cycle flux at the stationary state; the latter can have none.  \cite{hill1977,jiang2004mathematical}  Cyclic stationary flux(es) is responsible for dissipation.\cite{qian2016}


\subsection{Chemical Rate Equation in Macroscopic Limit}
\label{CMAE-kurtz-limit}

Both stochastic tra\-jectory with jumps given in Eq. (\ref{chemical-poisson-pro}), in terms of the Poisson process, and its corresponding chemical master equations that describe the probability, characterize the stochastic chemical reactions at a molecular level.  Together it constitutes a kinetic theory that is particularly suitable for systems with few particles, from single molecules to thousands. In contrast, classical chemical kinetics considers systems with particle numbers $\sim 10^{23}$ in a macroscopic volume $V$.  Recall that in the stochastic formulation, the $\vec{n}_V(t)$ in Eq. (\ref{chemical-poisson-pro}) and the $P_V(\vec{n},t)$ in Eq. (\ref{chemical-master-eq}) both have $V$ as a parameter.  If one considers the volume $V$ being large enough and the particle numbers of all chemical species are numerous, one can formally introduce particle number densities, or chemical concentrations, as 
\[ 
\vec{c}= \lim_{V\to \infty}  \left( \frac{\vec{n}_V}{V}\right). 
\]
This is known as the Kurtz limit in mathematical literature.  One expects that, under appropriate mathematical condition, the stochastic process $\vec{n}_V(t)/V$ characterized by the stochastic theory converges to a deterministic trajectory $\vec{c}(t)$.  More precisely, either $\vec{n}(t)/V$ as a stochastic process converges to $\vec{c}(t)$ as the solution to an ordinary differential equation (see Eq. \eqref{macro-rate-eq} below) for any finite time $t\in[0,T]$, or solution to the chemical master equation $P_V(V\vec{x},t)$ converges to an impulse function $\delta\big(\vec{x}-\vec{c}(t)\big)$. This convergence theorem was first proved by T. G. Kurtz in the 1970s.  \cite{kurtz1972, kurtz1978strong}

Before presenting  the rigorous theorem, an intuitive understanding comes from the asymptotic approximation of a Poission process. For a Poisson process $Y(\lambda t)$ with rate $\lambda>0$ and distribution $P(Y(\lambda t)=k)=\frac{(\lambda t)^k}{k!}e^{-\lambda t}$, $k\in \mathbb{N}$, it is well-known that $\mathbb{E}[Y(\lambda t)]=\text{Var}[Y(\lambda t)]=\lambda t$. Therefore, $Y(\lambda t)=\lambda t+\mathcal{O}\big(\sqrt{\lambda}\big)$ as $\lambda\rightarrow\infty$. Applying this formula to Eq. \eqref{chemical-poisson-pro}, we immediately have:
\begin{eqnarray}
	&&c_j(t)\\
	&=& c_j(0)+ \frac{1}{V}\sum_{i=1}^M \nu_{ij}\bigg[ Y_{i}^+\left( V\int_{0}^t \frac{r_i^+\big(\vec{n}(s);V\big)}{V}\, \rd s\right) - Y_i^-\left( V\int_{0}^t \frac{ r_i^-\big(\vec{n}(s);V\big)}{V}\,\rd s\right) \bigg]
	\nonumber\\
	&=& c_j(0)+\sum_{i=1}^M \nu_{ij}\bigg[\int_{0}^t \frac{ r_i^+\big(V\vec{c}(s);V\big)}{V}\,\rd s-\int_{0}^t \frac{r_i^-\big(V\vec{c}(s);V\big)}{V}\,\rd s\bigg] + \frac{ \mathcal{O}\big(\sqrt{V}\big) }{V} 
	\nonumber\\
	&\rightarrow&c_j(0)+\sum_{i=1}^M \nu_{ij}\bigg[\int_{0}^tR_i^+\big(\vec{c}(s)\big)\rd s-\int_{0}^tR_i^-\big(\vec{c}(s)\big)\rd s\bigg],\quad j=1,\cdots, N.\nonumber
	\label{kurtzlimit}
\end{eqnarray}
where
\[
\vec{c}:=\frac{\vec{n}}{V} \, \text{ and }\,  R_i^\pm\big(\vec{c}\big)=\lim_{V\rightarrow\infty}\frac{ r_i^\pm(V\vec{c};V)}{V}.
\]
Eq. (\ref{kurtzlimit}a) is the integral form of ordinary differential equation
\begin{equation}
	\frac{\rd c_j(t)}{\rd t} 
	= \sum_{i=1}^M \nu_{ij} \Big( 
	R_i^+(\vec{c}) - 
	R_i^-(\vec{c})\Big),\quad j=1,\cdots,N.
	\label{macro-rate-eq}
\end{equation}
This derivation, though only heuristic since the $n_j(t)$ is a random variable whose convergence requires advanced mathematical treatment, clearly demonstrates that the deterministic trajectory as described by a system of differential equations (\ref{macro-rate-eq}) is a mathematical consequence of the \textbf{\textit{Law of Large Numbers}}, by neglecting fluctuations on the order of $V^{-1/2}$.

The solution to Eq. (\ref{macro-rate-eq}), $\vec{c} = (c_1, c_2, \cdots, c_N )^T$, is the macroscopic concentrations of chemical species $\{X_j\}$. $R^+_i(\vec{c})$ and $R^-_i(\vec{c})$ are both non-negative functions with $c_j\geq0$ for all $j=1,\cdots,N$, representing the forward and backward reaction rate functions for the $i$th reaction. Eq. \eqref{macro-rate-eq} provides a macroscopic description of chemical reactions when the system size is sufficiently large.  

Equation (\ref{macro-rate-eq}) is traditionally understood as the ``definition'' for instantaneous rates of changes of chemical species due to $M$ reactions. $R_i^\pm\big(\vec{c}\big)$ are the number of reactions per unit volume per time; therefore they have wide applications in many disciplines.  It is a ``kinematic description'' of macroscopic population dynamics in which $R_i^\pm$ as a function of $\vec{c}$ is a ``constitutive relation'' to be supplied.

\begin{theorem}[LLN for Chemical Master Equations\cite{kurtz1972,kurtz1981clt,kurtz1978strong}]
	Assume the terms on the right-hand side of Eq. \eqref{macro-rate-eq} are Lipschitz continuous, and there exist constants $\epsilon_i$ and $B$ such that $$\frac{r_i^\pm(V\vec{c};V)}{V}\leq \epsilon_i(1+\|\vec{c}\|),\quad \bigg|\frac{r_i^\pm(V\vec{c};V)}{V}-R_i^\pm(\vec{c})\bigg|\leq \frac{\epsilon_iB(1+\|\vec{c}\|)}{V}.$$ Further assume $\sum_i\epsilon_i\|\vec{\nu}_i\|<\infty$, then the stochastic trajectories generated by Eq. \eqref{chemical-poisson-pro} converge in probability to the solution of Eq. \eqref{macro-rate-eq} for any finite time, i.e.
	\[
	\lim_{V\rightarrow\infty}\sup_{t\leq T}\bigg|\frac{\vec{n}(t)}{V}-\vec{c}(t)\bigg|=0,\ \text{a.s.} \ \text{ for } \  \forall T>0,
	\]
	provided the initial conditions coincide $\vec{n}_V(0)=V\vec{c}(0)$.
\end{theorem}

Analogous to the master equations, the stationary solutions to the ODEs in Eq. \eqref{macro-rate-eq} can be classified into several types. The solution satisfying the condition $\sum_{i=1}^{M}\nu_{ij}\big[R_i^+(\vec{c}^{\,ss})-R_i^-(\vec{c}^{\,ss})\big]$ ($j=1,\cdots,N$) defines the NESS ($\vec{c}^{\,ss}\geq0$), which however if exists is generally non-unique. A special kind of stationary solutions is called the equilibrium solution ($\vec{c}^{\,eq}\geq0$), provided $R_i^+(\vec{c}^{\,eq})=R_i^-(\vec{c}^{\,eq})$ hold for $\forall i\in[1,\cdots,M]$. This is the \textbf{\textit{condition of detailed balance}} for deterministic chemical kinetics.

The local existence and uniqueness of solutions to the first-order ODE system in Eq. \eqref{macro-rate-eq} has been well established, provided the reaction rates on the right-hand side are continuous functions and satisfy the Lipschitz continuity, \textit{i.e.} there exists a constant $B$ such that $|R_i^{\pm}(\vec{c}_2)-R_i^{\pm}(\vec{c}_1)|\leq B\|\vec{c}_2-\vec{c}_1\|$ for $\forall i=1,\cdots,M$. 

However, as the solutions of nonlinear ODEs can exhibit very complex behaviors, such as bistability, limit cycle oscillation, quasi-periodicity, even chaos, \textit{etc.}, to show the existence and uniqueness of a stationary solution, \textit{i.e.}, $T=\infty$, under proper conditions is not an easy task. We shall look into a particular class of ODEs, which is called the \textbf{\textit{chemical mass-action equations}}, by adopting the mass-action type rate law.

\textbf{\textit{The law of mass action}} gives one of the most widely used rate functions, with a single parameter for each reaction; it has a deep relation to ideal solution.  It has a polynomial functional form for a reaction rate,\cite{erdi1989mathematical}
\begin{eqnarray}
	r_i^{\pm}(\vec{n};V)=k_i^{\pm}V\prod_{j=1}^N\frac{n_j!}{\big(n_j-\nu_{ij}^{\pm}\big)!\cdot V^{\nu_{ij}^{\pm} }},
\end{eqnarray}
for $i=1,2\cdots,M$. $k_i^\pm$ denote the respective forward/backward reaction rate constants. The corresponding macroscopic reaction rate in the Kurtz limit reads
\begin{eqnarray}
	R_i^{\pm} (\vec{c})=\lim_{V\rightarrow\infty}\frac{r_i^\pm(V\vec{c};V)}{V}=k_i^{\pm} \prod_{j=1}^Nc_j^{\nu_{ij}^{\pm}}.
\end{eqnarray}
Inserting above formulas into Eq. \eqref{macro-rate-eq}, we arrive at the celebrated mass-action chemical kinetic equations. A notable mathematical properties of mass-action systems is that they do not admit negative solutions, once the initial concentrations are non-negative. Though it seems obvious to a chemist, this was first proven by Vol'pert.\cite{vol1972differential}

In 1972, Horn and Jackson\cite{horn1972general} have identified an important class of chemical reactions | the ``\textit{\textbf{complex-balanced systems}}'', whose stationary solution can be readily shown to exist, being unique and locally asymptotically stable. 
Generalizing the idea of detailed balance, 
they consider the whole reactants or products of each reaction as a ``complex'', characterized by the stoichiometric coefficient vector $\vec{\nu}_i^{\,+}$ or $\vec{\nu}_i^{\, -}$. Then a state $\vec{c}^{\, ss}$ of a chemical mass-action system is a \textit{\textbf{complex-balanced steady state}} if at each and every vertex of the corresponding reaction network, signified by the complexes,  the fluxes flowing into the vertex exactly balance the fluxes flowing out of the vertex.  One can check that there is a sequential inclusion relations, \textit{i.e.} \textbf{$\{\text{nonequilibrium steady state}\}\supset \{\text{complex-balanced steady state}\}$  $\supset\{\text{detailed-balance steady state, equilibrium}\}$}.  A complex-balanced system is a special chemical mass-action system that has at least one complex-balanced steady state. 

\begin{figure}[h]
	\centering
	\includegraphics[width=0.4\linewidth]{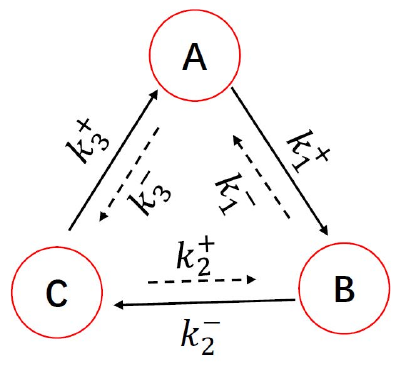}
	\caption{A cyclic reaction network. When $k_1^-=k_2^-=k_3^-=0$ and $k_1^+,k_2^+,k_3^+>0$, the system is complex balanced and contains a clockwise net flux even at the stationary state. If and only if $\frac{k_1^+k_2^+k_3^+}{k_1^-k_2^-k_3^-}=1$, the system satisfies the condition of detailed balance, and there is no net flux at the equilibrium state.}
	\label{complex balance}
\end{figure}

Whether a chemical reaction system is complex-balanced or not is closely related to two important topological properties of a reaction network. The first one is \textit{\textbf{weakly reversible}}\cite{erdi1989mathematical,feinberg,yu2018mathematical}, which states if there is a reaction path from complex A to complex B,  then there should be another path from complex B to complex A: There is a kinetic cycle.  The weakly reversible condition requires if two complexes are connected, they must be connected in both directions (it could be indirectly though other intermediate complexes). The second one is the \textit{\textbf{deficiency}}, which is an non-negative integer $m-\ell-s$ with $m$ as the total number of complexes, $\ell$ being the number of connected components (groups of connected complexes), and $s$ being the rank of the stoichiometric coefficient matrix\cite{horn1972necessary, feinberg1972complex,feinberg}. As an example, considering the reactions $2A\rightarrow A+B, 2B\rightarrow A+B$, we have $m=3,\ell=1,s=1$. 

\begin{theorem}[Zero Deficiency for Complex-Balanced System\cite{horn1972necessary, feinberg1972complex,feinberg}]
	\label{zero-deficiency}
	A chemical mass-action system is complex-balanced for all values of its reaction rate constants if and only if it is weakly reversible and has zero deficiency.
\end{theorem}
It is noted that depending on values of the rate constants $k_i^{\pm}$, a chemical mass-action system with deficiency greater than zero may also be complex-balanced.

\begin{theorem}[Uniqueness of NESS for Complex-Balanced System\cite{horn1972general}]
	\label{uniqueness-complex-balance}
	Given a chemical mass-action system is a complex-balanced. Then all of the following properties hold: 
	\begin{enumerate}[(i)]
		\item all positive steady states are complex-balanced, and there is exactly one steady state within every stoichiometric compatibility class;
		\item every positive steady state is locally asymptotically stable within its stoichiometric compatibility class.
	\end{enumerate}
\end{theorem}
Theorems \ref{zero-deficiency} and \ref{uniqueness-complex-balance}  together establish the existence, uniqueness and local asymptotic stability of a complex-balanced steady-state solution to the chemical mass-action equations.

Instead of above topological definition, a necessary and sufficient algebraic condition for a chemical mass-action system to be complex-balanced is 
\begin{equation}
	\label{complex-balance}
	\sum_{i=1}^M\bigg[R_i^+(\vec{c}^{\,ss})-R_i^-(\vec{c}^{\,ss})\bigg]\bigg[\prod_{j=1}^N\bigg(\frac{c_j}{c_j^{ss}}\bigg)^{\nu_{ij}^+}-\prod_{j=1}^N\bigg(\frac{c_j}{c_j^{ss}}\bigg)^{\nu_{ij}^-}\bigg]=0,
\end{equation}
for any solution $\vec{c}$  to the chemical mass-action equations, where $\vec{c}^{\,ss}$ denotes the steady-state solution. The physical meaning of ``complex balance'' is clearly seen from an alternative formulation of the above equality:
\begin{equation*}
	\left\{\sum_{i=1}^M\Big[\delta(\vec{\xi}-\vec{\nu}_{i}^{\,+})-\delta(\vec{\xi}-\vec{\nu}_{i}^{\,-})\Big]\Big[R_i^+(\vec{c}^{\,ss})-R_i^-(\vec{c}^{\,ss})\Big]\right\}\prod_{j=1}^N\bigg(\frac{c_j}{c_j^{ss}}\bigg)^{\xi_j}=0,
\end{equation*}
which states a complex balanced steady state is the one when all the influx and the efflux of a complex $\{\xi_1X_1+\cdots\xi_NX_N\}$, representing by the term $\prod_{j=1}^N(c_j/c_j^{ss})^{\xi_j}$, are precisely balanced. A special case is the detailed balanced state when $R_i^+(\vec{c}^{\,eq})=R_i^-(\vec{c}^{\,eq})$ for all $i=1,\cdots,M$.

\subsection{Diffusion Process and Fokker-Planck Equation for Concentration\\ 
	at Mesoscale}

In previous sections, we have discussed two closely related, different-level descriptions of chemical reactions. One is a molecular level trajectory view $\vec{n}(t)$ which describes each reaction occurs in a purely stochastic way with an exponentially distributed time and the changes in molecular number are discrete one-by-one; the other is a macroscopic  ensemble view, in which the discrete molecular numbers are replaced by continuous concentration functions which proceed in a deterministic fashion. The microscopic description clearly contains far more detailed information than the macroscopic description. However, as the state space spanned by molecule numbers is discrete and high-dimensional, it brings great difficulties in numerical simulations and applications. Is it possible to construct a simple continuous formulation which still inherits  some of the stochastic nature of chemical reactions?  In doing so one could easily derive useful information on the fluctuations when the system's volume is relatively large but finite.

In physics and theoretical chemistry, this issue has been discussed in connection to chemical Langevin equations or chemical Fokker-Planck equations, which provide a mesoscopic description of chemical reactions that is in between the aforementioned two levels.  It is important to point out that this formulation is not a rigorous mathematical deduction from the molecular level theory. 
A rigorous mathematical treatment on this subject is the {\em central limit theorem} that was also provided in the Kurtz work.\cite{kurtz1981clt} 

When all components of molecule numbers $\vec{n}$ are large in a large volume $V$,  in a continuous view the number density $\vec{n}/V$,  {\em e.g.},  concentration,  becomes non-negative real valued $\vec{x}$ as $V\to\infty$.  Correspondingly, the joint probability $P(\vec{n}(t);V)$ can be replaced by a continuous probability density function $p(\vec{x},t)\rd x$, where we have changed the notation of $\vec{c}$ in the last section to $\vec{x}$ $\in\mathbb{R}^{N}_{\geq 0}$, purely for aesthetics.  By applying the Kramers-Moyal expansion or a more rigorous system-size expansion \cite{van1983} to the probability fluxes $r^{\pm}_i(\vec{n}\mp\vec{\nu}_i;V)P(\vec{n}\mp\vec{\nu}_i;V)$ in Eq. \eqref{chemical-master-eq} and neglecting terms with order $\sim\mathcal{O}(V^{-2})$, we arrive at the chemical Fokker-Planck equation,
\begin{eqnarray}
	\label{chemical-FP-eq}
	\frac{\partial}{\partial t}p(\vec{x},t)
	&=&\sum_{i=1}^{M}\bigg\{-\sum_{j=1}^{N} \nu_{ij}\frac{\partial}{\partial x_j}\big[R^+_i(\vec{x}) p(\vec{x},t)-R^-_i(\vec{x}) p(\vec{x},t)\big]\\
	&&\qquad + \frac{1}{2} \sum_{l,j=1}^{N} \nu_{il}\nu_{ij}\frac{\partial^2}{\partial x_l\partial x_j}\big[R^+_i(\vec{x}) p(\vec{x},t)+R^-_i(\vec{x}) p(\vec{x},t)\big]\bigg\}\nonumber\\
	&=&-\sum_{j=1}^N \frac{\partial}{\partial x_j}\big[u_j(\vec{x}) p(\vec{x},t)\big] + \sum_{\ell,j=1}^{N} \frac{\partial^2}{\partial x_{\ell}\partial x_j}\big[ D_{\ell j}(\vec{x})p(\vec{x},t)\big],\nonumber
\end{eqnarray}
which provides a mesoscopic description for chemical reactions with small diffusion.  The drift vector {$\vec{u}=(u_1, u_2, \cdots, u_N)^T$} and the diffusion matrix {$\mathbf{D}=(D_{\ell j})_{N \times N}$} are given by
\begin{equation}\label{uD}
	u_j(\vec{x})=\sum_{i=1}^{M}\nu_{ij}[R^+_i(\vec{x})-R^-_i(\vec{x})], \quad
	D_{\ell j}(\vec{x})=\frac{1}{2V}\sum_{i=1}^{M}\nu_{i\ell}\nu_{ij}[R^+_i(\vec{x})+R^-_i(\vec{x})],
\end{equation}
where $R^{\pm}_i(\vec{x})$ are continuous versions of the original forward/backward reaction rates $r^{\pm}_i(\vec{n};V)$ correspondingly.  For example, with respect to the law of mass-action, we have  
\[
    R^{\pm}_i(\vec{x})=k_i^{\pm}\prod_{j=1}^N x_j^{\nu_{ij}^{\pm}}
\]
The diffusion matrix $\mathbf{D}$ is symmetric,  positive definite,  and it's order $\sim\mathcal{O}(V^{-1})$.

The chemical Fokker-Planck equation can be rewritten into equivalent forms that are often used.
\begin{eqnarray}
	\frac{\partial p}{\partial t}&=&-\nabla\cdot\big[\vec{u}(\vec{x})p\big]+\nabla^2:\big[\mathbf{D}(\vec{x})p\big]\qquad \text{(divergence form)}\\
	&=&-\nabla\cdot\big[\hat{\vec{u}}(\vec{x})p\big]+\nabla\cdot\big[\mathbf{D}(\vec{x})\nabla p\big]\hspace{0.3in} \text{(symmetric divergence form)}\\
	&=&\mathbf{D}(\vec{x}):\nabla^2 p+\tilde{\vec{u}}(\vec{x})\cdot\nabla\cdot p+a(\vec{x})p,\hspace{0.1cm} \text{(non-divergence form)}
	\label{divergence-forms}
\end{eqnarray}
where $\hat{\vec{u}}(\vec{x})=\vec{u}(\vec{x})-\nabla\cdot\mathbf{D}(\vec{x})$, $\tilde{\vec{u}}(\vec{x})=-\vec{u}(\vec{x})+2\nabla\cdot\mathbf{D}(\vec{x})$ and $a(\vec{x})=-\nabla\cdot\vec{u}(\vec{x})+\nabla^2:\mathbf{D}(\vec{x})$. 

In particular, with respect to the divergence form, the Fokker-Planck equation can be cast into a continuity equation from continuum mechanics,
\begin{equation}
	\label{continuity-eq}
	\frac{\partial p}{\partial t}+\nabla\cdot \vec{J}(\vec{x},t)=0,
\end{equation}
where $\vec{J}(\vec{x},t)=\vec{u}(\vec{x})p-\nabla\cdot\big[\mathbf{D}(\vec{x})p\big]$ denotes the \textbf{\textit{probability flux}}. The continuity equation is directly related to conservation laws. Integrating Eq. \eqref{continuity-eq} over $\mathbb{R}^N$ and using the divergence theorem yields
\begin{equation*}
	\frac{\rd}{\rd t}\int_{\mathbb{R}^N}p(\vec{x},t)\rd\vec{x}=0.
\end{equation*}
Consequently, 
\begin{equation*}
	\int_{\mathbb{R}^N}p(\vec{x},t)\rd\vec{x}=\int_{\mathbb{R}^N}p(\vec{x},0)\rd\vec{x}=1,
\end{equation*}
which states the total probability is conserved.

Similar to the NESS of master equations,  the solution to $\nabla\cdot \vec{J}(\vec{x})=0$ but non-vanishing $\vec{J}(\vec{x})$ defines a \textbf{\textit{nonequilibrium steady state}} $p^{ss}(\vec{x})$ of the Fokker-Planck equation.  If the probability flux itself vanishes,
\begin{equation*}
	\vec{J}(\vec{x})=\vec{u}(\vec{x})p^{eq}(\vec{x})-\nabla\cdot\big[\mathbf{D}(\vec{x})p^{eq}(\vec{x})\big]=0,
\end{equation*}
instead of only being a divergence-free vector field,  then one arrives at the \textbf{\textit{equilibrium state}} $p^{eq}(\vec{x})$ with detailed balance.

As a special class of linear parabolic partial differential equations, establishing the existence,  uniqueness and regularity of classical solutions to the Fokker-Planck equation is far more challenging than that for the master equation.  Here we only present results for Cauchy problems under the assumption of uniform ellipticity \cite{friedman1964, pavliotis2016stochastic}. For results based on alternative weak conditions, or with respect to different boundary conditions, please see Refs. \cite{pavliotis2016stochastic, stroock2008partial,qian_qian_tang2002}.

\begin{theorem}[Existence and Uniqueness of Classical Solution to FP Equation \cite{friedman1964}]
	Consider the general Fokker–Planck equation in Eq. \eqref{divergence-forms} equipped with the initial condition $p(\vec{x},0)=p_0(\vec{x})$.  Assume the coefficients are smooth and satisfy the following conditions:
	\begin{enumerate}[(i)]
		\item \textbf{Uniform ellipticity assumption:} there exists a constant $\alpha>0$ such that $\vec{\xi}^{\,T} \cdot\mathbf{D}(\vec{x})\cdot\vec{\xi}\geq\alpha\|\vec{\xi}\|^2$, $\forall \xi\in\mathbb{R}^N$, uniformly in $\vec{x}\in\mathbb{R}^N$;
		\item \textbf{Growth conditions:} $\|\mathbf{D}\|\leq B$, $\|\tilde{\vec{u}}(\vec{x})\|\leq B(1+\|\vec{x}\|)$, $\|a(\vec{x})\|\leq B(1+\|\vec{x}\|)$;
		\item \textbf{Bounded initial condition:} $|p_0(\vec{x})|\leq ce^{\alpha\|\vec{x}\|^2}$.
	\end{enumerate} 
	Then there exists a unique classical solution $p(\vec{x},t)$ to the Cauchy problem for the Fokker–Planck equation.   Furthermore,
	\begin{enumerate}[(1)]
		\item $p(\vec{x},t)\in C^{2,1}(\mathbb{R}^N,\mathbb{R}^+)$, $\lim_{t\rightarrow0}p(\vec{x},t)=p_0(\vec{x})$, and $\forall T>0$, there exists a constant $c>0$ such that $\|p(\vec{x},t)\|_{L^{\infty}(0,T)}\leq ce^{\alpha\|\vec{x}\|^2}$; 
		\item $p(\vec{x},t)$ is non-negative and normalized for all times:  
\[
    p(\vec{x},t)\geq0  \text{ and } 
    \int_{\mathbb{R}^N}p(\vec{x},t) \rd\vec{x}=1
\]
provided $p_0(\vec{x})$  is non-negative and normalized;
		\item there exist positive constants $K$ and $b$, such that all  following four functions are bounded:
\[
    |p|,  \,  \left|\frac{\partial p}{\partial t}\right|,   \,  
    \|\nabla p\|,   \,  \|\nabla^2p\| \leq Kt^{(-N+2)/2}\exp\bigg(-\frac{b\|\vec{x}\|^2}{2t}\bigg).
\]
\end{enumerate} 
\end{theorem}
From the last statement, it follows that all moments of the diffusion process whose $\mathbf{D}$ satisfies the uniform ellipticity condition exist.  In particular, we can multiply both sides of a Fokker-Planck equation by monomials $x_i^n$,  integrate over $\mathbb{R}^N$,  and perform integration by parts as many times as needed.

\subsection{Chemical Langevin Equation for Concentration Fluctuations}

It is well-known that the Fokker-Planck equation characterizes the time changes in the probability distribution of random variables, whose corresponding governing equations are stochastic differential equations, the Langevin equation to be exact. However, given a chemical Fokker-Planck equation, the corresponding Langevin equation is generally non-unique, interested readers are referred to the work of Schnoerr et al. \cite{schnoerr2014complex} for a comparison of various chemical Langevin equations. Among them, one widely adopted choice originated from Gillespie \cite{gillespie2000chemical} reads
\begin{equation}
	\label{chemical-Langevin-eq}
	\rd x_j =\sum_{i=1}^{M} \nu_{ij}\bigg[\tilde{r}^+_i(\vec{x})-\tilde{r}^-_i(\vec{x})\bigg] \rd t + \sum_{i=1}^{M} \nu_{ij}\bigg[\sqrt{\tilde{r}^+_i(\vec{x})}\rd B_{2i-1}-\sqrt{\tilde{r}^-_i(\vec{x})}\rd B_{2i}\bigg],
\end{equation}
where $j=1,2,\cdots, N$. $\rd B_1,\cdots,\rd B_{2M}$ stand for $2M$ independent Wiener processes. Its solution reads
\begin{eqnarray}
	x_j(t)&=&x_j(0)+\int_{0}^t\sum_{i=1}^{M} \nu_{ij}\bigg[\tilde{r}^+_i(\vec{x}(\tau))-\tilde{r}^-_i(\vec{x}(\tau))\bigg] \rd\tau\\
	&& + \int_0^t\sum_{i=1}^{M} \nu_{ij}\bigg[\sqrt{\tilde{r}^+_i(\vec{x}(\tau))}\rd B_{2i-1}(\tau)-\sqrt{\tilde{r}^-_i(\vec{x}(\tau))}\rd B_{2i}(\tau)\bigg],\nonumber
\end{eqnarray}
where $j=1,2,\cdots, N$. The last integral is defined in the sense of Ito's calculas.

The chemical Langevin equation is equivalent to the chemical Fokker-Planck equation in the sense of counting an ensemble of stochastic trajectories. It is not hard to verify that the probability density function of Eq. \eqref{chemical-Langevin-eq} satisfies the chemical Fokker-Planck equation in Eq. \eqref{chemical-FP-eq} according to Ito's calculus. Thus the chemical Langevin equation provides an alternative effective method for modeling chemical systems when the system size is not so large \cite{gillespie2000chemical, schnoerr2014complex}.  Based on the above formula, it is straightforward to see that the strength of randomness is proportional to the root square of the microscopic propensity function or the reaction rates.

\subsection{Reaction-Diffusion Equations for Spatial Heterogeneity}
All of our previous discussions are restricted to spacial homogenous chemical reactions, \textit{e.g.} reactions happening in a rapidly stirring vessel.  \cite{epstein_book}  In order to evaluate the influence of spatial heterogeneity,  reaction with diffusion is the simplest starting point. \cite{qian_razo}

Let $\partial V$ be an arbitrary surface enclosing a volume $V$. The conservation law says that the rate of change of the
amount of materials in $V$ is equal to the net transport of materials across $\partial V$ into $V$ plus the materials created/eliminated inside $V$. Thus
\begin{equation*}
	\frac{\partial}{\partial t}\int_Vc(\vec{x},t)\rd\vec{x}=-\int_{\partial V}\vec{J}\cdot \rd\vec{n}+\int_Vf\rd\vec{x},
\end{equation*}
where the material efflux $\vec{J}$ and source term $f$ are functions of concentration $c$, even for a single chemical species, at position $\vec{x}$ and time $t$. $\vec{n}$ denotes the outward pointing normal vector at the surface.  Applying the divergence theorem to the surface integral and assuming $c(\vec{x},t)$ is continuous, the last equation becomes
\begin{equation*}
	\int_V\bigg[\frac{\partial c}{\partial t}+\nabla\cdot\vec{J}-f(c,\vec{x},t)\bigg]\rd\vec{x}=0.
\end{equation*}
Since the volume $V$ is arbitrary, the integrand must be zero,
\begin{equation}
	\frac{\partial c}{\partial t}+\nabla\cdot\vec{J}=f(c,\vec{x},t).
\end{equation}
This is the conservation equation for concentration $c$ and holds for an arbitrary flux function $\vec{J}$.

In particular, by taking the classical Fick's law for normal diffusion, \textit{i.e.} $\vec{J}=-\textbf{D}(\vec{x},t)\cdot\nabla c$, we arrive at the reaction-diffusion equation,
\begin{eqnarray}
	\frac{\partial c}{\partial t}=\nabla\cdot(\textbf{D}\cdot\nabla c)+f,
\end{eqnarray}
where $\textbf{D}(\vec{x},t)$ stands for the diffusion matrix, $f$ can be chosen in accordance with laws of mass-action. In practice, many important models in chemistry and biology belong to this class.  See \cite{murray2003mathematical} for an extensive coverage of the applications of reaction diffusion equation in mathematical biology.

The phenomena that can be generated by reaction-diffusion equations,  a class of PDE,  are far richer than those by ODEs.   From the travelling waves,  shocks,  solitons,  to Belousov-Zhabotinsky oscillation,  and Turing pattern, \textit{etc.},  have all been extensively studied in terms of reaction-diffusion equations.   The mathematical struc\-tures of reaction-diffusion equations are far more complicated than the equations we discussed above.   We shall not discuss further on reaction-diffusion PDEs in the current review.  We refer readers to monographs on this topic \cite{aris1991,murray2003mathematical, grzybowski2009chemistry, volpert2014elliptic}.

\section{NESS Thermodynamics for Open Chemical Reaction Systems}
{Now we turn to the thermodynamics of chemical reaction systems. By utilizing the theory of steady-state thermodynamics with the chemical master equations and chemical Fokker-Planck equations as two concrete examples, we can see that the Gibbs' thermodynamic description of chemical reactions is also fully respected at the mesoscopic level. This will be the main focus of this section.}

\subsection{Equilibrium Chemical
	Thermodynamics in a Nutshell and Beyond}

Classical mechanics idealizes objects, large as planets and small as atoms, in terms of point masses, and describes their motion $\vx(t)$, where $\vx=(\text{ positions},\text{momenta })$ in terms of a set of ordinary differential equations.  The motion of all mechanical particles are encapsulated by a Hamiltonian function $H(\vx)$, $\vx\in\Omega$. The function $H(\vx)$ and its domain $\Omega$ have hidden dependence on the number of particles $N$ and the space that contains all the particles $V$. According to Newtonian law of motion, when the initial value $E$ is given, the system is forever constrained on the energy level $H(\vx)=E$, for the eternity. 

In classic thermodynamics, a state is the entire energy level set.  Thermodynamic processes, through work and heat, relate the different level set of a mechanical system with a given Hamiltonian $H(\vx)$. The first law of thermodynamics simply says that the work $\mathcal{W}$ and heat $\mathcal{Q}$ in a process that connects two energy levels satisfy $\mathcal{W}+\mathcal{Q}=\Delta E$.  Since the ``changing energy'' is the extra dimension of thermodynamic processes, thermodynamics is also widely known as energetics.  One can also express the multi-dimensional function $H(\vx)$ in terms of a $\Gamma(E)$, $E\in\mathbb{R}$:
\begin{equation}
	\Gamma(E) = \frac{\rd}{\rd E}\int_{H(\vx)\le E}
	\rd\vx.
\end{equation}
$k_B\ln \Gamma$ is called Boltzmann's entropy; $k_B$ is Boltzmann's constant.

\begin{figure}[h]
	\centerline{
		\begin{picture}(450,170)(-55,-15)
			\put(40,110){Mechanics}
			\put(120,110){Thermodynamics}
			\put(230,115){Generalized}
			\put(230,103){Gibbs' theory}
			\put(-20,65){state}
			\put(-20,35){process}
			\put(-20,5){system}
			\put(40,65){$\vx:=(\vec{q},\vec{p})$}
			\put(120,72){$\big\{\vx: \vx\in\Omega(V,N),$} 
			\put(150,59){$H(\vx;N)=E\big\}$}
			\put(230,72){$H(\cdot):=\big\{H(\vx),H\in\mathbb{R},$} 
			\put(307,59){$\vx\in\Omega\big\}$}
			\put(40,40){Newtonian}
			\put(40,30){motion}
			\put(130,40){work, heat}
			\put(130,30){quasi-static}
			\put(230,35){chemical reactions}
			\put(40,5){$\big\{\vx: H(\vx)=E\big\}$}
			\put(120,10){$\big\{F(T,V,N)\in\mathbb{R}\big\}$}
			\put(120,-5){or $\big\{\Gamma(E,V,N)\in\mathbb{R}\big\}$}
			\put(230,10){$\big\{F[H(\cdot)], H(\cdot)\in\mathcal{H}(\Omega)\big\}$ or} 
			\put(240,-5){$\big\{F[\Gamma(\cdot)], \Gamma(\cdot)\in\mathcal{H}(\mathbb{R})\big\}$}
			\put(-20,95){\line(1,0){370}}
			\put(-20,93){\line(1,0){370}}
			\put(-20,-15){\line(1,0){370}}
			\put(25,-15){\line(0,1){140}}
		\end{picture}
	}
	\caption{A mechanical state is $\vx=(\vec{q},\vec{p})$. A Newtonian system is determined by the initial total energy $E$; thus the entire level set $\{\vx: H(\vx)=E\}$.  A thermodynamic state is the totality of a such level set represented by $E$, together with $V$ and $N$ that are implicit in $H(\vx)$ and $\Omega$. A macroscopic thermodynamic system then is characterized by a Helmholtz free energy function $F(T,V,N)$, where $T$ is temperature; or Boltzmann's thermodynamic probability $\Gamma(E,V,N)$.  State changes are characterized by work, heat, and quasi-static processes. In Gibbs' theory that bridges mechanics and macroscopic $F$, the latter is recognized as a functional of the entire function $H(\cdot)$. The collection of all possible $H(\cdot)$ is a function space $\mathcal{H}(\Omega)$.  Chemical reactions joint two $H(\cdot)$'s into one or decompose one $H(\cdot)$ into two. }
	\label{figure2}
\end{figure}

Gibbsian chemical thermodynamics, \cite{guggenheim1933} however, deals with systems having different number of mechanical particles.  A chemical reaction can transform two molecules, as mechanical systems, into one molecule. Gibbs' theory in principle, therefore, deals with all possible Hamiltonian functions. The collection of all $H(\vx)$ is a space of functions $\mathcal{H}(\Omega)$.

A thermodynamic potential is a function with proper independent variables. In classical thermodynamics, free energy $F(T)$ is a function of temperature $T$ and entropy $S$ is a function of internal energy, $S(E)$.  They are related via Legendre-Fenchel transforms \cite{qian2022jctc}
\begin{equation} 
	S(E) = \inf_{T} \left\{ 
	\frac{E-F(T)}{T}\right\}  
	\text{ and }\, 
	F(T) = \inf_{E}\Big\{ 
	E-TS(E) \Big\},
\end{equation}
in which $F(T)$ is a concave function.  Significantly generalizing Gibbs' thermodynamic method \cite{qian2022jctc} one can also consider $F$ as a functional of an entire Hamiltonian function $H(\vx)$, $F[H(\cdot)]$, then correspondingly one has
\begin{subequations}
	\begin{equation}
		S[P(\cdot)] = \inf_{H} \left\{ \int_{\Omega} H(\vx)P(\vx)\rd \vx - F[H(\cdot)] \right\} = -
		\int_{\Omega} P(\vx)\ln P(\vx)\rd \vx,
	\end{equation}
	and in return if $F[H(\cdot)]$ is concave,
	\begin{equation}
		F[H(\cdot)] = \inf_{P}\left\{ 
		\int_{\Omega} H(\vx)P(\vx)\rd \vx - S[P(\cdot)]
		\right\} = -\ln\int_{\Omega} 
		e^{-H(\vx)} \rd \vx. 
	\end{equation}
\end{subequations} 
The very probability distribution $P(\vx)$ itself and the Shannon entropy $S[P(\cdot)]$ naturally arise in a Gibbsian chemical thermodynamic analysis of free energy as a functional of $H(\vx)$.

Beyond the equilibrium thermodynamics, there are many schools that advanced nonequilibrium thermodynamics for irreversible processes, see Refs. \cite{hill1977,berry1984_finite_time,keizer1987statistical,ZhuHongYangYong2015,qian2016}, and references cited within.  In recent years, following the work of Sekimoto, Jarzynski, Crooks, Qian, Seifert, Esposito, \textit{et al.}, the basic mathematical framework of stochastic thermodynamics, for systems with significant fluctuations, has been gradually established. Especially for Markov process modeled by master equations and Wiener process characterized through Langevin equations and Fokker-Planck equations, comprehensive thermodynamic theory from both the ensemble aspect and single trajectory level have been well formulated and are readily applied to chemical reactions. This will be the major part of our following discussions.

According to the general theory of nonequilibrium thermodynamics\cite{de1962},  for each dissipative system an entropy function $S=S(t)$ could be introduced, whose temporal change satisfies a balance equation,
\begin{equation}
\label{prigogine_eq}
	\frac{\rd S}{\rd t}=\frac{\dbar_eS}{\rd t}+ \frac{\dbar_iS}{\rd t},
\end{equation}
where $\dbar$ denotes the fact that it is not a total derivative in general; the subscripts `$e$' and `$i$' 
stand for {\em exchange} and {\em internal production} respectively.  This equation represents the fact that can be summarized as follows  \cite{prigogine-book}:
\vskip 0.1cm
\begin{quote}
\textit{The instantaneous rate of entropy change can be decomposed into two parts -- the rate of entropy exchange $\dbar_eS/\rd t$ and the rate of entropy production $\dbar_iS/\rd t$.   The entropy production rate is always non-negative, $\dbar_iS/\rd t$ $\geq 0$,  expressing the Second Law of Thermodynamics.   It is zero if and only if the system attains its equilibrium steady state.  In a nonequilibrium steady state $\rd S/\rd t=0$ while $\dbar_eS/\rd t=-\,\dbar_iS/\rd t<0$ representing steady dissipation.
}
\end{quote}
 \vskip 0.1cm\noindent
We shall now present,  separately,  concrete and detail mathematical formalisms of this  thermodynamics based on master equations and Fokker-Planck equations.

\subsection{Chemical Potential,  Affinity,  and Gibbs Free Energy for Nonequi\-librium States}
Since we are mainly dealing with homogenous chemical reaction mixtures in ideal dilute solutions, \textit{\textbf{the assumption of local equilibrium}}\cite{de1962} applies when chemical reactions happen in a much slower time scale than thermal equilibration: Chemical reactions are usually studied under isothermal,  isobaric condition.  As a result,  the intensive thermodynamic quantities,   temperature $T$ and pressure $p$ are set by the environment to which the solution is exposed. Then according to the \textit{\textbf{Gibbs relation}}, we have
\begin{eqnarray}
	\rd G=-S\rd T+V\rd p+\sum_{j=1}^N\mu_j\rd n_j,
\end{eqnarray}
where $G(T,p,n_1,\cdots,n_N)$ is the Gibbs free energy function, $n_j$ represents the molecular number of species $j$. 

The above Gibbs relation yields $\mu_j=(\partial G/\partial n_j)_{T,p}$, which is called the \textit{\textbf{chemical potential}} of species $j$. It is convenient to rewrite the chemical potential as a function of species concentration  ($c_j=n_j/V$) in the following way\cite{kondepudi2014modern}
\begin{eqnarray}
\label{chem-potential}
	\mu_j(c_j)=\mu_j^0+k_BT\ln a_j(c_j).
\end{eqnarray}
$\mu_j^0$ is called the standard chemical potential, $a_j$ is called the activity, which is related to concentration of species, as well as the system temperature and pressure. In general, the activity of a species in ideal solution is given by $a_j=\gamma_jc_j/c^0$, where $\gamma_j=\gamma_j(c_j)$ is the activity coefficient of species $j$. $c^0$ is the standard concentration (e.g. 1M) to make the activity be dimensionless. However, in the literature, people often takes a much simpler form $$\mu_j(c_j)=\mu_j^0+k_BT\ln c_j$$ for the chemical potential.

By its definition, the chemical potential measures the Gibbs free energy difference per infinitesimal number change of a chemical species. In the same way, the change in Gibbs free energy per single chemical reaction defines \textit{\textbf{the affinity}}, which is related to the chemical potential as
\begin{eqnarray}
	A_i(\vec{c}):=-\Delta_iG(\vec{c})=\sum_{j=1}^N(\nu_{ij}^+-\nu_{ij}^-)\mu_j(c_j),\quad  1\le i\le M.
\end{eqnarray}
$A_i$ stands for the affinity for the $i$th reaction, and $\Delta_iG(\vec{c})$ is the difference in Gibbs free energy before and after the reaction. Clearly, the affinity reflects the occurring tendency of a chemical reaction, which should be zero at the equilibrium ($A_i(\vec{c}^{\,eq})=0$). 

With respect to the form of chemical potential $\mu_j(c_j)=\mu_j^0+k_BT\ln c_j$ and laws of mass-action, we can reach an explicit formula for the affinity:
\begin{eqnarray}
	A_i(\vec{c})=k_BT\ln\bigg[\frac{R_i^+(\vec{c})}{R_i^-(\vec{c})}\bigg],\quad 1\le i\le M,
\end{eqnarray}
as the standard chemical potential is related to the reaction rate constants through
\begin{eqnarray*}
	k_BT\ln\bigg[\frac{k_i^+}{k_i^-}\bigg]=\sum_{j=1}^N(\nu_{ij}^+-\nu_{ij}^-)\mu_j^0,\quad 1\le i\le M.
\end{eqnarray*}
Meanwhile, the Gibbs free energy per unit volume of ideal dilute solutions is given by
\begin{eqnarray}
	G(\vec{c})=\sum_{j=1}^N \mu_j(c_j)c_j-k_BT\sum_{j=1}^{N}c_j+G^0,
\end{eqnarray}
in which the last two terms are associated with the contribution of solvent\cite{rao2016nonequilibrium}. It is straightforward to show that the Gibbs free energy possesses two significant mathematical properties as follows:
\begin{enumerate}[(1)]
	\item The Gibbs free energy is a monotonically decreasing function of time, i.e. $\rd G/\rd t\leq 0$;
	\item The Gibbs free energy reaches its minimum at the equilibrium, i.e. $G(\vec{c}(t))\geq G(\vec{c}^{\,eq})$.
\end{enumerate} 
\begin{proof} 
	(1) \begin{eqnarray*}
	\frac{\rd G}{\rd t}&=&\sum_{j=1}^N(\mu_j^0+k_BT\ln c_j)\frac{\rd c_j}{\rd t}=\sum_{j=1}^N(\mu_j^0+k_BT\ln c_j)\sum_{i=1}^M(\nu_{ij}^--\nu_{ij}^+)[R^+_i(\vec{c})-R^-_i(\vec{c})]\\
	&=&-\sum_{i=1}^MA_i(\vec{c})[R^+_i(\vec{c})-R^-_i(\vec{c})]=-k_BT\sum_{i=1}^M[R^+_i(\vec{c})-R^-_i(\vec{c})]\ln\bigg[\frac{R_i^+(\vec{c})}{R_i^-(\vec{c})}\bigg]\leq0.
\end{eqnarray*}

(2) \begin{eqnarray*}
	G(\vec{c}(t))-G(\vec{c}^{\,eq})&=&\sum_{j=1}^N \big[\mu_j(c_j)c_j-\mu_j(c_j^{eq})c_j^{eq}\big]-k_BT\sum_{j=1}^{N}(c_j-c_j^{eq})\\
	&=&\sum_{j=1}^N \mu_j(c_j^{eq})(c_j-c_j^{eq})+k_BT\sum_{j=1}^N\bigg[c_j\ln\bigg(\frac{c_j}{c_j^{eq}}\bigg)-(c_j-c_j^{eq})\bigg]\\
	&=&k_BT\sum_{j=1}^N\bigg[c_j\ln\bigg(\frac{c_j}{c_j^{eq}}\bigg)-(c_j-c_j^{eq})\bigg]\geq0.
\end{eqnarray*}
The last equality holds true by noticing $\sum_{j=1}^N \mu_j(c_j^{eq})(c_j-c_j^{eq})=0$. Due to the fact $A_i(\vec{c}^{\,eq})=0$, $\mu_j(c_j^{eq})$ must lie inside the linear space of conservation laws of chemical reactions, which therefore means the combination $\sum_{j=1}^N \mu_j(c_j^{eq})c_j(t)$ will remain unchanged during reactions and equal to $\sum_{j=1}^N \mu_j(c_j^{eq})c_j^{eq}$ for all the time.
\end{proof}

\subsection{Thermodynamics for General Discrete-State Markov Processes}
\label{sec:3.3}
We start with the entropy function, according to C. E. Shannon, for a general discrete-state Markov process characterized by the master equation given by Eq. \eqref{master-equation}:
\begin{equation}
	S(t)=S[P_n(t)]=-\sum_n P_n(t) \ln P_n(t).
\end{equation}
Then the instantaneous rate of entropy change at time $t$ reads
\begin{eqnarray*}
	&&\frac{\rd S(t)}{\rd t} = -\sum_n\frac{\rd P_n(t)}{\rd t}\ln P_n(t)=-\sum_{n,n'}(W_{n,n'}P_{n'}-W_{n',n}P_n)\ln P_n
 \\
	&&=\frac{1}{2}\sum_{n,n'}(W_{n,n'}P_{n'}-W_{n',n}P_n)\ln \bigg(\frac{P_{n'}}{P_n}\bigg)
	=-\frac{1}{2}\sum_{n,n'}(W_{n,n'}P_{n'}-W_{n',n}P_n)\ln \bigg(\frac{W_{n,n'}}{W_{n',n}}\bigg)
 \\
	&&\hspace{.4cm}+\,\frac{1}{2} \sum_{n,n'}(W_{n,n'}P_{n'}-W_{n',n}P_n)\ln \bigg(\frac{W_{n,n'}P_{n'}}{W_{n',n}P_n}\bigg).
\end{eqnarray*}
One immediately recognizes that the first logarithmic term has a uncanny resemblance to the standard chemical potential change $\Delta \mu^0$ of a unimolecular reaction, while the second logarithmic term remarkably matches the chemical potential change $\Delta\mu$ of the reaction (see Eq. \eqref{chem-potential}).  There is no longer any doubt that discrete-state continuous-time Markov process is the rigorous mathematical representation of chemical kinetics \cite{beard_qian_book,qian2021stochastic}. 

Through above analogy, the physical meanings of the two terms on the right-hand side become obvious. The first term is the entropy exchange rate with the environment. In the theory of nonequilibrium thermodynamics, it is often related to the heat exchange rate divided by temperature. Since the notion of temperature does not exist in the theory of Markov dynamics, we just simply set $T=1$, meaning our system/process is an isothermal one. While the second term is the instantaneous entropy production rate.
\begin{eqnarray}
	\label{entropy-balance-master}
	\frac{\dbar_eS}{\rd t}&=&\frac{1}{2}\sum_{n,n'}(W_{n,n'}P_{n'}-W_{n',n}P_n)\ln \bigg(\frac{W_{n',n}}{W_{n,n'}}\bigg),
	\\ 
	\frac{\dbar_iS}{\rd t}&=&\frac{1}{2}\sum_{n,n'}(W_{n,n'}P_{n'}-W_{n',n}P_n)\ln \bigg(\frac{W_{n,n'}P_{n'}}{W_{n',n}P_n}\bigg)\geq0.
\end{eqnarray}
It is easily seen that the entropy production rate is always non-negative, which equals to zero if and only if the system attains the equilibrium state ($W_{n,n'}P_{n'}=W_{n',n}P_n$ for all pairs of $n$ and $n'$). Thus within the framework of master equations, the second law of thermodynamics is respected. Furthermore, according to Onsager's theory of classical irreversible thermodynamics\cite{de1962}, \textbf{\textit{the total entropy production rate can be decomposed into the sum of many terms in the form ``nonequilibrium flux $\times$ nonequilibrium force'' representing different irreversible processes}}. This suggests that we can identify $J_{n,n'}=W_{n,n'}P_{n'}-W_{n',n}P_n$ as a flux and $\ln [(W_{n,n'}P_{n'})/(W_{n',n}P_n)]$ as the corresponding force, or affinity\cite{ge2010physical}.

According to the famous Kullback–Leibler divergence \cite{kullback1951information}, a relative entropy function (or free energy) can be introduced for master equations
\begin{equation}
	F(t)=F[P_n(t)|P_n^{ss}]=\sum_{n} P_n(t) \ln \bigg(\frac{P_n(t)}{P_n^{ss}}\bigg).
\end{equation}
\textbf{\textit{The relative entropy function enjoys two significant mathematical properties -- non-negativity, $F(t)\geq0$, and temporal monotonicity, $\rd F(t)/\rd t\leq0$. }} Therefore, according to the Lyapunov stability theorem, the Markov process characterized by master equations is \textit{\textbf{time irreversible}} and the NESS is globally asymptotically stable with respect to all possible initial conditions.
\begin{eqnarray*}
	F(t)&=&-\sum_{n} P_n(t) \ln \bigg(\frac{P_n^{ss}}{P_n(t)}\bigg)\geq-\sum_{n} P_n(t)\bigg(\frac{P_n^{ss}}{P_n(t)}-1\bigg)
    \\
    &=& \sum_{n} \Big[P_n^{ss}-P_n(t)\Big]=0,
	\\
	\frac{\rd F}{\rd t}&=&\sum_{n} \frac{\rd P_n(t)}{\rd t} \ln \bigg(\frac{P_n(t)}{P_n^{ss}}\bigg)=\sum_{n,n'}[W_{n,n'}P_{n'}(t)-W_{n',n}P_n(t)]\ln \bigg(\frac{P_n(t)}{P_n^{ss}}\bigg)\\
	&=&\frac{1}{2}\sum_{n,n'}[W_{n,n'}P_{n'}(t)-W_{n',n}P_n(t)]\ln \bigg(\frac{P_n(t)P_{n'}^{ss}}{P_{n'}(t)P_n^{ss}}\bigg)\\
	&=&\sum_{n,n'}W_{n,n'}P_{n'}(t)\ln \bigg(\frac{P_n(t)P_{n'}^{ss}}{P_{n'}(t)P_n^{ss}}\bigg)
	\leq\sum_{n,n'}W_{n,n'}P_{n'}(t)\bigg(\frac{P_n(t)P_{n'}^{ss}}{P_{n'}(t)P_n^{ss}}-1\bigg)\\
	&=&\sum_{n}\bigg(\frac{P_n(t)}{P_n^{ss}}\bigg)\sum_{n'} W_{n,n'}P_{n'}^{ss}-\sum_{n,n'}W_{n,n'}P_{n'}(t)\\
	&=&\sum_{n}\bigg(\frac{P_n(t)}{P_n^{ss}}\bigg)\sum_{n'} W_{n',n}P_n^{ss}-\sum_{n,n'}W_{n,n'}P_{n'}(t)=-\sum_{n}\frac{\rd P_n(t)}{\rd t}=0.
\end{eqnarray*}
During the derivation, we use the logarithmic inequality $\ln x\leq x-1$. Clearly, the relative entropy function attains its global minimum at the NESS. 

Furthermore, the relative entropy $F(t)$ satisfies an balance equation of its own, i.e.
\begin{eqnarray*}
	\frac{\rd F}{\rd t}&=&\frac{1}{2}\sum_{n,n'}[W_{n,n'}P_{n'}(t)-W_{n',n}P_n(t)]\ln \bigg(\frac{P_n(t)P_{n'}^{ss}}{P_{n'}(t)P_n^{ss}}\bigg)\\
	&=&\frac{1}{2}\sum_{n,n'}[W_{n,n'}P_{n'}(t)-W_{n',n}P_n(t)]\ln \bigg(\frac{W_{n,n'}P_{n'}^{ss}}{W_{n',n}P_n^{ss}}\bigg)\\
	&&-\,\frac{1}{2}\sum_{n,n'}[W_{n,n'}P_{n'}(t)-W_{n',n}P_n(t)]\ln \bigg(\frac{W_{n,n'}P_{n'}}{W_{n',n}P_n}\bigg).
\end{eqnarray*}
Actually, the first term in the second line has a definite sign:
\begin{eqnarray*}
	&&\frac{1}{2}\sum_{n,n'}[W_{n,n'}P_{n'}(t)-W_{n',n}P_n(t)]\ln \bigg(\frac{W_{n,n'}P_{n'}^{ss}}{W_{n',n}P_n^{ss}}\bigg)\\
    &=&-\sum_{n,n'}W_{n',n}P_n(t)\ln \bigg(\frac{W_{n,n'}P_{n'}^{ss}}{W_{n',n}P_n^{ss}}\bigg)
	\geq\sum_{n,n'}W_{n',n}P_n(t)\bigg(1-\frac{W_{n,n'}P_{n'}^{ss}}{W_{n',n}P_n^{ss}}\bigg)\\
    &=&\sum_{n,n'}W_{n',n}P_n(t)-\sum_{n}\bigg(\frac{P_n(t)}{P_n^{ss}}\bigg)\sum_{n'}W_{n,n'}P_{n'}^{ss}=0.
\end{eqnarray*}
There are also many other nonequilibrium thermodynamic quantities:
\begin{eqnarray}
    \hspace{.3in}
	\dot{E}_{hk}(t) &=&\frac{1}{2}\sum_{n,n'}
 \Big[ W_{n,n'}P_{n'}(t)-W_{n',n}P_n(t) \Big] \ln \bigg(\frac{W_{n,n'}P_{n'}^{ss}}{W_{n',n}P_n^{ss}}\bigg)\geq0,
\\
	  \frac{\dbar S_{ex}}{\rd t} &=&\frac{1}{2}\sum_{n,n'}\Big[W_{n,n'}P_{n'}(t)-W_{n',n}P_n(t)\Big]\ln \bigg(\frac{P_{n'}(t)P_n^{ss}}{P_n(t)P_{n'}^{ss}}\bigg)=-\frac{\rd F}{\rd t}\geq0,
\\
	\frac{\dbar S_{env}}{\rd t} &=&\frac{1}{2}\sum_{n,n'}\Big[W_{n,n'}P_{n'}(t)-W_{n',n}P_n(t)\Big]\ln \left(\frac{P_{n'}^{ss}}{P_n^{ss}}\right) = \frac{\rd S_{env}}{\rd t},
\end{eqnarray}
where $S_{env}(t)\equiv -\sum_{n}P_n(t)\ln P_n^{ss}$. They are called house-keeping energy input rate, $\dot{E}_{hk}$, excessive entropy production rate, $\dbar S_{ex}/\rd t$ and environment entropy change rate $\dbar S_{env}/\rd t$.  The excessive entropy production rate bears its name due to the fact that $\dbar S_{ex}/\rd t=0$ in the steady-state $P^{ss}_n$.
It has been also called free energy dissipation \cite{ge2010physical}. $\dot{E}_{hk}$ is clearly zero for all $t$ if detailed balance holds true; while the $\rd F/\rd t$ becomes zero if and only if at a steady state. These fine distinctions enable one to make a clear distinction between the NESS and the equilibrium state.

With these notions introduced above, we see that the following set of relations hold for the master equations,
\begin{eqnarray}
	\frac{\rd S}{\rd t}-\frac{\dbar_eS }{\rd t}&=&\frac{\dbar_iS}{\rd t}\geq0,\\
	\frac{\rd S}{\rd t}-\frac{\rd S_{env}}{\rd t}&=&-\frac{\rd F}{\rd t}\geq0,\\
	\frac{\rd S_{env}}{\rd t}-\frac{\dbar_eS}{\rd t}&=&\dot{E}_{hk}\geq0.
\end{eqnarray}
Esposito and van den Broeck called them \textit{\textbf{``three faces of the second law of thermodynamics''}}\cite{esposito2010three-1}. More importantly, we discover a cogent decomposition of the total entropy production rate into two non-negative sub-parts -- the house-keeping energy as the contribution from a driving force and free energy dissipation in connection to transient relaxation kinetics, 
\begin{equation}
	\frac{\dbar_iS}{\rd t}=\dot{E}_{hk}+\left(-\frac{\rd F}{\rd t}\right).
\label{3faces}
\end{equation}
The same conclusions have been obtained for the Fokker-Planck equation as well.\cite{van2010three}
When there is no active driving force, $\dot{E}_{hk}=0$, the rate of decreasing free energy is the total entropy production.  In an NESS, $\rd F/\rd t=0$ but not the other two in (\ref{3faces}); in an equilibrium all three are zero. 

\begin{remark}
Since we have shown that $\dbar_{ex}S/\rd t=-\rd F/\rd t$ and $\dbar_{env}S/\rd t=\rd S_{env}/\rd t$ with 
\[
    F(t) \equiv \sum_{n}P_n(t)\ln\left(\frac{P_n(t)}{P_n^{ss}}\right) \text{ and }
    S_{env}(t) \equiv -\sum_{n} P_n(t)\ln P_n^{ss},
\]
both are the total derivatives of functions w.r.t. $t$.  We shall use the more informative notations.   
\end{remark}

The non-driven system that satisfies the condition of detailed balance plays an important role in our thermodynamic theory of Markov process.  In this case, $F(t)$ and $S_{env}(t)$ become the free energy and internal energy in classical thermodynamics.  It has been proven that the detailed balance can be characterized by several rather difference statements.
\begin{theorem}[Condition of Detailed Balance\cite{jiang2004mathematical}]
	Assume an irreducible finite-state Markov process (characterized by master equations) with transition rate matrix $W=(W_{n,n'})$ and a stationary distribution $\{P_n^{ss}\}$. Then the following six statements are equivalent.
	\begin{enumerate}[(i)]
		\item The stationary distribution satisfies the condition of detailed balance, $W_{n,n'}P_{n'}^{ss}$ $=W_{n',n}P_n^{ss}$, $\forall n,n'=1,\cdots,L$;
		\item For any path connecting states $n$ and $n'$, $n=n_0,n_1,\cdots, n_k=n'$, a path independent potential function can be introduced, $$\ln\bigg(\frac{W_{n_0,n_1}W_{n_1,n_2}\cdots W_{n_{k-1},n_k}}{W_{n_k,n_{k-1}}\cdots W_{n_2,n_1}W_{n_1,n_0}}\bigg)=\ln P_{n_k}^{ss}-\ln P_{n_0}^{ss};$$
		\item It is a time-reversible, stationary Markov process;
		\item The matrix $W$ satisfies the Kolmogorov cycle condition, that is $$W_{n_0,n_1}W_{n_1,n_2}\cdots W_{n_{k-1},n_k}W_{n_k,n_0}=W_{n_0,n_k}W_{n_k,n_{k-1}}\cdots W_{n_2,n_1}W_{n_1,n_0}$$ for every sequence of distinct states $n_0,n_1,\cdots,n_k=1,\cdots,L$;
		\item There exists a positive diagonal matrix $\Pi$ such that matrix $W\Pi$ is symmetric;
		\item The stationary process has a zero entropy production rate, $\dbar_i S/\rd t=0$.
	\end{enumerate}
\end{theorem}

To reconstruct the first law of thermodynamics for master equations, let us suppose the condition of detailed balance holds. Now, according to above theorem, the transition rates $W_{n,n'}(\lambda_t)$, which can be time dependent via a slowly varying control variable $\lambda_t$ applied at time $t$ (much slower than the relaxation time of the corresponding Markov process), follow the canonical distribution:
\begin{equation}
	\frac{W_{n,n'}(\lambda_t)}{W_{n',n}(\lambda_t)}=\exp\{-\beta[\epsilon_n(\lambda_t)-\epsilon_{n'}(\lambda_t)]\},
\end{equation} 
where $\beta$ stands for the inverse temperature, and $\epsilon_n(\lambda_t)=-\beta^{-1}\ln P_n^{eq}(\lambda_t)$ is the energy of the system in state $n$ with control variable $\lambda_t$. Inserting this relation into the formula of entropy exchange rate, we obtain the heat exchange rate of the system
\begin{equation}
	\label{heat-ensemble-master}
	\frac{\dbar Q}{\rd t}=\beta^{-1}\frac{\dbar_e S}{\rd t}=\sum_{n,n'}(W_{n,n'}P_{n'}-W_{n',n}P_n)\epsilon_n(\lambda_t).
\end{equation}
Furthermore, since the system energy is obviously given by
\begin{equation}
	E(t)=\sum_{n}\epsilon_n(\lambda_t)P_n(t),
\end{equation}
we arrive at the first law of thermodynamics (or the law of energy conservation)
\begin{equation}
	\label{1st-law-ensemble-master}
	\frac{\rd E}{\rd t}=\frac{\dbar Q}{\rd t}+\frac{\dbar W}{\rd t},
\end{equation}
where the work rate is given by
\begin{equation}
	\label{work-ensemble-master}
	\frac{\dbar W}{\rd t}=\sum_{n}\frac{\rd\epsilon_n(\lambda_t)}{\rd t}P_n(t).
\end{equation}

\subsection{Nonequilibrium Thermodynamics for Continuous-Space Diffusion Processes}
Let us introduce the symbol for the space of $\vec{x}$ and use it for all the integrals below.
In this part, we are going to show all results obtained for master equations can be extended to the Fokker-Planck equation in parallel. With respect to the Gibbs entropy function in a continuous version,
\begin{equation}
	S(t)=S[p(\vec{x},t)]=-\int_{\mathbb{R}^N} \rd\vec{x}\, p(\vec{x},t)\ln p(\vec{x},t).
\end{equation}
where $p(\vec{x},t)$ denotes solutions to the Fokker-Planck equation in Eq. \eqref{chemical-FP-eq}. The entropy change rate is given by
\begin{eqnarray*}
	\frac{\rd S}{\rd t}&=&-\int_{\mathbb{R}^N}  \rd\vec{x}\left(\frac{\partial p(\vec{x},t)}{\partial t}\right)\ln p(\vec{x},t)
	=\int_{\mathbb{R}^N} \rd\vec{x} \bigg\{\nabla\cdot[\hat{\vec{u}}(\vec{x})p-\mathbf{D}(\vec{x})\cdot\nabla p]\bigg\}\ln p(\vec{x},t)\\
	&=&-\int \rd\vec{x}\frac{\vec{J}(\vec{x},t)\cdot\nabla p}{p}
	= \int \rd\vec{x}\bigg[\frac{\vec{J}(\vec{x},t)\cdot\mathbf{D}(\vec{x})^{-1}\cdot\vec{J}(\vec{x},t)}{p(\vec{x},t)}-\vec{J}(\vec{x},t)\cdot\mathbf{D}(\vec{x})^{-1}\cdot\hat{\vec{u}}(\vec{x})\bigg], 
\end{eqnarray*}
where the nonequilibrium flux $\vec{J}(\vec{x},t)=\vec{u}(\vec{x})p-\nabla\cdot[\mathbf{D}(\vec{x})p]=\hat{\vec{u}}(\vec{x})p-\mathbf{D}(\vec{x})\cdot\nabla p$.  The entropy exchange rate and the entropy production rate in Eq.  (\ref{prigogine_eq}) then are identified as 
\begin{subequations}
\label{entropy-balance-fp}
\begin{eqnarray}
	\frac{\dbar_eS}{\rd t}&=&-\int \rd\vec{x}\,\vec{J}(\vec{x},t)\cdot\mathbf{D}(\vec{x})^{-1}\cdot\hat{\vec{u}}(\vec{x}),\\ 
	\frac{\dbar_iS}{\rd t}&=&\int \rd\vec{x}\, \frac{\vec{J}(\vec{x},t)\cdot\mathbf{D}(\vec{x})^{-1}\cdot\vec{J}(\vec{x},t)}{p(\vec{x},t)}\geq0.
\end{eqnarray}
\end{subequations}
Due to the positive-definite nature of the diffusion matrix $\mathbf{D}(\vec{x})>0$, the entropy production rate is always non-negative, which equals to zero if and only if the system attains the equilibrium state with $\vec{J}(\vec{x},t)={\bf 0}$.  This is in complete agreement with the Second Law.   Furthermore, according to Onsager's theory, we identify $\vec{J}(\vec{x},t)=\vec{u}(\vec{x})p-\nabla\cdot[\mathbf{D}(\vec{x})p]$ as a flux and $\mathbf{D}(\vec{x})^{-1}\cdot\vec{J}(\vec{x},t)/p(\vec{x},t)$ as the corresponding force.

For the Fokker-Planck equation, the non-negativity and monotonicity of the following free energy function
\begin{equation}
	F(t)=F[p(\vec{x},t)|p^{ss}(\vec{x})]=\int \rd\vec{x}\, p(\vec{x},t)\ln \bigg(\frac{p(\vec{x},t)}{p^{ss}(\vec{x})}\bigg)
\end{equation}
could be verified. Clearly, the minimal free energy is obtained at the NESS. This reveals that it is indeed the relative entropy function (or Kullback-Leibler divergence) for the Fokker-Planck equation.
\begin{equation*}
	F(t)=-\int \rd\vec{x}\, p(\vec{x},t)\ln \bigg(\frac{p^{ss}(\vec{x})}{p(\vec{x},t)}\bigg)\geq -\int \rd\vec{x}\, p(\vec{x},t)\bigg(\frac{p^{ss}(\vec{x})}{p(\vec{x},t)}-1\bigg)=0,
\end{equation*}
and
\begin{eqnarray*}
	\frac{\rd F}{\rd t}&=&\int \rd\vec{x}
	\, \frac{\partial p(\vec{x},t)}{\partial t}\ln \bigg(\frac{p(\vec{x},t)}{p^{ss}(\vec{x})}\bigg)=-\int \rd\vec{x}\,  \nabla\cdot \vec{J}(\vec{x},t)\ln \bigg(\frac{p(\vec{x},t)}{p^{ss}(\vec{x})}\bigg)\\
	&=&\int \rd\vec{x}\,  \vec{J}(\vec{x},t)\cdot\nabla\ln \bigg(\frac{p(\vec{x},t)}{p^{ss}(\vec{x})}\bigg)=\int \rd\vec{x}\,  \vec{J}(\vec{x},t)\cdot\bigg(\frac{\nabla p(\vec{x},t)}{p(\vec{x},t)}-\frac{\nabla p^{ss}(\vec{x})}{p^{ss}(\vec{x})}\bigg)
 \\
	&=&-\int \rd\vec{x}\,  \vec{J}(\vec{x},t)\cdot\mathbf{D}(\vec{x})^{-1}\cdot\bigg(\frac{\vec{J}(\vec{x},t)}{p(\vec{x},t)}-\frac{\vec{J}^{ss}(\vec{x})}{p^{ss}(\vec{x})}\bigg)\\
	&=&-\int \rd\vec{x}\,  p(\vec{x},t)\bigg(\frac{\vec{J}(\vec{x},t)}{p(\vec{x},t)}-\frac{\vec{J}^{ss}(\vec{x})}{p^{ss}(\vec{x})}\bigg)\cdot\mathbf{D}(\vec{x})^{-1}\cdot\bigg(\frac{\vec{J}(\vec{x},t)}{p(\vec{x},t)}-\frac{\vec{J}^{ss}(\vec{x})}{p^{ss}(\vec{x})}\bigg)\leq0.
\end{eqnarray*}
During above derivation, we use integration by part and assume that the boundary term is zero (\textit{e.g.} infinite boundary condition). The following fact is also essential:
\begin{eqnarray*}
	&& \int \rd\vec{x}\, p(\vec{x},t) \frac{\vec{J}^{ss}(\vec{x})}{p^{ss}(\vec{x})}\cdot\mathbf{D}(\vec{x})^{-1}\cdot\bigg(\frac{\vec{J}(\vec{x},t)}{p(\vec{x},t)}-\frac{\vec{J}^{ss}(\vec{x})}{p^{ss}(\vec{x})}\bigg)
	\\
	&=& -\int \rd\vec{x}\,  \vec{J}^{ss}(\vec{x})\cdot\nabla\bigg(\frac{p(\vec{x},t)}{p^{ss}(\vec{x})}\bigg)=0.
\end{eqnarray*}

Further define the house-keeping energy input rate, excessive entropy production rate and environment entropy change rate as 
\begin{eqnarray}
	\dot{E}_{hk} &=&\int \rd\vec{x}\,  p(\vec{x},t)\bigg(\frac{\vec{J}^{ss}(\vec{x})}{p^{ss}(\vec{x})}\bigg)\cdot\mathbf{D}(\vec{x})^{-1}\cdot\bigg(\frac{\vec{J}^{ss}(\vec{x})}{p^{ss}(\vec{x})}\bigg)\geq0,
\label{329}\\
	-\frac{\rd F}{\rd t}&=&\int \rd\vec{x}\,  p(\vec{x},t)\bigg(\frac{\vec{J}(\vec{x},t)}{p(\vec{x},t)}-\frac{\vec{J}^{ss}(\vec{x})}{p^{ss}(\vec{x})}\bigg)\cdot\mathbf{D}(\vec{x})^{-1}\cdot\bigg(\frac{\vec{J}(\vec{x},t)}{p(\vec{x},t)}-\frac{\vec{J}^{ss}(\vec{x})}{p^{ss}(\vec{x})}\bigg)\geq0\\
    \hspace{.2in}
	\frac{\rd S_{env}}{\rd t}&=&-\int \rd\vec{x} \vec{J}(\vec{x},t)\cdot\frac{\nabla p^{ss}(\vec{x})}{p^{ss}(\vec{x})}.
 \label{331}
\end{eqnarray}
Similar to results for master equations, the house-keeping energy input rate becomes zero if the condition of detailed balance holds true; while the rate of decreasing free energy is zero if and only if at the NESS. More importantly, it is straightforward to verify that the sum of the rates of house-keeping energy and free energy dissipation gives the total entropy production rate, meaning the latter can be decomposed into two non-negative parts. 
\begin{equation*}
	\frac{\dbar_iS}{\rd t}=\dot{E}_{hk}+\left(-\frac{\rd F}{\rd t}\right).
\end{equation*}
With respect to above definitions, we see that the following relations hold for the Fokker-Planck equation,
\begin{eqnarray*}
	\frac{\rd S}{\rd t}-\frac{\dbar_e S }{\rd t}&=&\frac{\dbar_iS}{\rd t}\geq0,\\
	\frac{\rd S}{\rd t}-\frac{\rd S_{env}}{\rd t}&=&-\frac{\rd F}{\rd t}\geq0,\\
	\frac{\rd S_{env}}{\rd t}-\frac{\dbar_eS}{\rd t}&=&\dot{E}_{hk}\geq0,
\end{eqnarray*}
which are exactly the same as those for master equations.

A key step to recover the first law of thermodynamics for the Fokker-Planck equation is to suppose the validity of detailed balance condition, which states
\begin{equation}
	\label{DB-condition-FP}
	\mathbf{D}(\vec{x})^{-1}\cdot\hat{\vec{u}}(\vec{x};\lambda(t))=\nabla \ln p^{eq}(\vec{x};\lambda(t))=-\beta\nabla \epsilon(\vec{x};\lambda(t)),
\end{equation}
where the drift term $u(x)$ is assumed to depend on slowly varying external control variable $\lambda(t)$. The equilibrium state obeys the canonical distribution $p^{eq}(\vec{x};\lambda(t))=\exp[-\beta \epsilon(\vec{x};\lambda(t))]$ with $\epsilon(\vec{x};\lambda(t))$ as a potential function at state $\vec{x}$ with control variable $\lambda(t)$. With above formula in hand, the heat exchange rate reads
\begin{equation}
	\label{heat-ensemble-FP}
	\frac{\dbar Q}{\rd t}=\beta^{-1}\frac{\dbar_eS}{\rd t}=\int \rd\vec{x}\vec{J}(\vec{x},t)\cdot\nabla \epsilon(\vec{x};\lambda(t)).
\end{equation}
Introduce the system energy 
\begin{equation}
	E(t)=\int \rd\vec{x}\epsilon(\vec{x};\lambda(t))p(\vec{x},t).
\end{equation}
By using integration by parts and neglecting the boundary terms, we find that the law of energy conservation assumes the familiar form
\begin{equation*}
	\frac{\rd E}{\rd t}=\frac{\dbar Q}{\rd t}+\frac{\dbar W}{\rd t},
\end{equation*}
where the rate of work (power) is given by
\begin{equation}
	\label{work-ensemble-FP}
	\frac{\dbar W}{\rd t}=\int \rd\vec{x}\left(\frac{\rd\epsilon(\vec{x};\lambda(t))}{\rd t}\right)p(\vec{x},t).
\end{equation}

The formulations of steady-state thermodynamics for both master equations and Fokker-Planck equations are not a coincidence. Actually, it can be rigorously proven that there is an elegant one-to-one correspondence among all thermodynamic quanti\-ties in thermal physics and ``thermodynamic'' relations that we have discussed for both master equations and Fokker-Planck equations \cite{peng2018markov}.

Irreversible Markov processes with positive entropy production discussed above, play crucial role in a wide range of biological functions in living systems with the breakdown of detailed balance, such as high-fidelity transcription and translation, signaling, sensing, adaptation, pattern formation \cite{qianarpc07,gnesotto2018broken}.
Outside chemistry and biochemistry, in contrast to the widely employed Metropolis-Hastings algorithm in Markov chain Monte Carlo (MCMC) \cite{Liujun-book}, irreversible Markov chain have also found increasing applications\cite{Ma2015complete}.  It has been shown that for finite systems of interacting particles, irreversible processes converge more rapidly to their steady state than reversible ones\cite{kaiser2017acceleration}. Therefore, a host of recent literature focus on exploring the different ways in which the detailed balance condition can be broken and their consequences on the dynamics. For example, Sohl-Dickstein et al. proposed a method for performing Hamiltonian Monte Carlo without detailed balance that can largely eliminate sample rejection\cite{sohl2014hamiltonian}. Similar ideas were borrowed and developed by Michel et al.\cite{michel2020forward} for fast sampling by randomness control in irreversible Markov Chains. In yet another line of applications that were based on inferencing detailed balance breaking \cite{qianpnas04,sisan10,kimmel2018inferring,martinez2019inferring}, non-invasive methods were proposed to detect driven processes that occur in active matters, biological cells and tissue.

\section{Emergent Macroscopic Thermodynamics by Large Deviations Principle}
{Based on the steady-state thermodynamics for chemical master equations constructed in the last section, the famous Gibbs' thermodynamic theory for macroscopic chemical reactions can be derived as a natural consequence. The key lies on the large deviations principle. In addition, both the gradient structure for chemical reactions under the detailed balance condition, and the Kramers' formula for transitions across energy barriers can be better understood by utilizing the large deviations principle.}

\subsection{Large Deviations Principle for Chemical Master Equations}
\label{sec:4.1}
The mathematical theory of large deviations and its entropy description in terms of large deviation rate functions can be interpreted by the idea of multiple scales:  Formally the mathematics states the existence of the limit for an empirical mean value $\overline{X}^{(N)}$ of random variable $X$
\begin{subequations}
	\label{ldrf}
	\begin{equation}
		\varphi(x) = -\lim_{N\to\infty} 
		\frac{1}{N}\ln \Pr \Big\{ 
		\overline{X}^{(N)}\in (x,x+\rd x)\Big\}
\end{equation}
for each and every fixed $x\in\mathbb{R}$. This implies for a large $N$,
\begin{equation}
	    \Pr\Big\{ 
		\overline{X}^{(N)}\in (x,x+\rd x)\Big\} 
		\propto e^{-N\varphi(x)} 
		= e^{-N \inf_y\{xy-\psi(y)\}} \asymp 
		\int_{\mathbb{R}} e^{-N\{xy-\psi(y)\}} \rd y, 
\end{equation}
\end{subequations} 
in which $\psi(y)$ is the Legendre-Fenchel transform of $\varphi(x)$.  Note according to the law of large numbers, the $\overline{X}^{(N)}\to\mathbb{E}[X]$ as $N\to\infty$.  Therefore, the statement in (\ref{ldrf}b) necessarily contains a large parameter $N$; and
$\varphi(x)=\inf_y\{xy-\psi(y)\}$.

In Sec. \ref{CMAE-kurtz-limit}, we have shown the chemical mass-action equations emerge as a deterministic limit of the Markovian stochastic description. It means that, in the Kurtz limit, the probability solution to the chemical master equations becomes a Dirac's delta function, \textit{i.e.} $P(V\vec{x};V)\rightarrow \delta(\vec{x}-\vec{c}(t))$. In additional to this result by laws of large numbers, a more delicate analysis by the large deviations principle suggests an asymptotic expression as
\begin{equation}
	P(\vec{n}(t);V)=P(V\vec{c}(t);V)\simeq \exp[-V\varphi(\vec{c}(t),t)],
\end{equation}
which has been known as the WKB ansatz. $\varphi(\vec{c}(t),t)$ is called as the \textit{\textbf{large deviations rate function}} in literature, which possesses plenty of elegant properties in mathematics, such as convexity, non-negativity, \textit{etc.}\cite{touchette2009large}.

Substituting the WKB ansatz into the chemical master equations in Eq. \eqref{chemical-master-eq} and keeping the leading order term (see Appendix \ref{app_B}),  one arrives at
a \textit{\textbf{Hamilton-Jacobi equation}} (HJE) that has originated from classical Hamiltonian mechanics\cite{gao2022revisit,gao2023large}, 
\begin{subequations}
\label{HJE-Hu}
\begin{eqnarray}
	\frac{\partial \varphi(\vec{c},t)}{\partial t}&=&-H\left(\vec{c},\frac{\partial\varphi(\vec{c},t)}{\partial \vec{c}}\right),\\
	H(\vec{c},\vec{y})&=&\sum_{i=1}^{M}\bigg\{R^+_i(\vec{c})\big[\exp\big(\vec{\nu}_i\cdot\vec{y}\big)-1\big]+R^-_i(\vec{c})\big[\exp\big(-\vec{\nu}_i\cdot\vec{y}\big)-1\big]\bigg\}.
\end{eqnarray}
\end{subequations}
In some sense, the existence and uniqueness of the solutions to the HJE, a nonlinear first-order partial differential equation, are not well defined until more regularity conditions are provided \cite{evans_book,feng2006large}.

A time-independent $\varphi^{ss}(\vec{c})$ is called a stationary solution to (\ref{HJE-Hu}) when it satisfies the right hand side of above HJE being zero.  That is,
\begin{eqnarray}
\label{stationary-HJE}
	\sum_{i=1}^{M}\bigg\{R^+_i(\vec{c})\bigg[\exp\bigg(\vec{\nu}_i\cdot\frac{\partial\varphi^{ss}(\vec{c})}{\partial \vec{c}}\bigg)-1\bigg]
    \hspace{1in}
\\
	  +R^-_i(\vec{c})\bigg[\exp\bigg(-\vec{\nu}_i\cdot\frac{\partial\varphi^{ss}(\vec{c})}{\partial \vec{c}}\bigg)-1\bigg]\bigg\}=0.
\nonumber
\end{eqnarray}
The \textit{\textbf{stationary large deviations rate function}} is a Lyapunov function for the deterministic rate equations in Eq. \eqref{macro-rate-eq} \cite{gang1986lyapounov,Ge2016Mesoscopic}.  This can be verified as follows,
\begin{eqnarray*}
	&&\frac{d\varphi^{ss}(\vec{c})}{dt}=\frac{d\vec{c}}{dt}\cdot\bigg(\frac{\partial\varphi^{ss}(\vec{c})}{\partial \vec{c}}\bigg)=\sum_{i=1}^{M}\bigg[R^+_i(\vec{c})\bigg(\vec{\nu}_i\cdot\frac{\partial\varphi^{ss}(\vec{c})}{\partial \vec{c}}\bigg)-R^-_i(\vec{c})\bigg(\vec{\nu}_i\cdot\frac{\partial\varphi^{ss}(\vec{c})}{\partial \vec{c}}\bigg)\bigg]
 \\
	&&\leq\sum_{i=1}^{M}\bigg\{R^+_i(\vec{c})\bigg[\exp\bigg(\vec{\nu}_i\cdot\frac{\partial\varphi^{ss}(\vec{c})}{\partial \vec{c}}\bigg)-1\bigg]+R^-_i(\vec{c})\bigg[\exp\bigg(-\vec{\nu}_i\cdot\frac{\partial\varphi^{ss}(\vec{c})}{\partial \vec{c}}\bigg)-1\bigg]\bigg\}
 \\[6pt]
    && = 0.
\end{eqnarray*}
The inequality illustrates that a temporal monotonicity of $\varphi^{ss}(\vec{c}(t))$, which therefore can be used as a landscape (energy or relative entropy) function for chemical reactions at macroscopic scales.

Explicit, closed expressions for $\varphi^{ss}(\vec{c})$ can be obtained under the assumptions of both complex balance condition and detailed balance condition. Consider the following solution
\begin{equation}
\label{free-energy}
	\varphi^{ss}(\vec{c})=\sum_{j=1}^Nc_j\ln\bigg(\frac{c_j}{c_j^{ss}}\bigg)-c_j+c_j^{ss},
\end{equation}
where $\vec{c}^{\,ss}$ represents the steady-state solution. Since
\begin{equation*}
	\vec{\nu}_i\cdot\frac{\partial\varphi^{ss}(\vec{c})}{\partial\vec{c}}=\ln\prod_{j=1}^N\bigg(\frac{c_j}{c_j^{ss}}\bigg)^{\nu_{ij}}=\ln\bigg[\frac{R^-_i(\vec{c})R^+_i(\vec{c}^{\,ss})}{R^+_i(\vec{c})R^-_i(\vec{c}^{\,ss})}\bigg],
\end{equation*}
we insert it into the stationary HJE
\begin{eqnarray*}
	&&\sum_{i=1}^{M}\bigg\{\bigg[R^+_i(\vec{c})-R^-_i(\vec{c})\exp\bigg(-\vec{\nu}_i\cdot\frac{\partial\varphi^{ss}(\vec{c})}{\partial \vec{c}}\bigg)\bigg]\bigg[1-\exp\bigg(\vec{\nu}_i\cdot\frac{\partial\varphi^{ss}(\vec{c})}{\partial \vec{c}}\bigg)\bigg]\bigg\}\\
	&=&\sum_{i=1}^M\bigg[R_i^+(\vec{c}^{\,ss})-R_i^-(\vec{c}^{\,ss})\bigg]\bigg[\prod_{j=1}^N\bigg(\frac{c_j}{c_j^{ss}}\bigg)^{\nu_{ij}^+}-\prod_{j=1}^N\bigg(\frac{c_j}{c_j^{ss}}\bigg)^{\nu_{ij}^-}\bigg]=0,
\end{eqnarray*}
the second line is exactly the condition for complex balance in Eq. \eqref{complex-balance}. This results shows that for a chemical mass-action reaction system, a free energy function in Eq. \eqref{free-energy} can be derived under the condition of complex balance. Its convexity, non-negativity and monotonicity can all be verified. This provides a powerful justification for the general properties of the stationary large deviations rate function.
\begin{eqnarray*}
	\frac{\rd\varphi^{ss}(\vec{c})}{\rd t} &=& \sum_{j=1}^N\frac{dc_j}{dt}\ln\bigg(\frac{c_j}{c_j^{ss}}\bigg)=\sum_{j=1}^N\sum_{i=1}^M\nu_{ij}\bigg[R_i^+(\vec{c})-R_i^-(\vec{c})\bigg]\ln\bigg(\frac{c_j}{c_j^{ss}}\bigg)\\
	&=&\sum_{j=1}^N R_i^+(\vec{c})\ln\bigg[\frac{R^-_i(\vec{c})R^+_i(\vec{c}^{\,ss})}{R^+_i(\vec{c})R^-_i(\vec{c}^{\,ss})}\bigg]+R_i^-(\vec{c})\ln\bigg[\frac{R^+_i(\vec{c})R^-_i(\vec{c}^{\,ss})}{R^-_i(\vec{c})R^+_i(\vec{c}^{\,ss})}\bigg]\\
	&\leq&\sum_{j=1}^N R_i^+(\vec{c})\bigg[\frac{R^-_i(\vec{c})R^+_i(\vec{c}^{\,ss})}{R^+_i(\vec{c})R^-_i(\vec{c}^{\,ss})}-1\bigg]+R_i^-(\vec{c})\bigg[\frac{R^+_i(\vec{c})R^-_i(\vec{c}^{\,ss})}{R^-_i(\vec{c})R^+_i(\vec{c}^{\,ss})}-1\bigg]\\
	&=&-\sum_{j=1}^N\bigg[R^+_i(\vec{c}^{\,ss})-R^-_i(\vec{c}^{\,ss})\bigg] \bigg[\frac{R^+_i(\vec{c})}{R^+_i(\vec{c}^{\,ss})}-\frac{R^-_i(\vec{c})}{R^-_i(\vec{c}^{\,ss})}\bigg]\\
	&=&\sum_{i=1}^M\bigg[R_i^+(\vec{c}^{\,ss})-R_i^-(\vec{c}^{\,ss})\bigg]\bigg[\prod_{j=1}^N\bigg(\frac{c_j}{c_j^{ss}}\bigg)^{\nu_{ij}^+}-\prod_{j=1}^N\bigg(\frac{c_j}{c_j^{ss}}\bigg)^{\nu_{ij}^-}\bigg]=0
\end{eqnarray*}

Under the condition of complex balance, we see that the formula in Eq. \eqref{free-energy} satisfies the stationary HJE as a whole. In particular, if we further require
\begin{eqnarray}
	\bigg[R^+_i(\vec{c})-R^-_i(\vec{c})\exp\bigg(-\vec{\nu}_i\cdot\frac{\partial\varphi^{ss}(\vec{c})}{\partial \vec{c}}\bigg)\bigg]=0,
\end{eqnarray}
which is the weak detailed balance condition for deterministic chemical reactions, it yields
\begin{eqnarray*}	    
 \vec{\nu}_i\cdot\frac{\partial\varphi^{ss}(\vec{c})}{\partial \vec{c}} &=&-\ln\bigg[\frac{R^+_i(\vec{c})}{R^-_i(\vec{c})}\bigg] = -\ln\bigg[\frac{k_i^+}{k_i^-}\prod_{j=1}^Nc_j^{-\nu_{ij}}\bigg]
\\ 
  &=& \ln\bigg[\prod_{j=1}^N\bigg(\frac{c_j}{c_j^{eq}}\bigg)^{\nu_{ij}}\bigg]=\sum_{j=1}^N \nu_{ij}\ln\bigg(\frac{c_j}{c_j^{eq}}\bigg),
\end{eqnarray*}
by utilizing the equilibrium solution $\vec{c}^{\,eq}$. At this time, we arrive at a special solution to the stationary HJE,
\begin{equation}
	\varphi^{ss}(\vec{c})=\sum_{j=1}^Nc_j\ln\bigg(\frac{c_j}{c_j^{eq}}\bigg)-c_j+c_j^{eq},
\end{equation}
which has been widely known as the Gibbs free energy function for an ideal solution in equilibrium, which is consistent with chemical mass-action kinetic system under the detailed balance condition.

\subsection{Emergent Macroscopic Chemical Thermodynamics}
 
The three key mesoscopic nonequilibrium thermodynamic rates in Sec. \ref{sec:3.3} for the general discrete-state Markov process, Eqs. (\ref{329})-(\ref{331}), when applied to the more special integer values as discrete states and corresponding chemical master equation \eqref{chemical-master-eq}, become the total entropy production, free energy dissipation, and house-keeping energy input with the expressions:
\begin{eqnarray*}
	\frac{\dbar_i S^{\text{cme}}}{\rd t}&=&\sum_{i=1}^M\bigg\{  J_i\big[\vec{n}-\vec{\nu}_i\big]\ln\bigg[\frac{r^+_i(\vec{n}-\vec{\nu}_i; V)P(\vec{n}-\vec{\nu}_i;V)}{r^-_i(\vec{n}; V)P(\vec{n};V)}\bigg]\\
	&+& J_i\big[\vec{n}\big]\ln\bigg[\frac{r^+_i(\vec{n}; V)P(\vec{n};V)}{r^-_i(\vec{n}+\vec{\nu}_i; V)P(\vec{n}+\vec{\nu}_i;V)}\bigg]\bigg\},
 \\
	 \frac{\rd F^{\text{cme}}}{\rd t}&=& -\sum_{i=1}^M\bigg\{ J_i\big[\vec{n}-\vec{\nu}_i\big] \ln\bigg[\frac{P(\vec{n}-\vec{\nu}_i;V)P^{ss}(\vec{n};V)}{P(\vec{n};V)P^{ss}(\vec{n}-\vec{\nu}_i;V)}\bigg]\\
	&+& J_i\big[\vec{n}\big]\ln\bigg[\frac{P(\vec{n};V)P^{ss}(\vec{n}+\vec{\nu}_i;V)}{P(\vec{n}+\vec{\nu}_i;V)P^{ss}(\vec{n};V)}\bigg]\bigg\},
\\
	\dot{E}^{\text{cme}}_{hk} &=&\sum_{i=1}^M\bigg\{J_i\big[\vec{n}-\vec{\nu}_i\big]\ln\bigg[\frac{r^+_i(\vec{n}-\vec{\nu}_i; V)P^{ss}(\vec{n}-\vec{\nu}_i;V)}{r^-_i(\vec{n}; V)P^{ss}(\vec{n};V)}\bigg]
\\
	&+& J_i\big[\vec{n}\big]\ln\bigg[\frac{r^+_i(\vec{n}; V)P^{ss}(\vec{n};V)}{r^-_i(\vec{n}+\vec{\nu}_i; V)P^{ss}(\vec{n}+\vec{\nu}_i;V)}\bigg]\bigg\},
\end{eqnarray*}
where $J_i\big[\vec{n}\big]= r^+_i(\vec{n}; V)P(\vec{n};V)-r^-_i(\vec{n}+\vec{\nu}_i;V)P(\vec{n}+\vec{\nu}_i;V)$. These results, when combined with the stationary large deviations rate function $\varphi^{ss}(\vec{c})$ in Sec. \ref{sec:4.1}, give rise to a remarkable, general nonequilibrium thermodynamics for macroscopic chemical reaction systems as a mathematical consequence of the chemical master equation in the Kurtz limit. 
This unexpected discovery has been called {\bf\em stochastic kinetics dictates thermodynamics.}   One has the following important mathematical statement.

\begin{theorem}[Emergent macroscopic chemical thermodynamics \cite{Ge2016Mesoscopic,ge2016}]
For each time $t<\infty$, the three rates of total entropy production, house-keeping energy, and free energy dissipation for a chemical master equation converge to three corresponding deterministic functions of $\vec{c}$ for the macroscopic chemical reaction system in the Kurtz limit represented by rate equation (\ref{macro-rate-eq}) and its $\varphi^{ss}(\vec{c})$.  That is as $V$,  $\vec{n} \rightarrow\infty$ while $\vec{n}/V=\vec{c}$:
\begin{enumerate}[(i)]
\item 
\begin{equation*}			   
     \lim_{V\rightarrow\infty}\frac{1}{V}\frac{\dbar_i S^{{\rm cme}}[P(\vec{n};V)]}{\rd t} =\sum_{i=1}^M\bigg[R_i^+(\vec{c})-R_i^-(\vec{c})\bigg]\ln\bigg[\frac{R_i^+(\vec{c})}{R_i^-(\vec{c})}\bigg]\equiv \frac{\dbar_iS}{\rd t}\big[\vec{c}\,\big]\geq0,
\end{equation*}
which is a function of macroscopic concentration $\vec{c}$. A set of concentrations $\vec{c}=\big(c_1,\cdots,c_N\big)$ that satisfy $R_i^+(\vec{c})=R_i^-(\vec{c})$, for all $i=1,\cdots,M$ is called strongly detail balanced.
\item 
\begin{eqnarray*}			       
    \lim_{V\rightarrow\infty} \frac{\dot{E}_{hk}^{{\rm cme}}[P(\vec{n};V)]}{V} &=& \sum_{i=1}^M\bigg[R_i^+(\vec{c})-R_i^-(\vec{c})\bigg]\ln\bigg[\frac{R_i^+(\vec{c})}{R_i^-(\vec{c})}\exp\bigg(\vec{\nu}_i\cdot\frac{\partial\varphi^{ss}(\vec{c})}{\partial \vec{c}}\bigg)\bigg]
\nonumber\\
	&\equiv&\dot{E}_{hk}\big[\vec{c}\,\big] \geq0,
\end{eqnarray*}
which is also a function of $\vec{c}$.  A set of concentrations $\vec{c}$ that satisfy $\vec{\nu}_i\cdot(\partial\varphi^{ss}(\vec{c})/\partial \vec{c}\,)$ $=-\ln\big[R^+_i(\vec{c})/R^-_i(\vec{c})\big]$ for all $i=1,\cdots,M$ is called weakly detail balanced. 
\item 
\begin{equation*}			
    \lim_{V\rightarrow\infty}\frac{1}{V}\frac{\rd F^{{\rm cme}}[P(\vec{n};V)]}{\rd t}= \sum_{i=1}^M\bigg[R_i^+(\vec{c})-R_i^-(\vec{c})\bigg]\vec{\nu}_i\cdot\frac{\partial\varphi^{ss}(\vec{c})}{\partial \vec{c}}\equiv\frac{\rd F}{\rd t}\big[\vec{c}\,\big] \leq 0.
\end{equation*}
A set of concentrations $\vec{c}$ that satisfy $\vec{\nu}_i\cdot(\partial\varphi^{ss}(\vec{c})/\partial \vec{c})=0$ for all $i=1,\cdots,M$ is a macroscopic kinetic steady states.
\item The total entropy production rate for macroscopic chemical reactions can be decomposed into two non-negative terms -- its excessive and house-keeping parts. 
	\begin{equation*}
		\frac{\dbar_i S}{\rd t}[\vec{c}\,]=\dot{E}_{hk}[\vec{c}\,] +\left(-\frac{\rd F}{\rd t}[\vec{c}\,]\right).
	\end{equation*}
\end{enumerate}
\end{theorem}
\begin{proof} 
Here we are going to show the non-negativity of these three rates only.
\begin{enumerate}[(i)]
\item Its validity is straightforward.
\item By noting the inequality $x\geq 1-\exp(-x)$ for $\forall x\in \mathbb{R}$ and the stationary HJE in Eq. \eqref{stationary-HJE}, we have
\begin{eqnarray*}
		&&-\frac{\rd F}{\rd t}\big[\vec{c}\,\big] =-\sum_{i=1}^M\bigg[R_i^+(\vec{c})-R_i^-(\vec{c})\bigg]\vec{\nu}_i\cdot\frac{\partial\varphi^{ss}(\vec{c})}{\partial \vec{c}}\\
		&\geq&\sum_{i=1}^M\bigg\{R_i^+(\vec{c})\bigg[1-\exp\bigg(\vec{\nu}_i\cdot\frac{\partial\varphi^{ss}(\vec{c})}{\partial \vec{c}}\bigg)\bigg]+R_i^-(\vec{c})\bigg[1-\exp\bigg(-\vec{\nu}_i\cdot\frac{\partial\varphi^{ss}(\vec{c})}{\partial \vec{c}}\bigg)\bigg]\bigg\}=0.
\end{eqnarray*}
\item By applying the inequality $\ln x\geq 1-1/x$ for $\forall x>0$, we have
	\begin{eqnarray*}
		\dot{E}_{hk}\big[\vec{c}\,\big] &=&\sum_{i=1}^M\bigg[R_i^+(\vec{c})-R_i^-(\vec{c})\bigg]\ln\bigg[\frac{R_i^+(\vec{c})}{R_i^-(\vec{c})}\exp\bigg(\vec{\nu}_i\cdot\frac{\partial\varphi^{ss}(\vec{c})}{\partial \vec{c}}\bigg)\bigg]\\
		&\geq&\sum_{i=1}^M\bigg\{R_i^+(\vec{c})\bigg[1-\frac{R_i^-(\vec{c})}{R_i^+(\vec{c})}\exp\bigg(-\vec{\nu}_i\cdot\frac{\partial\varphi^{ss}(\vec{c})}{\partial \vec{c}}\bigg)\bigg]\\
		&&+R_i^-(\vec{c})\bigg[1-\frac{R_i^+(\vec{c})}{R_i^-(\vec{c})}\exp\bigg(\vec{\nu}_i\cdot\frac{\partial\varphi^{ss}(\vec{c})}{\partial \vec{c}}\bigg)\bigg]\bigg\}=0.
	\end{eqnarray*}
\end{enumerate}
\end{proof}

Considering laws of mass-action and under the condition of complex balance we have more specific expressions for the macroscopic rates of total entropy production, house-keeping energy, and free energy dissipation.  That is,
\begin{eqnarray}
	&&\frac{\dbar_i S}{\rd t}\big[\vec{c}\,\big]=\sum_{i=1}^M\bigg[k_i^+\prod_{j=1}^Nc_j^{\nu^+_{ij}}-k_i^-\prod_{j=1}^Nc_j^{\nu^-_{ij}}\bigg]\ln\bigg[\frac{k_i^+}{k_i^-}\prod_{j=1}^Nc_j^{-\nu_{ij}}\bigg],\\
	&&\frac{\rd F}{\rd t}\big[\vec{c}\,\big]= -\sum_{i=1}^M\bigg[k_i^+\prod_{j=1}^Nc_j^{\nu^+_{ij}}-k_i^-\prod_{j=1}^Nc_j^{\nu^-_{ij}}\bigg]\ln\bigg[\prod_{j=1}^N\bigg(\frac{c_j}{c_j^{ss}}\bigg)^{-\nu_{ij}}\bigg],\\
	&&\dot{E}_{hk}\big[\vec{c}\,\big]= \sum_{i=1}^M\bigg[k_i^+\prod_{j=1}^Nc_j^{\nu^+_{ij}}-k_i^-\prod_{j=1}^Nc_j^{\nu^-_{ij}}\bigg]\ln\bigg[\frac{k_i^+}{k_i^-}\prod_{j=1}^N(c_j^{ss})^{-\nu_{ij}}\bigg].
\end{eqnarray}
From above formulas, it is straightforward to see that if strong detailed balance holds, the house-keeping energy rate becomes zero, $\dot{E}_{hk}=0$, due to the fact that $k_i^+\prod_{j=1}^N(c_j^{eq})^{\nu_{ij}^+}=k_i^-\prod_{j=1}^N(c_j^{eq})^{\nu_{ij}^-}$ for $\forall i\in[1,\cdots, M]$. Meanwhile, we also have $\dbar_i S/\rd t=-\rd F/\rd t$.

Beside the deterministic kinetic limit as a consequence of the law of large numbers (LLN), the large deviations principle allows the characterization of fluctuations, with both small quadratic deviations and large deviations. J. Keizer first discovered the relation between steady state quadratic fluctuations and responses of general stochastic chemical reaction kinetics in the 1980s through diffusion approximation\cite{keizer1982nonequilibrium,keizer1987statistical}.  This is analogous to the central limit theorem (CLT) in the theory of probability; we now give a rigorous statement and proof of this result. 

\begin{theorem}[Fluctuation-dissipation theorem\cite{qian2021stochastic}]
Let $\vec{c}^{\,ss}$ be a stable stationary solution to macroscopic chemical rate equation \eqref{macro-rate-eq}. Assuming $\varphi^{ss}(\vec{c})$ is twice differentiable and further denote matrices ${\bf \Xi}$, ${\bf A}$, and ${\bf\hat{B}}$ with elements:
\begin{eqnarray*}
	\Xi_{ij}=\frac{\partial^2 \varphi^{ss}(\vec{c}^{\,ss})}{\partial c_i\partial c_j}, \ A_{ij}=\sum_{\ell=1}^M \nu_{\ell i}\nu_{\ell j}\big[R^+_{\ell}(\vec{c}^{\,ss})+R^-_{\ell}(\vec{c}^{\,ss})\big], \ {\rm and}
\\
    \hat{B}_{ij}=\frac{\partial \vec{B}_i(\vec{c}^{\,ss})}{\partial c_j},
    \  {\rm where } \ \vec{B}_i(\vec{c}) = \sum_{\ell=1}^M \nu_{\ell i}\big[R^+_{\ell}(\vec{c})-R^-_{\ell}(\vec{c})\big].
\end{eqnarray*} 
Then we have
\begin{equation}
		\bm{\Xi A\Xi}=-\bm{\Xi\hat{B}}-\bm{\hat{B}\Xi}.
\end{equation}
Furthermore, if $\mathbf{\Xi}$ is invertible, then $\bm{A}=-\bm{\Xi^{-1}\hat{B}}-\bm{\hat{B}\Xi^{-1}}$.
\end{theorem}
\begin{proof} 
Taking the second derivative of the left-hand side of stationary HJE in Eq. \eqref{stationary-HJE}, we arrive at
\begin{eqnarray*}
	&&\sum_{l=1}^{M}\bigg\{\bigg[\frac{\partial R^+_l(\vec{c}^{\,ss})}{\partial c_i}-\frac{\partial R^-_l(\vec{c}^{\,ss})}{\partial c_i}\bigg]\bigg(\sum_{k=1}^N\vec{\nu}_{lk}\frac{\partial^2\varphi^{ss}(\vec{c}^{\,ss})}{\partial c_k\partial c_j}\bigg)\\
	&&\qquad+\bigg[\frac{\partial R^+_l(\vec{c}^{\,ss})}{\partial c_j}-\frac{\partial R^-_l(\vec{c}^{\,ss})}{\partial c_j}\bigg]\bigg(\sum_{k=1}^N\vec{\nu}_{lk}\frac{\partial^2\varphi^{ss}(\vec{c}^{\,ss})}{\partial c_k\partial c_i}\bigg)\\
	&&\qquad+\big[R^+_l(\vec{c}^{\,ss})+R^-_l(\vec{c}^{\,ss})\big]\bigg(\sum_{k=1}^N\vec{\nu}_{lk}\frac{\partial^2\varphi^{ss}(\vec{c}^{\,ss})}{\partial c_k\partial c_i}\bigg)\bigg(\sum_{k=1}^N\vec{\nu}_{lk}\frac{\partial^2\varphi^{ss}(\vec{c}^{\,ss})}{\partial c_k\partial c_j}\bigg)\bigg\}=0,
\end{eqnarray*}
$\forall i,j=1,\cdots,N$, by noticing the identity $\vec{\nu}_i\cdot(\partial\varphi^{ss}(\vec{c}^{\,ss})/\partial \vec{c})=0$. This is exactly the equality for which one is looking.
\end{proof}

\subsection{Gradient Structure on a Manifold for Chemical Reaction Kinetics}
The $\varphi^{ss}(\vec{c}\,)$ in Eq. (\ref{stationary-HJE}) is a Lyapunov function of the nonlinear kinetics in Eq. (\ref{macro-rate-eq}), whose right hand side usually is a non-gradient vector field even for most chemical systems with detailed balance \cite{Li-Qian-Yi-jcp}. The mathematical concept of a gradient, however, is dependent upon the choice of a metric.  It is now clear, according to the celebrated Jordan-Kinderlehre-Otto scheme \cite{jordan1998variational,otto2001geometry}, that the Fokker-Planck equation with detailed balance has an elegant gradient structure in a ``metric space''; it corresponds to an optimal transport driven by a free energy function: The reversible diffusion process follows the direction of the steepest descent of the Gibbs-Boltzmann entropy in terms of the Wasserstein distance, which offers a modern geometric description of the second law of thermodynamics. Similar gradient structures also has been discovered for master equations with a discrete state space and a counterpart of the Wasserstein distance \cite{chow2012fokker,maas2011gradient,mielke2013geodesic}.

For a general reversible chemical reaction system with the law of mass-action, Mielke et. al. have shown that its kinetics can be reformulated into a generalized gradient structure \cite{mielke2017non} that reads
\begin{equation}
\label{chem-gradient}
	\frac{\rd\vec{c}\,(t)}{\rd t}= \Big[ \nabla_{\vec{\xi}}\,\Psi^*\big(\vec{c}\,(t),\vec{\xi}\ \big)\Big]_{\vec{\xi}=-\nabla F(\vec{c}\,(t))},
\end{equation}
along the macroscopic dynamic trajectory $\vec{c}\,(t)$, $t\ge 0$.  The $F(\vec{c}\,)$ in (\ref{chem-gradient}) is a half of the Gibbs free energy in Eq. (\ref{free-energy}) for ideal chemical solution.  Notably the dissipation potential $\Psi^*$ satisfies the following set of conditions:
\begin{itemize}
	\item $\Psi^*(\vec{c},\vec{\xi}\,)$ is a convex function of $\vec{\xi}$ for each fixed $\vec{c}$;
	\item $\Psi^*$ is non-negative;
	\item $\Psi^*(\vec{c},\vec{0})=0$;
	\item $\Psi^*(\vec{c},\vec{\xi})=\Psi^*(\vec{c},-\vec{\xi})$ for all $\vec{c}$ and $\vec{\xi}$.
\end{itemize}
One example of such $\Psi^*$, found in  \cite{mqw_paper_2}, is $\ln\big(1+p(c)e^{-\xi}-2p(c)+p(c)e^{\xi} \big)$ where $0<p(c)<\frac{1}{2}$.  

By introducing the Legendre transform 
\begin{equation}
\label{Psi}
\Psi(\vec{c},\vec{\zeta}\ )=\sup_{\vec{\xi}\in \mathbb{R}^n}\Big\{ \vec{\xi}\cdot\vec{\zeta}-\Psi^*(\vec{c},\vec{\xi}\ )\Big\},
\end{equation} 
Eq. \eqref{chem-gradient} indicates that $\dot{\vec{c}}\big(\vec{c}\,\big)$ is the conjugate to $-\nabla F\big(\vec{c}\,\big)$.  Thus the Fenchel-Young inequality \cite{qian2022jctc} can be cast into an energy-dissipation balance equation,
\begin{equation}
	\Psi\Big(\vec{c}\,(t),\dot{\vec{c}}\,(t) \Big)+\Psi^*\Big(\vec{c}\,(t),-\nabla F\big(\vec{c}\,(t)\big)\Big)+\frac{\rd}{\rd t}F\big(\vec{c}\,(t)\big)=0
\label{gen_dissipation}
\end{equation}
for all $t$.  Eq. (\ref{gen_dissipation}) suggests a balance between the free energy change $\rd F/\rd t$ and the dissipation $\Psi+\Psi^*$. By utilizing the mathematical properties required above for $\Psi^*$, one has $\Psi+\Psi^*\geq0$, which means the free energy along the dynamics $\vec{c}(t)$ must be non-increasing function of time $t$, $\rd F\big(\vec{c}(t)\big)/\rd t\leq0$.

This generalized gradient structure has a close relation to the large deviations principle. Recall that on a microscopic level, we use $\vec{n}(t)=(n_1(t),\cdots, n_N(t))$ to denote the numbers of chemical species $X_1,\cdots,X_N$ at time $t$. Then it is shown that the trajectory $\{\vec{n}(t), 0\le t\le T\}$ satisfies a large-deviation principle of the form
\begin{equation}
	\Pr\bigg\{\frac{\vec{n}(t)}{V}\simeq \vec{c}(t)\bigg\}\propto 
	\exp\left\{-V\int_0^TL\big(\, \vec{c}(t),\dot{\vec{c}}(t) \, \big) \rd t\right\},
	\text{ as }\, V\rightarrow\infty
\end{equation}
where the Lagrangian is given by the Legendre-Fenchel transform of the Hamiltonian function \cite{Ge2016Mesoscopic}
\begin{eqnarray}
	L\big(\vec{c},\vec{\zeta}\, \big)&=&\sup_{\vec{\xi}\in\mathbb{R}^N}
	\left\{ \vec{\xi}\cdot\vec{\zeta}-H\big(\vec{c},\vec{\xi}\ \big) \right\},
\\
	H\big(\vec{c},\vec{\xi}\,\big)&=&\sum_{i=1}^M\left[k_i^+\prod_{j=1}^Nc_j^{\nu_{ij}^+}\big(e^{\vec{\nu}_i\cdot\vec{\xi}}-1\big)+k_i^-\prod_{j=1}^Nc_j^{\nu_{ij}^-}\big(e^{-\vec{\nu}_i\cdot\vec{\xi}}-1\big)\right].
\end{eqnarray}
The Lagrangian function $L$ can be immediately connected to the free-energy dissipation via Fenchel-Young inequality, which has also been identified with {\em entropy production} \cite{qian2022jctc}, for an arbitrary $\vec{\zeta}$:
\begin{equation}	L\big(\vec{c},\vec{\zeta}\, \big)=\Psi\big(\vec{c},\vec{\zeta}\ \big)+\Psi^*\Big(\vec{c},-\nabla F(\vec{c}\,) \Big)+ \vec{\zeta}\cdot \nabla F(\vec{c}\,) \ge 0,
\end{equation}
in which the equality holds if and only if $\vec{\zeta}=\dot{\vec{c}}$ as a function of $\vec{c}$ through the rate equations.  It is straightforward to see that the optimal curve for which $L\big(\vec{c}(t),\dot{\vec{c}}(t)\big)=0$ is the solution of the chemical mass-action equations. In this sense, the gradient structure agrees with the large deviations theory via Legendre-Fenchel duality. 

Given $L$ and its Legendre transform $H$, the $F$ and $\Psi, \Psi^*$ can be found through
\begin{equation}
\nabla F\big(\vec{c}\,\big)=\nabla_{\zeta}L(\vec{c},\vec{\zeta}=0),\   \Psi^*\big(\vec{c},\vec{\xi}\,\big)=H\big(\vec{c},\nabla F(\vec{c})+\vec{\xi}\,\big)-H\big(\vec{c},\nabla F(\vec{c}\,)\big),
\end{equation}
where $\Psi$ is the Legendre transform of $\Psi^*$. In Ref. \cite{mielke2014relation}, it was discovered that the existence of such potentials is equivalent to the time-reversibility of $L$ or $H$, i.e.
\begin{equation}
	L(\vec{c},\vec{\zeta})-L(\vec{c},-\vec{\zeta})=2\nabla F(\vec{c})\cdot\vec{s},\quad \forall\vec{c},\vec{\zeta},
\end{equation}
or equivalently
\begin{equation}
	H(\vec{c},\nabla F(c)+\vec{\xi})=H(\vec{c},\nabla F(c)-\vec{\xi}),\quad \forall\vec{c},\vec{\xi}.
\end{equation}
Observe that this condition fixes the driving
force $\nabla F(c)$ uniquely. Furthermore, under the condition of detailed balance, the uniqueness of a convex dual pair $(\Psi,\Psi^*)$ can be established\cite{mielke2017non}, which gives
\begin{eqnarray}
	F(\vec{c})&=&\frac{1}{2}\sum_{j=1}^N c_j\ln\bigg(\frac{c_j}{c_J^{eq}}\bigg)-c_j+c_j^{eq},\\
	\Psi^*(\vec{c},\vec{\xi})&=&\sum_{i=1}^M 2\left(k_i^+k_i^-\prod_{j=1}^Nc_j^{\nu_{ij}^++\nu_{ij}^-}\right)^{\frac{1}{2}} \Big(\cosh(\vec{\nu}_i\cdot\vec{\xi})-1\,\Big),\\	\Psi(\vec{c},\vec{\zeta})&=&\sup_{\vec{\xi}\in \mathbb{R}^n}\Big\{\vec{\xi}\cdot\vec{\zeta}-\Psi^*(\vec{c},\vec{\xi}\,)\Big\}.
\end{eqnarray}

\subsection{Kramers' Formula for Transition Rate}
By choosing an appropriate reaction coordinate, an elementary chemical reaction can be viewed as a nonlinear dynamical motion between two potential wells $A$ and $B$, corresponding to reactants and products respectively, separated by a barrier. Once the system moves across the energy barrier and out of the potential energy well $A$, we will say the forward chemical reaction happens; likewise if the system moves out of the potential energy well $B$, the backward reaction happens.

In the presence of a deterministic potential force field only, the system would never have a chance to cross the energy barrier when starting from a potential well. Thus to make chemical reactions possible, a key step is to introduce random perturbations. By extending the physical picture from deterministic to stochastic, the system will occasionally move against the deterministic force field and even cross the barrier, which however is a rare event and occurs in a very long time scale.  H. A. Kramers first formulated the mathematical theory for this type of problem in the 1940s\cite{kramers1940brownian}, which correlates rate constants of chemical reactions with the height of energy barriers.

A central issue is to calculate the expectation of the \textit{\textbf{first passage time}}, which reflects the duration that a stochastic process takes to jump out of the potential well and crossing the energy barrier for the first time. To be concrete, let us consider a random variable $X(t)$ characterizing the state of the system. Further suppose its time evolution is governed by 
\begin{equation}
	\label{linear-evo-eq}
	\frac{\partial p(x,t|x_0,0)}{\partial t}=\mathcal{L}(x)p(x,t|x_0,0),
\end{equation}
where $\mathcal{L}(x)$ is the evolution operator for the transition probability density from state $x_0$ at time $0$ to state $x$ at time $t$. A widely used form of $\mathcal{L}(x)$ is the one for the Fokker-Planck equation, i.e. $\mathcal{L}(x)=-\frac{\partial}{\partial x}u(x)+\frac{\partial^2}{\partial x^2}D(x)$. We consider that whenever a stochastic trajectory $X(t)$ crosses the boundary of a potential well $\partial\Omega$, it is removed, which is implemented by imposing an absorbing boundary condition, $p(x_A\in\partial\Omega,t|x_0,0)=0$.

Now the mean first passage time of a stochastic process given the initial value $x_0$ is
\begin{eqnarray}
	\tau(x_0)=\int_0^\infty t\bigg[-\frac{\rd}{\rd t}\int_{\Omega}\rd x \, p(x,t|x_0,0)\bigg]\rd t=\int_0^\infty \rd t\int_{\Omega}\rd x \, p(x,t|x_0,0),
\end{eqnarray}
in which $\int_{\Omega}\rd xp(x,t|x_0,0)$ gives the survival probability that a stochastic process starting at state $x_0$ remains within the region $\Omega$ bounded by the absorbing boundary $\partial\Omega$ during time $t$. During the derivation, an integration by parts is performed by assuming the survival probability tends to be zero faster than $t^{-1}$
as $t\rightarrow\infty$. Such a fast decay can be attained due to the absorbing boundary condition.

Inserting the solution to the linear evolution equation in \eqref{linear-evo-eq} and applying the adjoint operator $\mathcal{L}^+$ on both sides, we have
\begin{eqnarray*}
	&&\mathcal{L}^+(x_0)\tau(x_0)=\mathcal{L}^+(x_0)\int_0^\infty \rd t\int_{\Omega}\rd x\,e^{t\mathcal{L}(x)}\delta(x-x_0)\\
	&=&\int_0^\infty \rd t\int_{\Omega}\rd x\,e^{t\mathcal{L}(x)}\mathcal{L}^+(x_0)\delta(x-x_0)=\int_0^\infty \rd t\int_{\Omega}\rd x\mathcal{L}(x)e^{t\mathcal{L}(x)}\delta(x-x_0)\\
	&=&\int_0^\infty \rd t\int_{\Omega}\rd x\frac{\partial p(x,t|x_0,0)}{\partial t}	=\int_{\Omega}\rd x\big[p(x,\infty|x_0,0)-\delta(x-x_0)\big],
\end{eqnarray*}
by performing transposition and integration on the right-hand side. Since $p(x,\infty|x_0,0)$ $=0$, we arrive at the famous \textit{\textbf{Dynkin's formula for the mean first passage time}}\cite{klebaner2012introduction}.
\begin{equation}
	\mathcal{L}^+(x_0)\tau(x_0)=-1.
\end{equation}

For operator of the Fokker-Planck equation with a non-vanishing potential $U(x)$, we have 
\begin{eqnarray}
	 \hspace{.3in} \left\{
	\begin{aligned}
		&D(x_0)\frac{\rd^2\tau(x_0)}{\rd x_0^2}-\frac{\rd U(x_0)}{\rd x_0}\frac{\rd\tau(x_0)}{\rd x_0}=e^{\phi(x_0)}\frac{\rd}{\rd x_0}\bigg[D(x_0)e^{-\phi(x_0)}\frac{\rd\tau(x_0)}{\rd x_0}\bigg]=-1,\\
		&\tau(x_A)=0,\frac{\rd\tau(x_0)}{\rd x_0}\bigg|_{x_0=x_R}=0,
	\end{aligned}
	\right.
\end{eqnarray}
associated with an absorbing boundary condition at $x_A$ and a reflecting boundary condition at $x_R$, where $\phi(x_0)=\ln D(x_0)-\int^{x_A}_{x_0}D^{-1}(z)\frac{\rd U(z)}{\rd z}\rd z$ corresponds to the dimensionless free energy of the stochastic process. Its solution can be solved as \cite{sung2018statistical}
\begin{equation}	\tau(x)=\int_{x}^{x_A}\rd z\,e^{\phi(z)}D^{-1}(z)\int_{x_R}^{z}\rd y\,e^{-\phi(y)}.
\end{equation}
Mathematically, the Fokker-Planck equation with $D(x)$ can be transformed into that with constant $D$ \cite{risken1996fokker}. Therefore, one can always assume a constant diffusion coefficient $D$ without loss of generality.

\begin{figure}[h]
	\centering
	\includegraphics[width=0.7\linewidth]{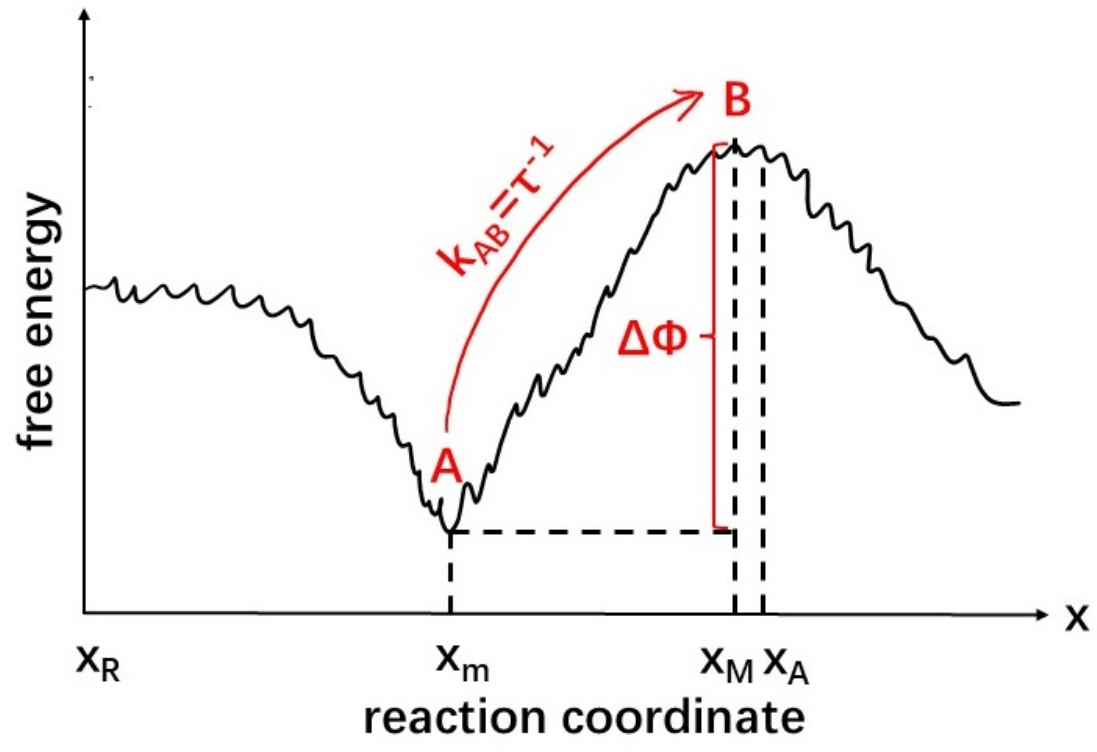}
	\caption{Relation between chemical reaction rate and activation energy barrier in the picture of Kramers' theory.}
	\label{Kramers theory}
\end{figure}

Based on the above formula, we are ready to calculate the mean first passage time for the Kramers' escape problem -- the duration for a stochastic process to escapes a potential well by surmounting an energy barrier. To this end, an absorbing boundary is placed at the target point $x_A$ somewhere after the energy barrier top, and the reflecting boundary condition at $x_R=-\infty$. For the free energy barrier height $\Delta \phi=\phi(x_M)-\phi(x_m)$ is much larger than unity, where $x_M$ and $x_m$ denote the maximum and minimum points of the potential well, integrals in the formula for the mean first passage time can be solved analytically (see Fig. \ref{Kramers theory}). By adopting the quadratic approximation by expanding $\phi(y)\simeq \phi(x_m)+\frac{1}{2}\phi''(x_m)(y-x_m)^2$ around the minimum point $x_m$ and $\phi(z)\approx \phi(x_M)-\frac{1}{2}|\phi''(x_M)|(z-x_M)^2$ around the maximum point $x_M$ respectively, the mean first passage time reads
\begin{eqnarray}
	\hspace{.6cm} 
    \tau&\simeq&D^{-1}\int_{-\infty}^{x_M}\rd z\,e^{\phi(x_M)-\frac{1}{2}|\phi''(x_M)|(z-x_M)^2}\int_{-\infty}^{\infty}\rd y\,e^{-\phi(x_m)-\frac{1}{2}\phi''(x_m)(y-x_m)^2}\\
	&=&\frac{\pi}{D\sqrt{|\phi''(x_M)|\phi''(x_m)}}e^{\Delta \phi}.
\nonumber
\end{eqnarray}
This is the famous Kramers' formula for the reaction rate -- an inverse of the mean first passage time. It, for the first time, provides a rigorous justification on the Arrhenius's equation $k=A\exp[-E_a/(k_BT)]$, which states the activation energy is given by the height of free energy barrier $E_a= k_BT\Delta\phi$, while the prefactor $A$ is correlated with curvatures of the potential well and the energy barrier.

Here we would like to explore the relation between the reaction rate and the mean first passage time a bit more. Imagine an ensemble of i.i.d. particles. Due to the adsorption boundary, certain amount of particles will be removed from the ensemble as soon as they leave the domain $\Omega$ for the first time. To balance the loss, a particle source is added to allow new particles joining the ensemble too. Due to our assumption that the stochastic dynamics is time homogeneous, the particle density $p(x,t)$ approaches a steady state $p^{ss}(x)$ in the long time limit. In other words, the average number of particles leaving $\Omega$ becomes equal to those injected by the source. According to Kramers\cite{kramers1940brownian}, the rate constant $k$ is then defined as the constant net flux out of $\Omega$ normalized by the population inside it.
\begin{equation}
	k(x_0)=\frac{J^{ss}(x_0)}{\displaystyle 
 \int_{\Omega}p^{ss}(x)\rd x},
\end{equation}
where $J^{ss}(x_0)=J^{ss}$ denotes the stationary net flux at $x_0$, which is assumed to be a constant. For the Fokker-Planck equation, the constant flux reads\cite{sung2018statistical} 
\begin{eqnarray}
	J^{ss}&=&-\frac{\rd U(x_0)}{\rd x_0}p^{ss}(x_0)-\frac{\rd}{\rd x_0}\big[D(x_0)p^{ss}(x_0)\big]\\
	&=&-D(x_0)e^{-\phi(x_0)}\frac{\rd}{\rd x_0}\big[e^{\phi(x_0)}p^{ss}(x_0)\big].\nonumber
\end{eqnarray}
From this formula, it is straightforward to show that 
\begin{eqnarray*}
	J^{ss}\int_{x_A}^x\rd x_0D^{-1}(x_0)e^{\phi(x_0)}=-\int_{x_A}^x\rd x_0\frac{\rd}{\rd x_0}\big[e^{\phi(x_0)}p^{ss}(x_0)\big]=-e^{\phi(x)}p^{ss}(x),
\end{eqnarray*}
due to $p^{ss}(x_A)=0$. Performing the integration once more, we have
\begin{eqnarray*}	k^{-1}=\frac{1}{J^{ss}}\int_{x_R}^{x_A}p^{ss}(x)\rd x=\int_{x_R}^{x_A}\rd xe^{-\phi(x)}\int^{x_A}_x\rd x_0D^{-1}(x_0)e^{\phi(x_0)}.
\end{eqnarray*}
Then following the same procedure of quadratic approximation, we will arrive at \textit{\textbf{an inverse relation between Kramers' rate and the mean first passage time}}: 
\begin{equation}
	k=\tau^{-1},
\end{equation}
which holds in a much general sense\cite{reimann1999universal}.

The Kramers' formula can be derived under the weak-noise limit by using the large deviations principle alternatively. Consider a one-dimensional nonlinear dynamic system with weak random perturbations.
\begin{equation}
	\label{weak-noise-limit}
	\rd X^\epsilon(t)=b(X^\epsilon(t))\rd t+\sqrt{\epsilon}\,\rd B_t,
\end{equation}
where $\epsilon\ll1$ is a small parameter. $B_t$ represents a standard Wiener process. Now we are most interested in the first passage time from a potential energy well (say $\Omega$), which is defined as
\begin{equation}
	\tau_\epsilon=\inf\{t|X^\epsilon(t)\in\partial \Omega\}
\end{equation}
where $\partial \Omega$ denotes the boundary of the energy well $\Omega$.

In the limit of weak perturbations $\epsilon\rightarrow0$, the Freidlin-Wentzell theory\cite{freidlin1998random, dembo2009large} shows that the trajectories $\gamma_{[0,T]}=\{X^{\epsilon}(t)|t\in[0,T]\}$ generated by Eq. \eqref{weak-noise-limit} from time $0$ to $T$  satisfy the large deviations principle
\begin{equation*}
	P_{\epsilon}(\gamma_{[0,T]})\simeq \exp\bigg[-\frac{I(\gamma_{[0,T]})}{\epsilon}\bigg],
\end{equation*}
with the rate function $I(\gamma_{[0,T]})=\frac{1}{2}\int_0^T|dX^{\epsilon}(t)/dt-b(X^{\epsilon}(t))|^2dt$. Then for all trajectories starting from $x_0$ at time $0$ to $x$ at time $\tau$, the large deviations principle states
\begin{equation*}
	P_{\epsilon}(x,\tau|x_0,0)\simeq \exp\bigg[-\frac{V(x,\tau|x_0,0)}{\epsilon}\bigg],
\end{equation*}
where the quasi-potential $V(x,\tau|x_0,0)=\inf_{\gamma_{[0,T]}:X^{\epsilon}(0)=x_0,X^{\epsilon}(\tau)=x}I(\gamma_{[0,T]})$ is given by the contraction principle\cite{touchette2009large}. 

Further define the \textit{\textbf{activation energy}}
\begin{equation}
	V^*=\inf_{x\in\partial\Omega}\inf_{\tau\geq0}V(x,\tau|x_0,0),
\end{equation}
which is called ``the principle of minimum available energy'' in according with the contraction principle. It reflects a simple fact that the system will exit the energy well along the most probable path at the most probable time and position. Now we have the following conclusions for the first passage time:
\begin{eqnarray}
	&&\mathbb{E}(\tau_{\epsilon})\simeq \exp\bigg(\frac{1}{\epsilon} V^*\bigg),\\
	&&\lim_{\epsilon\rightarrow0}\mathbb{P}\bigg\{\exp\bigg(\frac{V^*-\delta}{\epsilon}\bigg)<\tau_{\epsilon}<\exp\bigg(\frac{V^*+\delta}{\epsilon}\bigg)\bigg\}=1,\quad \forall \delta>0,\\
	&&\lim_{\epsilon\rightarrow0}\mathbb{P}\bigg\{\frac{\tau}{\mathbb{E}(\tau_{\epsilon})}>s\bigg\}=e^{-s},\quad \forall s\in \mathbb{R}^+.
\end{eqnarray}
The first one states the Kramers' formula. The second formula shows the first passage time converges in probability to its expectation in the limit of $\epsilon\rightarrow0$, a version of weak laws of large numbers. The last formula shows the \textit{\textbf{asymptotic distribution of the first passage time is exponential}}. As a result, we can safely treat the transition between two stable state separated by an energy barrier in a long time scale as an elementary chemical reaction, and adopt the discrete Markovian chain as its mathematical description. This in some way justifies the universality of taking the language of chemical reactions for describing various complex phenomena embedding a time scale separation.

\section{Stochastic Thermodynamics of Nonequilibrium Chemical Reaction Systems}
{As we have claimed, when only a few molecules are considered, a stochastic description in the form of chemical Langevin equations would be more appropriate. Thanks to the stochastic thermodynamics, we can not only analyze the entropy production and entropy flux along single trajectory during chemical reactions, but also establish its rigorous mathematical connections with the steady-state thermodynamics. In this way, self-consistent thermodynamic formulations can be constructed for chemical reactions happening at each level, which provide an alternative yet quite fundamental understanding about chemical reaction systems.}

\subsection{Stochastic Thermodynamics for Markovian Jumps}

The concept of entropy production is not limited to the ensemble formulation; it has been introduced as a trajectory dependent stochastic quantity in the theory of stochastic thermodynamics. Restricted to Markovian jumps on a set of discrete states $1\leq n\leq L$, we can consider the probability $P[\gamma]$ for a forward path $\gamma=\{(n_0,t_0),(n_1,t_1),\cdots,(n_M,t_M),(n_M,T)\}$ starting from an initial state $n_0$ at time $t_0$, and making a sequence of random jumps from state $n_{j-1}$ (prior to jump) to state $n_j$ (after jump) at time $t_j$, until reaching a final state $n_M$ at time $T$.  This convention is known as {\em right-continuous with left limits}, or {\em c\`{a}dl\`{a}g} in French.  An explicit expression for $P[\gamma]$ can be obtained by solving the master equations\cite{peliti2021stochastic}, which reads
\begin{eqnarray}
	P[\gamma]&=&\exp\left( 
	\int_{t_M}^{T}W_{n_{M},n_{M}}(\lambda_s)\rd s\right)\\
	&&\times\prod_{j=M}^{1}\left[W_{n_j,n_{j-1}}(\lambda_{t_j})\exp\left( 
	\int_{t_{j-1}}^{t_j}W_{n_{j-1},n_{j-1}}(\lambda_s)\rd s\right)\right]P_{n_0}(t_0), \nonumber
\end{eqnarray}
where $P_{n_0}(t_0)$ denotes the initial probability of the system being in state $n_0$. We have supposed here that the transition probability is ``controlled'' by an external parameter $\lambda_s$ at time $s$. The factor $\exp\int_{t_{j}}^{t_{j+1}}W_{n_j,n_j}\rd s$ gives the survival probability for the system remaining in state $n_j$ during the time interval $[t_j,t_{j+1})$. Summing over all possible trajectories, meaning integrating over all possible jump time points $t_1,t_2,\cdots, t_M$, all possible states $n_1,n_2,\cdots,n_M$ and all possible number of jumps $M\geq0$, we have $\int P[\gamma]D[\gamma]=1$, in which $D[\gamma]$ denotes the Lebesgue measure of a stochastic trajectory in infinite-dimensional function space. The normalization condition is guaranteed by the master equations.

Motivated by the condition of detailed balanced, $W_{n,n'}P^{eq}_{n'}=W_{n',n}P^{eq}_{n}$, which suggests the equilibrium distribution follows the Gibbs-Boltzmann distribution $P^{eq}_{n}=e^{\beta F-\beta \epsilon_{n}}$ with an equilibrium free energy $F=-\beta^{-1}\ln\big(\sum_{n}e^{-\beta\epsilon_{n}}\big)$. As a consequence, the condition of detailed balance can be rewritten as
\begin{eqnarray}
	\frac{W_{n,n'}}{W_{n',n}}=\frac{P^{eq}_{n}}{P^{eq}_{n'}}=e^{\beta(\epsilon_{n'}-\epsilon_n)}.
\end{eqnarray}
In general, we may define $\zeta_{n,n'}=\beta^{-1}\ln(W_{n,n'}/W_{n',n})+\epsilon_n-\epsilon_{n'}$, which is antisymmetric in exchange of $n$ and $n'$, i.e. $\zeta_{n,n'}=-\zeta_{n',n}$. Particularly, if $\zeta_{n,n'}$ can be written as $\zeta_{n,n'}=\psi_n-\psi_{n'}$, where $\psi_n$ is a function of state $n$, we could be able to recover the condition of detailed balance as 
\begin{eqnarray*}
	\frac{W_{n,n'}}{W_{n',n}}=e^{\beta(\tilde{\epsilon}_{n'}-\tilde{\epsilon}_n)},
\end{eqnarray*}
where $\tilde{\epsilon}_n=\epsilon_n-\psi_n$. 

Otherwise, if $\zeta_{n,n'}$ can not be cast into the difference between two functions solely depending on single state, we may define $q_{n,n'}=\epsilon_n-\epsilon_{n'}-\zeta_{n,n'}$ as the heat absorbed by the system from the heat bath when jumping from state $n'$ to state $n$. It means  
\begin{eqnarray}
	\frac{W_{n,n'}}{W_{n',n}}=e^{-\beta q_{n,n'}}.
\end{eqnarray}
We note that this relation remains valid even when the condition of detailed balance breaks down, and thus will be called \textit{\textbf{the condition of local detailed balance}}.

With respect to above defined quantities, it is straightforward to see that $\epsilon_n$ and $\zeta_{n,n'}$ can not be uniquely determined from the transition probabilities $W_{n,n'}$, since we may introduce a new set of parameters
\begin{eqnarray}
	\tilde{\epsilon}_n&=&\epsilon_n-\psi_n,\\
	\tilde{\zeta}_{n,n'}&=&\zeta_{n,n'}-\psi_n+\psi_{n'}.
\end{eqnarray}
Consequently, the heat transfer $q_{n,n'}$ can be decomposed in a different way
\begin{eqnarray}
	q_{n,n'}=\tilde{\epsilon}_n-\tilde{\epsilon}_{n'}-\tilde{\zeta}_{n,n'}.
\end{eqnarray}
This is well-known in physics as the \textbf{\textit{gauge freedom}}. \textbf{\textit{Each gauge transformation corresponds to a particular decomposition of the total force into a conservative force and a dissipative force.}} Since \textbf{\textit{all physical measurable quantities must be independent of a gauge choice}}, our above reformulation suggests that the energy $\epsilon_n$ and dissipative work $\zeta_{n,n'}$ are not measurable (since they are gauge dependent), while the heat transfer $q_{n,n'}$ is gauge independent and thus measurable.

\begin{figure}[h]
	\centering
	\includegraphics[width=1\linewidth]{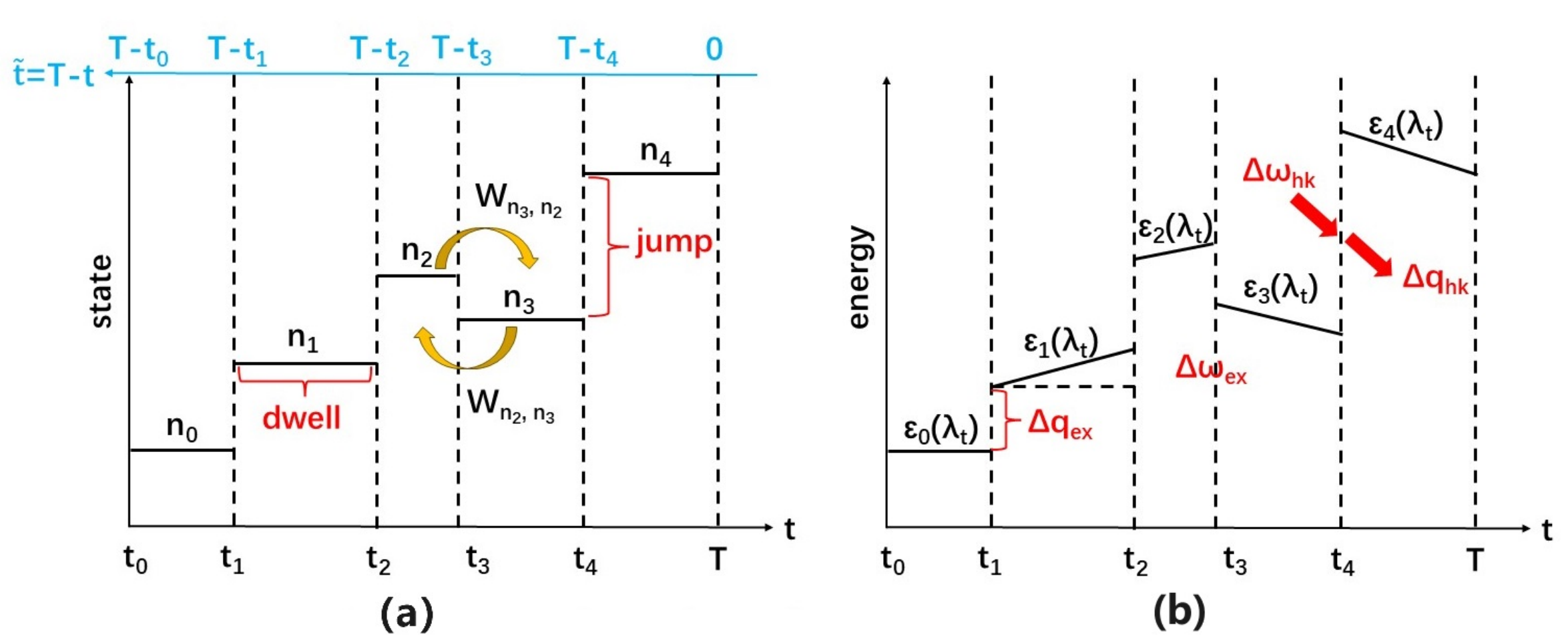}
	\caption{Illustration of (a) Markov jumps among discrete states and (b) corresponding energy changes under external controls. Time coordinates for forward and backward trajectories are marked separately.}
	\label{Markov jumps}
\end{figure}

Now we are ready to construct the first and second laws of thermodynamics at the trajectory level, which in turn will be shown to be fully consistent with our previous formulations at the ensemble level after taking averaging. In agreement with our previous setup, we consider the energy of each state depends on the external control parameter, $\epsilon_n=\epsilon_n(\lambda_t)$, which is slowly varied during the dynamics. As illustrated in Fig. \ref{Markov jumps}, the system energy changes in two different ways. Firstly, inside each dwell, since the system state remains unchanged, the energy is continuously changed as a result of variation of external control parameter $\lambda(t)$, which obviously is some kind of work; secondly, at the $k$th jump from state $n_{k-1}$ to state $n_k$, the system exchanges heat with the heat bath at an amount of $q_{n_k,n_{k-1}}$. Overall, the total heat absorbed by the system along the trajectory is given by
\begin{eqnarray}
&&\Delta q_{tot}[\gamma]=\Delta q_{ex}[\gamma]+\Delta q_{hk}[\gamma],
 \\
	&&\Delta q_{ex}[\gamma]=\epsilon_{n_1}(t_1)-\epsilon_{n_0}(t_1)+\epsilon_{n_2}(t_2)-\epsilon_{n_1}(t_2)\cdots+\epsilon_{n_M}(t_M)-\epsilon_{n_{M-1}}(t_{M}),
 \\
	&&\Delta q_{hk}[\gamma]=-\zeta_{n_1,n_0}-\zeta_{n_2,n_1}\cdots-\zeta_{n_M,n_{M-1}},
\end{eqnarray}
which can be separated into two distinct contributions -- the excessive heat $\Delta q_{ex}$ and the house-keeping heat $\Delta q_{hk}$. 

Similarly, the work along the trajectory can also be decomposed into an excessive term due to the variation of the control parameter and a house-keeping term raised by nonservative forces.
\begin{eqnarray}
	\Delta 	w_{tot}[\gamma]&=&\Delta w_{ex}[\gamma]+\Delta w_{hk}[\gamma],
 \\
	\Delta w_{ex}[\gamma]&=&\epsilon_{n_0}(t_1)-\epsilon_{n_0}(t_0)+\epsilon_{n_1}(t_2)-\epsilon_{n_1}(t_1)\cdots+\epsilon_{n_M}(T)-\epsilon_{n_M}(t_M),\\
	\Delta 	w_{hk}[\gamma]&=&\zeta_{n_1,n_0}+\zeta_{n_2,n_1}\cdots+\zeta_{n_M,n_{M-1}}.
\end{eqnarray}
We notice that except for a minus sign, the house-keeping work and heat are exactly the same $\Delta w_{hk}[\gamma]=-\Delta q_{hk}[\gamma]$. Now the first law of thermodynamics is regained along the trajectory
\begin{eqnarray}
	\epsilon_{n_M}(T)-\epsilon_{n_0}(t_0)&=&\Delta q_{tot}[\gamma]+\Delta w_{tot}[\gamma]=\Delta q_{ex}[\gamma]+\Delta w_{ex}[\gamma].
\end{eqnarray}

It is natural to expect that, after taking average over all possible trajectories, above formulas introduced along single Markovian trajectory will be fully consistent with those defined at the ensemble level.
\begin{eqnarray}
	\frac{\dbar Q_{ex}(t)}{\rd t}&=&\lim_{\rd t\rightarrow0}\frac{1}{\rd t}\mathbb{E}\big[\Delta q_{ex}[\gamma(t,\rd t)]\big]\\
	&=&\sum_{n,n'}W_{n,n'}P_{n'}(t)(\epsilon_{n}-\epsilon_{n'})=\frac{1}{2}\sum_{n,n'}J_{n,n'}(t)(\epsilon_{n}-\epsilon_{n'}),\nonumber\\
	\frac{\dbar Q_{hk}(t)}{\rd t}&=&\lim_{\rd t\rightarrow0}\frac{1}{\rd t}\mathbb{E}[\Delta q_{hk}[\gamma(t,\rd t)]]\\
	&=&-\sum_{n,n'}W_{n,n'}P_{n'}(t)\zeta_{n,n'}=-\frac{1}{2}\sum_{n,n'}J_{n,n'}(t)\zeta_{n,n'},\nonumber\\
	\frac{\dbar Q_{tot}(t)}{\rd t}&=&\frac{\dbar Q_{ex}(t)}{\rd t}+\frac{\dbar Q_{hk}(t)}{\rd t}=\frac{1}{2}\sum_{n,n'}J_{n,n'}(t)q_{n,n'},\\
	\frac{\dbar W_{ex}(t)}{\rd t}&=&\lim_{\rd t\rightarrow0}\frac{1}{\rd t}\mathbb{E}[\Delta w_{ex}[\gamma(t,\rd t)]]=\sum_{n}P_{n}(t)\frac{\rd\epsilon_{n}(\lambda_t)}{\rd\lambda_t},\\
	\frac{\dbar W_{hk}(t)}{\rd t}&=&\lim_{\rd t\rightarrow0}\frac{1}{\rd t}\mathbb{E}[\Delta w_{hk}[\gamma(t,\rd t)]]\\
	&=&\sum_{n,n'}W_{n,n'}P_{n'}(t)\zeta_{n,n'}=\frac{1}{2}\sum_{n,n'}J_{n,n'}(t)\zeta_{n,n'},\nonumber\\
	\frac{\dbar W_{tot}(t)}{\rd t}&=&\frac{\dbar W_{ex}(t)}{\rd t}+\frac{\dbar W_{hk}(t)}{\rd t},
\end{eqnarray}
where $\gamma(t,\rd t)=\{(n,t),(n',t+\rd t)\}$ denotes an infinitely short trajectory starting at state $n$ and ending at state $n'$ during time interval $[t,t+\rd t)$. Compared to formulas for steady-state thermodynamics at the ensemble level (see Eqs.\eqref{heat-ensemble-master}-\eqref{work-ensemble-master}), we notice that without the constraint of detailed balance condition, the total heat/work are further separated into two parts due to their different physical origins. However, the first law of thermodynamics are still fully respected, as
\begin{eqnarray}
	\frac{\rd E(t)}{\rd t}=\frac{\dbar Q_{tot}(t)}{\rd t}+\frac{\dbar W_{tot}(t)}{\rd t}=\frac{\dbar Q_{ex}(t)}{\rd t}+\frac{\dbar W_{ex}(t)}{\rd t},
\end{eqnarray}
where the system energy $E(t)=\sum_{n}\epsilon_{n}(t)P_n(t)$.

\subsection{Stochastic Entropy and Fluctuation Theorems}

In the previous section, we give an explicit formula for the forward trajectory of a Markov chain, based on which the heat and work along the trajectory can be well defined separately. Physically, it would also be of great interest to see what will happen to a Markovian process when the time or the applied forces are reversed separately and even together. Thus with respect to a given forward path $\gamma=\{(n_0,t_0),(n_1,t_1),\cdots,(n_M,t_M),(n_M,T)\}$, we can introduce the corresponding backward path by time reversal as the one starting from state $n_M$ at time $T$ and ending in state $n_0$ at time $0$ and with the external control protocol $\tilde{\lambda}(t)=\lambda(T-t)$. It is straightforward to show that the probability of the backward path reads
\begin{eqnarray*}
	P[\tilde{\gamma}]&=&\prod_{j=0}^{M-1}\bigg[e^{\int_{T-t_{j+1}}^{T-t_{j}}W_{n_{j},n_{j}}(\tilde{\lambda}_{s})\rd s}W_{n_{j},n_{j+1}}(\tilde{\lambda}_{T-t_{j+1}})\bigg]e^{\int_{0}^{T-t_M}W_{n_{M},n_{M}}(\tilde{\lambda}_{s})\rd s}P_{n_M}(T),\\
	&=&\prod_{j=0}^{M-1}\bigg[e^{\int_{t_{j}}^{t_{j+1}}W_{n_{j},n_{j}}(\lambda_{s})\rd s}W_{n_{j},n_{j+1}}(\lambda_{t_{j+1}})\bigg]e^{\int_{t_M}^{T}W_{n_{M},n_{M}}(\lambda_{s})\rd s}P_{n_M}(T),\nonumber
\end{eqnarray*}
by making variable transform $s\rightarrow T-s$ and by making use of the fact $\tilde{\lambda}(t)=\lambda(T-t)$.

To proceed, we introduce the \textbf{\textit{entropy production along a stochastic trajectory}} $\gamma$:
\begin{eqnarray}
	\Delta_i s[\gamma]=\ln\frac{P[\gamma]}{P[\tilde{\gamma}]}.
\end{eqnarray}
Inserting explicit formulas for $P[\gamma]$ and $P[\tilde{\gamma}]$, one finds that all dwelling probabilities among forward and backward paths are exactly cancelled. Only contributions from jumps among different states survive, as well as a term accounting for the ratio between the initial and final state distributions.
\begin{eqnarray}
	\Delta_i s[\gamma]=\sum_{j=1}^M\ln\frac{W_{n_j,n_{j-1}}(\lambda_{t_j})}{W_{n_{j-1},n_j}(\lambda_{t_j})}+\ln\frac{P_{n_0}(t_0)}{P_{n_M}(T)}\equiv-\Delta_e s[\gamma]+\Delta s[\gamma],
\end{eqnarray}
where $\Delta_e s[\gamma]=-\sum_{j=1}^M\ln[{W_{n_j,n_{j-1}}(\lambda_{t_j})}/{W_{n_{j-1},n_j}(\lambda_{t_j})}]$ represents the entropy exchange between the system and the environment along the specified trajectory. And $\Delta s[\gamma]=\ln[P_{n_0}(t_0)/P_{n_M}(T)]$ is the trajectory entropy change.

According to Seifert\cite{seifert2005entropy}, the \textit{\textbf{stochastic entropy}} of state $n$ at time $t$ is given by
\begin{eqnarray}
	s(n,t)=-\ln P_n(t)
\end{eqnarray}
whose ensemble average leads to the classical entropy function
\begin{eqnarray}
	\mathbb{E}[s(n,t)]=-\sum_nP_n(t)\ln P_n(t)=S(t).
\end{eqnarray}
A major difference between the stochastic entropy and the ensemble entropy is that the former is a random variable in nature. With this definition in hand, the trajectory entropy change can be rewritten as the difference between stochastic entropies of the initial and final states, i.e. $\Delta s[\gamma]=s(n_M,T)-s(n_0,0)$.

Furthermore, we can consider the instantaneous stationary distribution $P^{ss}_n(\lambda_t)$, which represents a steady state when the transition probability is frozen in time to the value $W_{n,n'}(\lambda_t)$, such that
\begin{eqnarray*}
	\sum_{n'}W_{n,n'}(\lambda_t)P^{ss}_{n'}(\lambda_t)=0.
\end{eqnarray*}
Now the total entropy production along a trajectory can be further split into two parts:\index{trajectory entropy}
\begin{eqnarray}
	\Delta_i s[\gamma]=\Delta s_{hk}[\gamma]+\Delta s_{ex}[\gamma],
\end{eqnarray}
with a house-keeping part and an excessive part as
\begin{eqnarray}
	&&\Delta s_{hk}[\gamma]=\sum_{j=1}^M\ln\frac{W_{n_j,n_{j-1}}(\lambda_{t_j})P^{ss}_{n_{j-1}}(\lambda_{t_j})}{W_{n_{j-1},n_j}(\lambda_{t_j})P^{ss}_{n_{j}}(\lambda_{t_j})},\\
	&&\Delta s_{ex}[\gamma]=\ln\frac{P_{n_0}(t_0)}{P_{n_M}(T)}+\sum_{j=1}^M\ln\frac{P^{ss}_{n_{j}}(\lambda_{t_j})}{P^{ss}_{n_{j-1}}(\lambda_{t_j})}=\sum_{j=1}^M\ln\frac{P_{n_{j-1}}(\lambda_{t_j})P^{ss}_{n_{j}}(\lambda_{t_j})}{P_{n_{j}}(\lambda_{t_j})P^{ss}_{n_{j-1}}(\lambda_{t_j})}.
\end{eqnarray}
Note if the system satisfies the condition of detailed balance, $\Delta s_{hk}[\gamma]=0$. On the other hand, if we focus on the so-called adiabatic process, which means the probability distribution takes the steady state form $P^{ss}_n(\lambda_t)$ during the whole time, we have $\Delta s_{ex}[\gamma]=0$.

The house-keeping and excessive entropy production along a trajectory could be expressed through an alternative more physically meaningful form. Consider a dual Markov process with the adjoint transition matrix given by\cite{Esposito2010Three}
\begin{eqnarray}
	W_{n,n'}^{\dagger}(\lambda_t)=\frac{W_{n',n}(\lambda_t)P^{ss}_n(\lambda_t)}{P^{ss}_{n'}(\lambda_t)}.
\end{eqnarray}
Apparently, this process has the same steady state as the original one. And $W^{\dagger}$ coincides with $W$ if and only if under the condition of detailed balance. Let $P[\gamma^{\dagger}]$ be the probability of trajectories generated by the Markov process $W^{\dagger}(\lambda_t)$ and $P[\tilde{\gamma}^{\dagger}]$ be that for time reversal trajectories generated by $W^{\dagger}(\lambda_{T-t})$. Then, with respect to explicit expressions for $P[\gamma]$ and $P[\tilde{\gamma}]$, it can be verified that
\begin{eqnarray}
	\Delta s_{hk}[\gamma]=\ln\frac{P[\gamma]}{P[\gamma^{\dagger}]},\qquad \Delta s_{ex}[\gamma]=\ln\frac{P[\gamma]}{P[\tilde{\gamma}^{\dagger}]}.
\end{eqnarray}

It is remarkable that the trajectory entropy production together with its house-keeping and excessive parts defined above obey \textit{\textbf{integral fluctuation theorems}}\cite{seifert2012stochastic}. To see this, we take expectations over all possible trajectories on the exponential form of each quantity as
\begin{eqnarray}
	\mathbb{E}[e^{-\Delta_i s[\gamma]}]&=&\int D[\gamma] P[\tilde{\gamma}]=\int D[\tilde{\gamma}] P[\tilde{\gamma}]=1,\\
	\mathbb{E}[e^{-\Delta s_{hk}[\gamma]}]&=&\int D[\gamma] P[\gamma^{\dagger}]=\int D[\gamma^{\dagger}] P[\gamma^{\dagger}]=1,\\
	\mathbb{E}[e^{-\Delta s_{ex}[\gamma]}]&=&\int D[\gamma] P[\tilde{\gamma}^{\dagger}]=\int D[\tilde{\gamma}^{\dagger}] P[\tilde{\gamma}^{\dagger}]=1.
\end{eqnarray}
The second equality in each line holds since the Lebesgue measures of forward, backward and their dual trajcotories are all the same (there is a one-to-one correspondence among each kind of trajectories). Consequently, by Jensen's inequality, we arrive at the strengthened second laws of thermodynamics for Markov processes at the ensemble level
\begin{eqnarray}
	\mathbb{E}[\Delta_i s[\gamma]]&=&\int D[\gamma] P[\gamma]\ln\frac{P[\gamma]}{P[\tilde{\gamma}]}\geq0,\\
	\mathbb{E}[\Delta s_{hk}[\gamma]]&=&\int D[\gamma] P[\gamma]\ln\frac{P[\gamma]}{P[\gamma^{\dagger}]}\geq0,\\
	\mathbb{E}[\Delta s_{ex}[\gamma]]&=&\int D[\gamma] P[\gamma]\ln\frac{P[\gamma]}{P[\tilde{\gamma}^{\dagger}]}\geq0.
\end{eqnarray}

Especially, consider an infinitely short trajectory $\gamma(t,dt)=\{(n,t),(n',t+dt)\}$ starting from state $n$ and ending at state $n'$ during $dt$. We have
\begin{eqnarray}
	&&\lim_{\rd t\rightarrow0}\frac{1}{\rd t}\mathbb{E}[\Delta_i s[\gamma(t,\rd t)]]\\
	&=&\sum_{n,n'}W_{n,n'}(\lambda_{t})P_{n'}(\lambda_{t})\ln\frac{W_{n,n'}(\lambda_{t})P_{n'}(\lambda_{t})}{W_{n',n}(\lambda_{t})P_{n}(\lambda_{t})}=\frac{\dbar_i S}{\rd t}\geq0,\nonumber\\
	&&\lim_{\rd t\rightarrow0}\frac{1}{\rd t}\mathbb{E}[\Delta s_{hk}[\gamma(t,\rd t)]]\\
	&=&\sum_{n,n'}W_{n,n'}(\lambda_{t})P_{n'}(\lambda_{t})\ln\frac{W_{n,n'}(\lambda_{t})P^{ss}_{n'}(\lambda_{t})}{W_{n',n}(\lambda_{t})P^{ss}_{n}(\lambda_{t})}=\dot{E}_{hk}(t)\geq0,\nonumber\\
	&&\lim_{\rd t\rightarrow0}\frac{1}{\rd t}\mathbb{E}\big[\Delta s_{ex}[\gamma(t,\rd t)]\big]\\
	&=&\sum_{n,n'}W_{n,n'}(\lambda_{t})P_{n'}(\lambda_{t})\ln\frac{P_{n'}(\lambda_{t})P^{ss}_{n}(\lambda_{t})}{P_{n}(\lambda_{t})P^{ss}_{n'}(\lambda_{t})}=-\frac{\rd F}{\rd t}\geq0,\nonumber
\end{eqnarray}
by taking average over all possible state pairs $n$ and $n'$. Above results confirm the trajectory entropy production is a well-defined stochastic generalization of its corresponding physical quantity at the ensemble level, so are its house-keeping and excessive parts.

With respect to the trajectory entropy production, a much stronger relation can be constructed, which is known as the \textit{\textbf{detailed fluctuation theorem}}\cite{seifert2012stochastic}. Similar to the derivation of integral fluctuation theorems, this time we take expectations over those trajectories which generate the same trajectory entropy production. This can be easily achieved by introducing Dirac' delta function.
\begin{eqnarray*}
	&&\mathbb{E}[e^{-\Delta_i s[\gamma]}\delta(\Delta_i s[\gamma]-\Sigma)]=e^{-\Sigma}\int D[\gamma] P[\gamma]\delta(\Delta_i s[\gamma]-\Sigma)\\
	&=&\int D[\gamma] P[\gamma]\frac{P[\tilde{\gamma}]}{P[\gamma]}\delta(\Delta_i s[\gamma]-\Sigma)=\int D[\gamma] P[\tilde{\gamma}]\delta(\Delta_i s[\gamma]-\Sigma)\\
	&=&\int D[\tilde{\gamma}] P[\tilde{\gamma}]\delta(-\Delta_i s[\tilde{\gamma}]-\Sigma).
\end{eqnarray*}
In the last equality, we use the fact that the backward of a backward trajectory is the original forward trajectory ($\tilde{\tilde{\gamma}}=\gamma$), such that there is an opposite sign between the trajectory entropy production of the two, i.e. $\Delta_i s[\gamma]=-\Delta_i s[\tilde{\gamma}]$. From above formula, we obtain the \textit{\textbf{detailed fluctuation theorems}}:
\begin{eqnarray}
	P_F(\Sigma)e^{-\Sigma}=P_B(-\Sigma),
\end{eqnarray}
where $P_F(\Sigma)=\int D[\gamma] P[\gamma]\delta(\Delta_i s[\gamma]-\Sigma)$ and $P_B(-\Sigma)=\int D[\tilde{\gamma}] P[\tilde{\gamma}]\delta(\Delta_i s[\tilde{\gamma}]+\Sigma)$ represent the probability density for all forward and backward trajectories whose trajectory entropy production equal to $\Sigma$ and $-\Sigma$ respectively. 

In the same way, by noticing that both time reversal and duality operations are an involution ($\tilde{\tilde{\gamma}}=\gamma, (\gamma^{\dagger})^{\dagger}=\gamma$), we could verify that the detailed fluctuation theorems hold for the house-keeping and excessive trajectory entropy production too. They can be cast into a unified form as
\begin{eqnarray}
	P_F(\Delta_i s)e^{-\Delta_i s}&=&P_B(-\Delta_i s),\\
	P_F(\Delta s_{hk})e^{-\Delta s_{hk}}&=&P^{\dagger}(-\Delta s_{hk}),\\
	P_F(\Delta s_{ex})e^{-\Delta s_{ex}}&=&P_B^{\dagger}(-\Delta s_{ex}),
\end{eqnarray}
where $P_B(-\Delta_i s)$, $P^{\dagger}(-\Delta s_{hk})$ and $P_B^{\dagger}(-\Delta s_{ex})$ denote the probability densities for backward, dual, dual-backward trajectories whose total, house-keeping and excessive trajectory entropy production equal to marked values respectively. So are the $P_F(\cdot)$s. By taking further average over all possible values of trajectory entropy production, it is easy to show the detailed fluctuation theorem contains the corresponding integral fluctuation theorem as a special case.

As a direct application, let us consider a Markov process starting from an equilibrium state with an initial control parameter $\lambda(t_0)$ and ending at another equilibrium state corresponding to the final control parameter $\lambda(T)$. Then we have $P_{n_0}(t_0)=e^{\beta F(\lambda_{t_0})-\beta\epsilon_{n_0}(\lambda_{t_0})}$ and $P_{n_M}(T)=e^{\beta F(\lambda_{T})-\beta\epsilon_{n_M}(\lambda_{T})}$. Inserting these formulas into the expression of trajectory entropy production, we have $\Delta_i s[\gamma]=\beta(\Delta w_{tot}[\gamma]-\Delta F)$, where $\Delta F=F(\lambda_{T})-F(\lambda_{t_0})$ denotes the free energy difference, and $\Delta w_{tot}[\gamma]$ is the total work. In this case, the detailed and integral fluctuation theorems for the trajectory entropy production become the famous Crooks' relation\cite{crooks1999entropy}
\begin{eqnarray}
	P_F(\Delta w_{tot})e^{\beta(\Delta F-\Delta w_{tot})}&=&P_B(-\Delta w_{tot}),
\end{eqnarray}
and Jarzynski equality\cite{jarzynski1997nonequilibrium} as
\begin{eqnarray}
	\mathbb{E}[e^{-\beta\Delta w_{tot}[\gamma]}]&=&e^{-\beta\Delta F}.
\end{eqnarray}

\subsection{Stochastic Thermodynamics for Langevin Dynamics}
In this part, in analogy to discrete Markovian jumps, we are going to construct the stochastic thermodynamics for continuous Langevin dynamics. Here we consider a general Langevin dynamics characterized by $d\vec{x}_t=\vec{u}(\vec{x}_t;\lambda_t)dt+\mathbf{\sigma}(\vec{x}_t)\bullet d\vec{B}_t$ in the sense of Ito's calculus. $\vec{B}_t$ represents multi-dimensional Wiener processes. In addition, the drift term is modulated by an external control $\lambda_t$. Its corresponding Fokker-Planck equation reads $\partial p(\vec{x},t)/\partial t=-\nabla\cdot \vec{J}(\vec{x},t)$ with probability flux $\vec{J}(\vec{x},t)=\vec{u}(\vec{x};\lambda_t)p(x,t)-\nabla\cdot[\mathbf{D}(\vec{x})p(\vec{x},t)]$, where the diffusion matrix $\mathbf{D}(\vec{x})=\sigma^T(\vec{x})\cdot\sigma(\vec{x})/2$ is symmetric and positive definite.

Now, we are ready to construct the first law of thermodynamics for Langevin dynamics at the trajectory level. Referring to the state energy $\epsilon(\vec{x};\lambda_t)$ defined through detailed balance condition as in Eq. \eqref{DB-condition-FP}, the total heat absorbed by the system along a given trajectory per unit time can be separated into an excessive part and a house keeping part, i.e.
\begin{eqnarray}
	\frac{\dbar q_{tot}(\vec{x}_t,t)}{\rd t}&=&\frac{\dbar q_{ex}(\vec{x}_t,t)}{\rd t}+\frac{\dbar q_{hk}(\vec{x}_t,t)}{\rd t},\\
	\frac{\dbar q_{ex}(\vec{x}_t,t)}{\rd t}&=&\nabla_{x}\epsilon(\vec{x}_t;\lambda_t)\circ \frac{\rd\vec{x}_t}{\rd t},\\
	\frac{\dbar q_{hk}(\vec{x}_t,t)}{\rd t}&=&-\hat{\vec{u}}(\vec{x}_t;\lambda_t)\cdot\mathbf{D}^{-1}(\vec{x}_t)\circ \frac{\rd \vec{x}_t}{\rd t},
\end{eqnarray}
where $\hat{\vec{u}}(\vec{x}_t;\lambda_t)=\vec{u}(\vec{x}_t;\lambda_t)-\nabla\cdot\mathbf{D}(\vec{x}_t)$. Note the last two formulas are interpreted in the sense of Stratonovich calculus as indicated by ``$\circ$''. 

Meanwhile, the instantaneous work along the trajectory can also be decomposed into an excessive term due to the variation of the control parameter and a house-keeping term raised by drifts.
\begin{eqnarray}
	\frac{\dbar w_{tot}(\vec{x}_t,t)}{\rd t}&=&\frac{\dbar w_{ex}(\vec{x}_t,t)}{\rd t}+\frac{\dbar w_{hk}(\vec{x}_t,t)}{\rd t},\\
	\frac{\dbar w_{ex}(\vec{x}_t,t)}{\rd t}&=&\nabla_{\lambda}\epsilon(\vec{x}_t;\lambda_t)\cdot\frac{\rd\lambda_t}{\rd t},\\
	\frac{\dbar w_{hk}(\vec{x}_t,t)}{\rd t}&=&\hat{\vec{u}}(\vec{x}_t;\lambda_t)\cdot\mathbf{D}^{-1}(\vec{x}_t)\circ \frac{\rd\vec{x}_t}{\rd t},
\end{eqnarray}
Again all stochastic calculus is interpreted in the Stratonovich sense. We notice that except for a minus sign, the instantaneous house-keeping work and heat are exactly the same $\dbar q_{hk}(x_t,t)/dt=-\dbar w_{hk}(x_t,t)/dt$. Now the first law of thermodynamics is regained along every trajectory
\begin{eqnarray}
	\frac{\rd\epsilon(\vec{x}_t;\lambda_t)}{\rd t}&=&\frac{\dbar q_{tot}(x_t,t)}{\rd t}+\frac{\dbar w_{tot}(x_t,t)}{\rd t}=\frac{\dbar q_{ex}(x_t,t)}{\rd t}+\frac{\dbar w_{ex}(x_t,t)}{\rd t}.
\end{eqnarray}

To show their counterparts at the ensemble level -- formulas for steady-state thermodynamics in Eqs.\eqref{heat-ensemble-FP}-\eqref{work-ensemble-FP}, we need to calculate
\begin{eqnarray}
	\frac{\dbar Q_{ex}(t)}{\rd t}&=&\mathbb{E}\bigg[\frac{\dbar q_{ex}(x_t,t)}{\rd t}\bigg]=\int \rd\vec{x}\nabla_{x} \epsilon(\vec{x},\lambda_t)\cdot \vec{J}(\vec{x},t),\\
	\frac{\dbar Q_{hk}(t)}{\rd t}&=&\mathbb{E}\bigg[\frac{\dbar q_{hk}(x_t,t)}{\rd t}\bigg]=-\int \rd\vec{x}\big[\hat{\vec{u}}(\vec{x};\lambda_t)\cdot\mathbf{D}^{-1}(\vec{x})\cdot\vec{J}(\vec{x},t)\big],\\
	\frac{\dbar Q_{tot}(t)}{\rd t}&=&\frac{\dbar Q_{ex}(t)}{\rd t}+\frac{\dbar Q_{hk}(t)}{\rd t},\\
	\frac{\dbar W_{ex}(t)}{\rd t}&=&\mathbb{E}\bigg[\frac{\dbar w_{ex}(x_t,t)}{\rd t}\bigg]=\int \rd\vec{x}\frac{\partial \epsilon(\vec{x},\lambda_t)}{\partial\lambda}\frac{\rd\lambda_t}{\rd t}p(\vec{x},t),\\
	\frac{\dbar W_{hk}(t)}{\rd t}&=&\mathbb{E}\bigg[\frac{\dbar w_{hk}(x_t,t)}{\rd t}\bigg]=\int \rd\vec{x}\big[\hat{\vec{u}}(\vec{x};\lambda_t)\cdot\mathbf{D}^{-1}(\vec{x})\cdot\vec{J}(\vec{x},t)\big],\\
	\frac{\dbar W_{tot}(t)}{\rd t}&=&\frac{\dbar W_{ex}(t)}{\rd t}+\frac{\dbar W_{hk}(t)}{\rd t}.
\end{eqnarray}
Consequently, the first law of thermodynamics are fully respected at the ensemble level, i.e.
\begin{eqnarray*}
	\frac{dE(t)}{\rd t}=\frac{\dbar Q_{tot}(t)}{\rd t}+\frac{\dbar W_{tot}(t)}{\rd t}=\frac{\dbar Q_{ex}(t)}{\rd t}+\frac{\dbar W_{ex}(t)}{\rd t},
\end{eqnarray*}
where the system energy $E(t)=\int \epsilon(\vec{x},\lambda_t)d\vec{x}$. In this part, we only need to show 
\begin{eqnarray*}
	&&\mathbb{E}\bigg[\frac{\dbar q_{ex}(x_t,t)}{\rd t}\bigg]=\mathbb{E}_{\vec{x}\sim p(\vec{x},t)}\bigg[\mathbb{E}_{\vec{x}_t=\vec{x}}\bigg[\nabla_{x}\epsilon(\vec{x}_t;\lambda_t)\circ \frac{\rd \vec{x}_t}{\rd t}\bigg]\bigg]\\
	&=&\mathbb{E}_{\vec{x}\sim p(\vec{x},t)}\bigg[\mathbb{E}_{\vec{x}_t=\vec{x}}\bigg[\nabla_{x}\epsilon(\vec{x}_t;\lambda_t)\bullet\frac{\rd\vec{x}_t}{\rd t}\bigg]+\mathbf{D}(\vec{x}):\nabla^2_{x}\epsilon(\vec{x};\lambda_t)\bigg]\\
	&=&\mathbb{E}_{\vec{x}\sim p(\vec{x},t)}\bigg[\mathbb{E}_{\vec{x}_t=\vec{x}}\bigg[\nabla_{x}\epsilon(\vec{x}_t;\lambda_t)\cdot\vec{u}(\vec{x}_t;\lambda_t)+\nabla_{x}\epsilon(\vec{x}_t;\lambda_t)\cdot\mathbf{\sigma}(\vec{x}_t)\bullet\frac{\rd\vec{B}_t}{\rd t}\bigg]+\mathbf{D}(\vec{x}):\nabla^2_{x}\epsilon(\vec{x};\lambda_t)\bigg]\\
	&=&\mathbb{E}_{\vec{x}\sim p(\vec{x},t)}\bigg[\nabla_{x}\epsilon(\vec{x};\lambda_t)\cdot\vec{u}(\vec{x};\lambda_t)+\mathbf{D}(\vec{x}):\nabla^2_{x}\epsilon(\vec{x};\lambda_t)\bigg]\\
	&=&\int \rd\vec{x}\nabla_{x}\epsilon(\vec{x};\lambda_t)\cdot\big[\vec{u}(\vec{x};\lambda_t)p(\vec{x},t)-\nabla\cdot[\mathbf{D}(\vec{x})p(\vec{x},t)]\big]=\int \rd\vec{x}\nabla_{x}\epsilon(\vec{x};\lambda_t)\cdot\vec{J}(\vec{x},t).
\end{eqnarray*}

Next, we are going to explore the second law of thermodynamics. With respect to the stochastic entropy $s(\vec{x}_t,t)=-\ln p(\vec{x}_t,t)$, it is straightforward to calculate its entropy change rate along a given trajectory $\gamma=\{\vec{x}_{t}|t\in[0,T]\}$ as
\begin{flalign*}
	&\frac{\rd s(\vec{x}_t,t)}{\rd t}=-\frac{1}{p(\vec{x}_t,t)}\frac{\partial p(\vec{x}_t,t)}{\partial t}-\frac{1}{p(\vec{x}_t,t)}\nabla p(\vec{x}_t,t)\circ\frac{\rd\vec{x}_t}{\rd t}=\frac{\dbar_is(\vec{x}_t,t)}{\rd t}+\frac{\dbar_e s(\vec{x}_t,t)}{\rd t},
\end{flalign*}
which have been separated into two distinct contributions -- the instantaneous trajectory entropy production rate and entropy exchange rate as
\begin{eqnarray}
	\frac{\dbar_i s(\vec{x}_t,t)}{\rd t}&=&-\frac{1}{p(\vec{x}_t,t)}\frac{\partial p(\vec{x}_t,t)}{\partial t}+\frac{\vec{J}(\vec{x}_t,t)\cdot\mathbf{D}^{-1}(\vec{x}_t)}{p(\vec{x}_t,t)}\circ \frac{\rd \vec{x}_t}{\rd t},\\
	\frac{\dbar_e s(\vec{x}_t,t)}{\rd t}&=&-\hat{\vec{u}}(\vec{x}_t;\lambda_t)\cdot\mathbf{D}^{-1}(\vec{x}_t)\circ\frac{\rd \vec{x}_t}{\rd t},
\end{eqnarray}
by using the Fokker-Planck equation. 

Upon ensemble averaging, the entropy production rate has to become non-negative as required by the second law of thermodynamics. The explicit procedure proceeds in two successive steps\cite{seifert2005entropy}: (1) average over all possible stochastic trajectories $\{\vec{x}_t\}$ at a given position $\vec{x}$ and time $t$; (2) average over all $\vec{x}$ with probability density $p(\vec{x},t)$. 
This leads to
\begin{eqnarray}
	\mathbb{E}\bigg[\frac{\dbar_is}{\rd t}\bigg]&=&\mathbb{E}_{\vec{x}\sim p(\vec{x},t)}\bigg[\mathbb{E}_{\vec{x}_t=\vec{x}}\bigg[\frac{\dbar_is}{\rd t}\bigg]\bigg]\\
	&=&\int \rd\vec{x}\frac{\vec{J}(\vec{x},t)\cdot\mathbf{D}^{-1}(\vec{x})\cdot\vec{J}(\vec{x},t)}{p(\vec{x},t)}=\frac{\dbar_iS(t)}{\rd t}\geq 0,\nonumber\\
	\mathbb{E}\bigg[\frac{\dbar_es}{\rd t}\bigg]&=&\mathbb{E}_{\vec{x}\sim p(\vec{x},t)}\bigg[\mathbb{E}_{\vec{x}_t=\vec{x}}\bigg[\frac{\dbar_es}{\rd t}\bigg]\bigg]\\
	&=&-\int \rd\vec{x}\vec{J}(\vec{x},t)\cdot\mathbf{D}^{-1}(\vec{x})\cdot\hat{\vec{u}}(\vec{x};\lambda_t)=\frac{\dbar_eS(t)}{\rd t}.\nonumber
\end{eqnarray}
Comparing these results with those presented for the steady-state thermodynamics of Fokker-Planck equations previously, we can clearly see that the ensemble results are recovered upon properly averaging.

To show the validity of above formulas, we refer to several novel results about Ito's calculus and Stratonovich calculus for Langevin equations, i.e. 
\begin{eqnarray*}
	\frac{\rd f(\vec{x}_t,t)}{\rd t}&=&\frac{\partial f(\vec{x}_t,t)}{\partial t}+\nabla f(\vec{x}_t,t)\circ \frac{\rd\vec{x}_t}{\rd t}\\
	&=&\frac{\partial f(\vec{x}_t,t)}{\partial t}+\nabla f(\vec{x}_t,t)\bullet\frac{\rd\vec{x}_t}{\rd t}+\mathbf{D}(\vec{x}_t):\nabla^2 f(\vec{x}_t,t),
\end{eqnarray*} 
as well as the fact for Ito's integral $\mathbb{E}_{\vec{x}_t=\vec{x}}[f(\vec{x}_t,t)\bullet dB_t]=0$. Consequently, we have
\begin{eqnarray*}
	&&\mathbb{E}\bigg[\frac{\dbar_is}{\rd t}\bigg]=-\mathbb{E}_{\vec{x}\sim p(\vec{x},t)}\bigg[\frac{1}{p(\vec{x},t)}\frac{\partial p(\vec{x},t)}{\partial t}\bigg]+\mathbb{E}_{\vec{x}\sim p(\vec{x},t)}\bigg[\mathbb{E}_{\vec{x}_t=\vec{x}}\bigg[\frac{\vec{J}(\vec{x}_t,t)\cdot\mathbf{D}^{-1}(\vec{x}_t)}{p(\vec{x}_t,t)}\circ \frac{\rd\vec{x}_t}{\rd t}\bigg]\bigg]\\
	&=&\mathbb{E}_{\vec{x}\sim p(\vec{x},t)}\bigg[\mathbb{E}_{\vec{x}_t=\vec{x}}\bigg[\frac{\vec{J}(\vec{x}_t,t)\cdot\mathbf{D}^{-1}(\vec{x}_t)}{p(\vec{x}_t,t)}\bullet \frac{\rd\vec{x}_t}{\rd t}\bigg]\bigg]+\mathbb{E}_{\vec{x}\sim p(\vec{x},t)}\bigg[\mathbf{D}(\vec{x}):\nabla\bigg[\frac{\vec{J}(\vec{x},t)\cdot\mathbf{D}^{-1}(\vec{x})}{p(\vec{x},t)}\bigg]\bigg]\\
	&=&\mathbb{E}_{\vec{x}\sim p(\vec{x},t)}\bigg[\mathbb{E}_{\vec{x}_t=\vec{x}}\bigg[\frac{\vec{J}(\vec{x}_t,t)\cdot\mathbf{D}^{-1}(\vec{x}_t)}{p(\vec{x}_t,t)}\cdot \vec{u}(\vec{x}_t;\lambda_t)+\frac{\vec{J}(\vec{x}_t,t)\cdot\mathbf{D}^{-1}(\vec{x}_t)}{p(\vec{x}_t,t)}\cdot \mathbf{\sigma}(\vec{x}_t)\bullet \frac{\rd\vec{B}_t}{\rd t}\bigg]\bigg]\\
	&+&\mathbb{E}_{\vec{x}\sim p(\vec{x},t)}\bigg[-\frac{\vec{J}(\vec{x},t)\cdot\mathbf{D}^{-1}(\vec{x})\cdot(\nabla\cdot\mathbf{D}(\vec{x}))}{p(\vec{x},t)}+\frac{\nabla\cdot\vec{J}(\vec{x},t)}{p(\vec{x},t)}-\frac{\vec{J}(\vec{x},t)\cdot\nabla p(\vec{x},t)}{p^2(\vec{x},t)}\bigg]\\
	&=&\mathbb{E}_{\vec{x}\sim p(\vec{x},t)}\bigg[\frac{\vec{J}(\vec{x},t)\cdot\mathbf{D}^{-1}(\vec{x})}{p^2(\vec{x},t)}\cdot \big[\vec{u}(\vec{x};\lambda_t)p(\vec{x},t)-\nabla\cdot\mathbf{D}(\vec{x})p(\vec{x},t)-\mathbf{D}(\vec{x})\cdot\nabla p(\vec{x},t)\big]\bigg]\\
	&=&\int \rd\vec{x}\frac{\vec{J}(\vec{x},t)\cdot\mathbf{D}^{-1}(\vec{x})\cdot\vec{J}(\vec{x},t)}{p(\vec{x},t)},
\end{eqnarray*}
and
\begin{eqnarray*}
	&&\mathbb{E}\bigg[\frac{\dbar_es}{\rd t}\bigg]=-\mathbb{E}_{\vec{x}\sim p(\vec{x},t)}\bigg[\mathbb{E}_{\vec{x}_t=\vec{x}}\bigg[\hat{\vec{u}}(\vec{x}_t;\lambda_t)\cdot\mathbf{D}^{-1}(\vec{x}_t)\circ\frac{\rd\vec{x}_t}{\rd t}\bigg]\bigg]\\
	&=&-\mathbb{E}_{\vec{x}\sim p(\vec{x},t)}\bigg[\mathbb{E}_{\vec{x}_t=\vec{x}}\bigg[\hat{\vec{u}}(\vec{x}_t;\lambda_t)\cdot\mathbf{D}^{-1}(\vec{x}_t)\bullet\frac{\rd\vec{x}_t}{\rd t}\bigg]\bigg]-\mathbb{E}_{\vec{x}\sim p(\vec{x},t)}\bigg[\mathbf{D}(\vec{x}):\nabla\big[\hat{\vec{u}}(\vec{x};\lambda_t)\cdot\mathbf{D}^{-1}(\vec{x})\big]\bigg]\\
	&=&-\mathbb{E}_{\vec{x}\sim p(\vec{x},t)}\bigg[\mathbb{E}_{\vec{x}_t=\vec{x}}\bigg[\hat{\vec{u}}(\vec{x}_t;\lambda_t)\cdot\mathbf{D}^{-1}(\vec{x}_t)\cdot\vec{u}(\vec{x}_t;\lambda_t)+\hat{\vec{u}}(\vec{x}_t;\lambda_t)\cdot\mathbf{D}^{-1}(\vec{x}_t)\cdot\mathbf{\sigma}(\vec{x}_t)\bullet \frac{\rd\vec{B}_t}{\rd t}\bigg]\bigg]\\
	&-&\mathbb{E}_{\vec{x}\sim p(\vec{x},t)}\bigg[\mathbf{D}(\vec{x}):\nabla\big[\hat{\vec{u}}(\vec{x};\lambda_t)\cdot\mathbf{D}^{-1}(\vec{x})\big]\bigg]\\
	&=&-\mathbb{E}_{\vec{x}\sim p(\vec{x},t)}\bigg[\hat{\vec{u}}(\vec{x};\lambda_t)\cdot\mathbf{D}^{-1}(\vec{x})\cdot\vec{u}(\vec{x};\lambda_t)+\mathbf{D}(\vec{x}):\nabla\big[\hat{\vec{u}}(\vec{x};\lambda_t)\cdot\mathbf{D}^{-1}(\vec{x})\big]\bigg]\\
	&=&-\int \rd\vec{x}\big[\hat{\vec{u}}(\vec{x};\lambda_t)\cdot\mathbf{D}^{-1}(\vec{x})\cdot\vec{J}(\vec{x},t)\big].
\end{eqnarray*}
During the derivation, we adopt integration by parts and the infinite boundary condition, as well as the fact $\int \rd\vec{x}\partial p(\vec{x},t)/\partial t=\rd[\int \rd\vec{x}p(\vec{x},t)]/dt=0$.

More importantly, the fluctuation theorems are valid for the Langevin dynamics too. To see this, we consider a forward trajectory $\gamma=\{\vec{x}_t|t\in[0,T]\}$ generated by the Langevin dynamics $d\vec{x}_t=\vec{u}(\vec{x}_t;\lambda_t)dt+\mathbf{\sigma}(\vec{x}_t)\bullet d\vec{B}_t$, whose backward trajectory $\tilde{\gamma}=\{\tilde{\vec{x}}_t|t\in[0,T]\}$ associating with both a reversed trajectory $\tilde{\vec{x}}_t=\vec{x}_{T-t}$ and a reversed protocol $\tilde{\lambda}_t=\lambda_{T-t}$ in time. The initial and final positions of forward and backward trajectories satisfy $\tilde{\vec{x}}_0=\vec{x}_T$ and $\tilde{\vec{x}}_T=\vec{x}_0$. Clearly, the Langevin dynamics for the backward trajectory is given by $d\tilde{\vec{x}}_t=\vec{u}(\tilde{\vec{x}}_t;\tilde{\lambda}_t)dt+\mathbf{\sigma}(\tilde{\vec{x}}_t)\bullet d\vec{B}_t$. 

The probability of the forward trajectory can be explicitly expressed through a path integral formula\cite{peliti2021stochastic, cugliandolo2017rules}
\begin{eqnarray}
	p[\gamma]&=&e^{-A[\vec{x}_t;\lambda_t]}p(\vec{x}_0,0),\quad\\ A[\vec{x}_t;\lambda_t]&=&\frac{1}{4}\int_{0}^T \rd t \bigg\{\mathbf{D}^{-1}(\vec{x}_t):\bigg[\frac{\rd\vec{x}_t}{\rd t}-\hat{\vec{u}}(\vec{x}_t;\lambda_t)\bigg]^2+2\nabla_x\cdot\vec{u}(\vec{x}_t;\lambda_t)\bigg\},\nonumber
\end{eqnarray}
where $A[\vec{x}_t;\lambda_t]$ is known as the Onsager-Machlup action\cite{onsager1953fluctuations, machlup1953fluctuations}. At the same time, the probability of the backward trajectory reads
\begin{eqnarray}
	p[\tilde{\gamma}]&=&e^{-A[\tilde{\vec{x}}_t;\tilde{\lambda}_t]}p(\vec{x}_T,T),\\
	A[\tilde{\vec{x}}_t;\tilde{\lambda}_t]&=&\frac{1}{4}\int_{0}^T \rd t \bigg\{\mathbf{D}^{-1}(\tilde{\vec{x}}_t):\bigg[\frac{\rd\tilde{\vec{x}}_t}{\rd t}-\hat{\vec{u}}(\tilde{\vec{x}}_t;\tilde{\lambda}_t)\bigg]^2+2\nabla_x\cdot\vec{u}(\tilde{\vec{x}}_t;\tilde{\lambda}_t)\bigg\}\nonumber\\
	&=&\frac{1}{4}\int_{0}^T \rd t \bigg\{\mathbf{D}^{-1}(\vec{x}_t):\bigg[\frac{\rd\vec{x}_t}{\rd t}+\hat{\vec{u}}(\vec{x}_t;\lambda_t)\bigg]^2+2\nabla_x\cdot\vec{u}(\vec{x}_t;\lambda_t)\bigg\}.\nonumber
\end{eqnarray}
The last equality holds since only the time derivative term has an odd parity under time reversal\cite{ding2022covariant,ding2022unified}.

Accordingly, the logarithm of the probability ratio between the forward and backward trajectories gives
\begin{eqnarray*}
	\ln\bigg(\frac{p[\gamma]}{p[\tilde{\gamma}]}\bigg)&=&\ln\bigg(\frac{p(\vec{x}_0,0)}{p(\vec{x}_T,T)}\bigg)+A[\tilde{\vec{x}}_t;\tilde{\lambda}_t]-A[\vec{x}_t;\lambda_t]\\
	&=&\ln\bigg(\frac{p(\vec{x}_0,0)}{p(\vec{x}_T,T)}\bigg)+\int_{t=0}^T \rd t \hat{\vec{u}}(\vec{x}_t;\lambda_t)\cdot\mathbf{D}^{-1}(\vec{x}_t)\circ\frac{\rd\vec{x}_t}{\rd t}.
\end{eqnarray*}
The above formula has a significant thermodynamic interpretation. Referring to the definition of stochastic entropy, we recognize $\Delta s[\gamma]=\ln[p(\vec{x}_0,0)/p(\vec{x}_T,T)]$ as the entropy change along the forward trajectory $\gamma$ starting from state $x_0$ at time $0$ and ending in state $x_T$ at time $T$. $p(\vec{x}_0,0)$ and $p(\vec{x}_T,T)$ denote arbitrary normalized initial and final distributions. Furthermore, the entropy exchange along the forward trajectory can be calculated as
\begin{eqnarray}
	\Delta_e s[\gamma]&=&\int_{t=0}^T\frac{\dbar_es(\vec{x}_t,t)}{\rd t}\rd t=-\int_{t=0}^T\rd t \hat{\vec{u}}(\vec{x}_t;\lambda_t)\cdot\mathbf{D}^{-1}(\vec{x}_t)\circ\frac{\rd\vec{x}_t}{\rd t}\\
	&=&-\ln\bigg(\frac{p[\gamma|x_0]}{p[\tilde{\gamma}|\tilde{x}_0]}\bigg),\nonumber
\end{eqnarray}
as the logarithmic difference between conditional probabilities of the forward path $p[\gamma|x_0]$ and backward path $p[\tilde{\gamma}|\tilde{x}_0]$. 

As a result, we have the trajectory entropy production as\cite{peliti2021stochastic, cugliandolo2017rules}
\begin{eqnarray}
	\Delta_i s[\gamma]=\Delta s[\gamma]-\Delta_e s[\gamma]=\ln\bigg(\frac{p[\gamma]}{p[\tilde{\gamma}]}\bigg).
\end{eqnarray}
Then by averaging over all possible stochastic trajectories $\gamma$, we obtain an integral fluctuation theorem
\begin{eqnarray*}
	\mathbb{E}_{\gamma}[e^{-\Delta_i s[\gamma]}]=\int D[\gamma] p[\gamma]e^{-\Delta_i s[\gamma]}=\int D[\tilde{\gamma}] p[\tilde{\gamma}]=1.
\end{eqnarray*}
Furthermore, the Jensen's inequality implies $\mathbb{E}_{\gamma}[\Delta_i s[\gamma]]\geq0$, as a manifestation of the second law of thermodynamics. Contrarily, if we make an average over those trajectories with the same entropy production only, a detailed fluctuation theorem is obtained, i.e.
\begin{eqnarray*}
	P_F(\Sigma)e^{-\Sigma}=P_B(-\Sigma),
\end{eqnarray*}
where $P_F(\Sigma)=\int D[\gamma] p[\gamma]\delta(\Delta_i s[\gamma]-\Sigma)$ and $P_B(-\Sigma)=\int D[\tilde{\gamma}] p[\tilde{\gamma}]\delta(\Delta_i s[\tilde{\gamma}]+\Sigma)$.


\section{Molecular Diffusion and Theory of Stochastic Reaction Rate}

The physics of stochastic diffusion of individual molecules in aqueous solution, as established by Einstein \cite{einstein1905motion}, Langevin \cite{langevin1908theorie}, and the modern theory of stochastic processes in continuous time and space \cite{van1992stochastic}, has provided a satisfactory theoretical framework for studying the rate of chemical reactions in great details. 

To illustrate this class of models, we first focus on simple association between an $A$ molecule in a sea of $B$ molecules; both are immersed in an aqueous solution at temperature $T$ and there is an attractive interaction between $A$ and $B$ with force $\vec{f}=-\nabla U$.   

\subsection{Canonical kinetic ensemble}
\label{sec:5A}

\begin{equation}
	\label{fpe4ABpair}
	\frac{\partial \rho(\vec{x},t)}{\partial t}
	= D\nabla\cdot
	\left(\nabla\rho(\vec{x},t)
	- \frac{\vec{f}(\vec{x})}{k_BT}\rho(\vec{x},t)\right), 
\end{equation}
where $\vec{x}=\vec{0}$ represents the center of mass of the single $A$, and $\rho(\vec{x},t)$ stands for the concentration of $B$ at position $\vec{x}$ and time $t$.  We have assumed that the diffusion coefficients of $A$ and $B$ in the aqueous solution are $D_A$ and $D_B$. Therefore the relative diffusion coefficient of $B$ w.r.t. $A$ is
$D=D_A+D_B$.  Let the $\vec{x}\in\Omega\subset\mathbb{R}^3$ and 
\[
\int_{\Omega} \rd\vec{x} = V
\]
be the volume of the reaction vessel, in which the total number of $B$ molecules is
\[
\int_{\Omega} \rho(\vec{x},t) \rd\vec{x} = N.
\]
Thus, $N/V=c_{tot}$ is the total concentration of $B$ in the vessel. The boundary condition for the $\rho(\vec{x},t)$ is
\begin{equation} 
	\left[\, \nabla\rho(\vec{x},t)
	- \frac{\vec{f}(\vec{x})}{k_BT}\rho(\vec{x},t)\right]_{\partial\Omega}  = 0,
\end{equation} 
where $\partial\Omega$ denotes the boundary of $\Omega$.  

It is easy to verify that the solution to Eq. \ref{fpe4ABpair} with given $N$ has the form $N\rho_V^o(\vec{x},t)=(Vc_{tot})\rho_V^o(\vec{x},t)$ where $\rho_V^o(\vec{x},t)$ is the solution with given $V$ and $N=1$.

\subsection{Grand canonical kinetic ensemble} 

Alternatively, we let $\vec{x}\in\mathbb{R}^3$, and a boundary condition at infinity $\tilde{\rho}(\vec{x},t) = c_{free}$ as $\vec{x}\to\infty$.  We see that in this case the solution to (\ref{fpe4ABpair}) with given $c_{free}$ has the form
$c_{free}\,\tilde{\rho}^o(\vec{x},t)$. 

The solution $\rho^o_V(\vec{x},t)$ in Sec. \ref{sec:5A} is a function of $V$. When the potential function $U(\vec{x})$ having the property
\[
\lim_{\vec{x}\to\infty} U(\vec{x}) = 0 
\]
sufficiently fast as $\vec{x}\to\infty$, the stationary $\rho^o_V(\vec{x})\to 0$ as $V\to\infty$.  The mathematical relation between $\rho^o_{V}(\vec{x},t)$ and $\tilde{\rho}^o(\vec{x},t)$ is
\begin{equation}
	\lim_{V\to\infty} V\rho^o_V(\vec{x},t)  
	= \tilde{\rho}^o(\vec{x},t).
\end{equation}
Actually,
\begin{equation} 
	\int_{\mathbb{R}^3} 
	\Big(\tilde{\rho}^o(\vec{x},t) - 1
	\Big) \rd\vec{x} = \frac{\text{concentration of bound }B}{\text{concentration of free }B}.
\end{equation}

\section{Conclusion}
As a typical multiscale phenomenon, the chemical reactions have attracted wide attentions from communities of chemistry, biology and engineering. At the molecular level, each reaction is a highly random event; while at the macroscopic level, chemical reactions are well-known for their nonlinearity. Considering these intrinsic difficulties, mathematical models at different levels of descriptions have been developed, whose deep connections have been clarified in the past years. 

From the kinetic aspect, chemical master equations for the molecular number distribution of few reactive molecules and chemical rate equations for massive molecules within a macroscopic bulk are closely related in the Kurtz limit, as a direct consequence of laws of large numbers. In addition, we also have the chemical Langevin equations and the corresponding chemical Fokker-Planck equations to account for the fluctuations in the molecular number or concentration.

Alternatively, from the thermodynamic aspect, steady-state thermodynamics can be well established for chemical reactions occurring in open systems, with respect to either chemical master equations or chemical Fokker-Planck equations. Gibbs' thermodynamics and Kramers' formula for transtion rate emerge naturally in the macroscpic limit, in which the large deviations principle plays a key role. With the information of either Markovian jumps or trajectories of Langevin dynamics in hand, the latest results of stochastic thermodynamics can be applied to chemical reactions too. It is noteworthy that for chemical reaction systems, significant generalizations of the second law of thermodynamics can be made. They are deeply related to the following key questions -- Is it an equilibrium or a nonequilibrium system? Whether is the system open or closed? Whether is the condition of detailed balance fulfilled or not?

As we have stated in the introduction, the theory of chemical reactions has already become a primary language of chemistry. This situation is exactly the same as that of the Newtonin mechanics to classical physics, though the latter is far more appreciated by most scientists. Thus, we hope through our presentation the representativeness and significance of chemical reactions in the study of nonequilibrium thermodynamics, multiscale phenomena, etc, could be recognized by more and more readers.

\appendix

\section{Pure and mixed chemical kinetic state}

A kinetic state with an exponentially distributed sojourn time $T$,  $f_T(t)=re^{-rt}$,   if and only if it is memoryless:
\[
	\frac{\mathbb{P}\{ T>\tau+t \big\} }{\mathbb{P}\{
	         T>\tau \} } = \frac{e^{-r(\tau+t)} }{e^{-r\tau} } = e^{-rt}
	         = \mathbb{P}\{T>t\}.
\]
Such a state is ``unaware of'' the elapsed time $\tau$; it is without life history.\cite{exponential_prl,exponential_nature}  There is an uncanny resemblance between the concept of a stochastic elementary reaction that occurs to a pure chemical kinetic state with a single exponential waiting time and R.  A.  Fisher's notion of a population species with a single Malthusian parameter \cite{fisher-book} for fitness.  For a kinetic state with a sojourn time $T$ that consists of multiple $K$ exponentials,  
\begin{equation}
    f_T(t) = \sum_{k=1}^K p_k r_k e^{-r_kt},   \ \frac{\displaystyle   \mathbb{P}\big\{ T > \tau+t \big\} }{\displaystyle   \mathbb{P}\big\{ T > \tau \big\}   } =  \frac{\displaystyle \sum_{k=1}^K  p_k e^{-r_k(\tau+t)}  }{\displaystyle  \sum_{k=1}^K p_ke^{-r_k\tau} },
\end{equation}
where $p_1+\cdots+p_K=1$.   The waiting time cannnot memoryless; in fact one has
\begin{eqnarray}
\label{heterokinetics}
       \frac{\rd}{\rd\tau} \mathbb{E}\big[  T \big|\, T>\tau \big] &=& \frac{\rd}{\rd\tau}\left(\frac{\sum_{k=1}^K  \frac{p_k}{r_k}\, e^{-r_k\tau}  }{ \sum_{k=1}^K p_ke^{-r_k\tau} }\right)
 = \frac{\overline{ \big(r^{-1}-\overline{r^{-1}}\ \big)^2 } }{\overline{r^{-1}}^{\,2}} > 0,
\end{eqnarray} 
where $\overline{[\cdots]}$ denotes average over $p_kr_ke^{-r_kt}$.  Eq.  (\ref{heterokinetics}) should be compared and contrasted with a mixed population of organisms,  with $K$ subpopu\-lations each with its own per capita growth rate $g_k$,   whose overall per capita growth rate is the mean growth rate averaged over the subpopulations, $\overline{g}$:
\begin{equation}
	 x_{tot} = \sum_{k=1}^K x_k,  \  \overline{g}=\frac{1}{x_{tot}}\left( \frac{\rd x_{tot}}{\rd t} \right) = \frac{1}{x_{tot}}  \sum_{k=1}^K g_kx_k,
\end{equation}
where $x_k$ is the population size of the $k^{th}$ subpopulation.
Thus,
\begin{equation}
\frac{\rd \overline{g}}{\rd t} = \frac{\rd}{\rd t} \left( \frac{\sum_{k=1}^Kg_kx_k }{\sum_{k=1}^K x_k } \right)  =  \overline{ (g- \overline{g} \, )^2 } > 0,
\end{equation}
where $\overline{[\cdots]}$ is average over $x_k$.

\section{Emergent kinetics of chemical reactions in the asymptotic limits}
\label{app_B}
In Sec. IID, we derive the chemical Fokker-Planck equation from the chemical master equations in the Kurtz limit. However, as clearly denoted through the pre-factor $V^{-1}$ in Eq. \eqref{uD}, the drift and diffusion terms can not simultaneously exist in the mathematical limit, which means the chemical Fokker-Planck equation in Eq. \eqref{chemical-FP-eq} is pathological.

To address this issue, here we start from a Markov-chain model for chemical reactions (or a discrete version of chemical master equations in Eq. \eqref{chemical-master-eq}).
\begin{eqnarray*}
	P(\vec{n},t+\Delta t)-P(\vec{n},t)=&&\sum_{i=1}^M\bigg[r^+_i(\vec{n}-\vec{\nu}_i; V)P(\vec{n}-\vec{\nu}_i,t)-r^+_i(\vec{n}; V)P(\vec{n},t)\nonumber\\
	&&\qquad-r^-_i(\vec{n}; V)P(\vec{n},t)+r^-_i(\vec{n}+\vec{\nu}_i;V)P(\vec{n}+\vec{\nu}_i,t)\bigg]\Delta t,
\end{eqnarray*}
where $\Delta t\rightarrow0$ represents a small time step. By replacing the discrete particle number vector $\vec{n}$ by the concentrations of chemical reactants $\vec{c}=\vec{n}/V$ in a continuous version and further assuming the probability density function $P(\vec{c},t)$ are many times differentiable, then we have
\begin{eqnarray*}
	&&P(\vec{n},t+\Delta t)=P(V\vec{c},t+\Delta t)=P(V\vec{c},t)+\frac{\partial P(V\vec{c},t)}{\partial t}\Delta t+O((\Delta t)^2),\\
	&&r^\pm_i(\vec{n}\mp\vec{\nu}_i; V)P(\vec{n}\mp\vec{\nu}_i,t)=r^\pm_i(V\vec{c}; V)P(V\vec{c},t)\mp \frac{1}{V}\sum_{j=1}^N \nu_{ij}\frac{\partial [r^\pm_i(V\vec{c}; V)P(V\vec{c},t)]}{\partial c_j}\\
	&&\qquad\qquad\qquad\qquad\qquad\qquad+\frac{1}{2V^2}\sum_{l,j=1}^N \nu_{il}\nu_{ij}\frac{\partial^2 [r^\pm_i(V\vec{c}; V)P(V\vec{c},t)]}{\partial c_l\partial c_j}+O(V^{-3}).
\end{eqnarray*}
for sufficiently small time step $\Delta t$ and large volume size $V$. Inserting above formulas into the discrete chemical master equations, we arrive at
\begin{eqnarray*}
	&&\frac{\partial P(V\vec{c},t)}{\partial t}+O(\Delta t)=-\sum_{j=1}^N\frac{\partial}{\partial c_j} \bigg\{\sum_{i=1}^M\nu_{ij} \bigg[\frac{r^+_i(V\vec{c}; V)}{V}-\frac{r^-_i(V\vec{c}; V)}{V}\bigg]P(V\vec{c},t)\bigg\}\\
	&&+\frac{1}{2V}\sum_{l,j=1}^N\frac{\partial^2}{\partial c_l\partial c_j} \bigg\{\sum_{i=1}^M\nu_{il}\nu_{ij} \bigg[\frac{r^+_i(V\vec{c}; V)}{V}-\frac{r^-_i(V\vec{c}; V)}{V}\bigg]P(V\vec{c},t)\bigg\}+O(V^{-2}).
\end{eqnarray*}

Clearly, in the limit of $\Delta t\rightarrow0$ and $V\rightarrow\infty$, two terms on the right-hand side of above equation can not exist simultaneously. In the Kurtz limit, i.e. $\Delta t\sim V^{-1}\rightarrow0$, we have 
\begin{eqnarray*}
	\frac{\partial p(\vec{c},t)}{\partial t}=-\sum_{j=1}^N\frac{\partial}{\partial c_j} \big[u_j(\vec{c})p(\vec{c},t)\big],
\end{eqnarray*}
where $u_j(\vec{c})=\sum_{i=1}^M\nu_{ij}[R_i^+(\vec{c})-R_i^-(\vec{c})]$ and $R_i^\pm(\vec{c})=\lim_{V\rightarrow\infty}r^\pm_i(V\vec{c}; V)/V$. Actually, we obtain a transport equation without the diffusion term, which is known as the Liouville equation in classical mechanics.

To correctly take the contribution of diffusion into consideration, we turn to the WKB method. By adopting the following ansatz $P(V\vec{c},t)\simeq\exp[-V\varphi(\vec{c}(t),t)]$ and inserting it into the discrete chemical master equations, we get
\begin{eqnarray*}
	&&\exp[-V\varphi(\vec{c},t+\Delta t)+V\varphi(\vec{c},t)]-1=\exp\bigg[-\frac{\partial\varphi(\vec{c},t)}{\partial t}(V\Delta t)+O(V(\Delta t)^2)\bigg]-1\\
	&=&\sum_{i=1}^{M}\bigg\{\frac{r^+_i(V\vec{c}-\vec{\nu}_i;V)}{V}\exp\big[-V\varphi(\vec{c}-\vec{\nu}_i/V,t)+V\varphi(\vec{c},t)\big]-\frac{r^+_i(V\vec{c};V)}{V}\\
	&&\qquad-\frac{r^-_i(V\vec{c};V)}{V}+\frac{r^-_i(V\vec{c}+\vec{\nu}_i;V)}{V}\exp\big[-V\varphi(\vec{c}+\vec{\nu}_i/V,t)+V\varphi(\vec{c},t)\big]\bigg\}\\
	&=&\sum_{i=1}^{M}\bigg\{\bigg[\frac{r^+_i(V\vec{c};V)}{V}+O(V^{-1})\bigg]\exp\bigg[\vec{\nu}_i\cdot\frac{\partial\varphi(\vec{c},t)}{\partial \vec{c}}+O(V^{-1})\bigg]-\frac{r^+_i(V\vec{c};V)}{V}\\
	&&\qquad-\frac{r^-_i(V\vec{c};V)}{V}+\bigg[\frac{r^-_i(V\vec{c};V)}{V}+O(V^{-1})\bigg]\exp\bigg[-\vec{\nu}_i\cdot\frac{\partial\varphi(\vec{c},t)}{\partial \vec{c}}+O(V^{-1})\bigg]\bigg\}\\
	&=&\sum_{i=1}^{M}\bigg\{\bigg[R^+_i(\vec{c})+O(V^{-1})\bigg]\exp\bigg[\vec{\nu}_i\cdot\frac{\partial\varphi(\vec{c},t)}{\partial \vec{c}}\bigg]\bigg[1+O(V^{-1})\bigg]-R^+_i(\vec{c})\\
	&&\qquad-R^-_i(\vec{c})+\bigg[R^-_i(\vec{c})+O(V^{-1})\bigg]\exp\bigg[-\vec{\nu}_i\cdot\frac{\partial\varphi(\vec{c},t)}{\partial \vec{c}}\bigg]\bigg[1+O(V^{-1})\bigg]\bigg\}\\
	&=&\sum_{i=1}^{M}\bigg\{R^+_i(\vec{c})\bigg[\exp\bigg(\vec{\nu}_i\cdot\frac{\partial\varphi(\vec{c},t)}{\partial \vec{c}}\bigg)-1\bigg]+R^-_i(\vec{c})\bigg[\exp\bigg(-\vec{\nu}_i\cdot\frac{\partial\varphi(\vec{c},t)}{\partial \vec{c}}\bigg)-1\bigg]+O(V^{-1})\bigg\}.
\end{eqnarray*}
This yields the following Hamilton-Jacobi equation
\begin{eqnarray*}	&&\frac{\partial\varphi(\vec{c},t)}{\partial t}\\
	&=&-\ln\bigg\{1+\sum_{i=1}^{M}\bigg\{R^+_i(\vec{c})\bigg[\exp\bigg(\vec{\nu}_i\cdot\frac{\partial\varphi(\vec{c},t)}{\partial \vec{c}}\bigg)-1\bigg]+R^-_i(\vec{c})\bigg[\exp\bigg(-\vec{\nu}_i\cdot\frac{\partial\varphi(\vec{c},t)}{\partial \vec{c}}\bigg)-1\bigg]\bigg\}\bigg\}.
\end{eqnarray*}
in the Kurtz limit. A further Taylor expansion of the logarithmic term gives the one considered in Eq. \eqref{HJE-Hu}.

\section{Statistical counting in discrete and continuous time}
\label{app_C}
We use a simple example to illustrate the wide range of subtleties of stochastic countings in the limit of $\Delta x$, $\Delta t\to 0$.  Consider a Markov counting process $X(t)$ taking independent increment $+\Delta x$, $-\Delta x$, $0$ with respective probabilities $p, q, r$ in each $\Delta t$ time step \cite{mqw_paper_2}.   Then in the limit of $\epsilon\to 0$ with $\Delta x=\eta\epsilon$ and $\Delta t=\gamma^{-1}\epsilon$,  the law of total probability with probabilistic causality 
\[
	\mathbb{P}\{ X(t+\Delta t)= x \} = 
	 p\,\mathbb{P}\{ X(t)= x-\Delta x \}  + r\,\mathbb{P}\{ X(t)= x \}  + q\,\mathbb{P}\{ X(t)= x+\Delta x \}, 
\]
yields an Hamilton-Jacobi type nonlinear partial differential equation for the large deviations rate function 
$\varphi(x,t)=-\lim_{\epsilon\to 0}\epsilon\ln\mathbb{P}\{X(t)\in (x,x+\rd x)\}$, that is
\begin{equation}
\label{HJEa5}
 \frac{\partial\varphi}{\partial t} + H\left(x,\frac{\partial\varphi}{\partial x}\right) = 0, \ H(x,y) = \gamma\ln\Big\{pe^{\eta y}  + r + qe^{-\eta y} \Big\},
\end{equation}
in which $H(x,y)$ is a Hamiltonian function.   The corresponding Hamiltonian dynamics reads 
\begin{eqnarray}
\label{xdot}
 \dot{x} &=&\frac{\partial H}{\partial y}=\frac{\gamma\eta\big( pe^{\eta y} - q e^{-\eta y}\big) }{ pe^{\eta y} + r + q e^{-\eta y} },\\
 \dot{y} &=&-\frac{\partial H}{\partial x}= 0.
\end{eqnarray}
$\dot{x}$ is monotonically dependent upon $y$; with $\dot{x}\to\pm\gamma\eta$ as $y\to\pm\infty$.  For isotropic movements with $p=q=(1-r)/2$ and when $\eta y\ll 1$, (\ref{xdot}) becomes 
\begin{equation}
  y \simeq \frac{\tilde{v}}{\eta(1-r)}\left\{1+\frac{2-3r}{3(1-r)^2}\frac{\tilde{v}^2}{2}\right\} \simeq
  \frac{m\tilde{v}}{\sqrt{1-\big(\frac{\tilde{v}}{c}\big)^2 }},
\label{m_and_c}
\end{equation}
where $\tilde{v}\ll 1$, $m=[\,\eta(1-r)\,]^{-1}$ and  $c=\sqrt{3/(2-3r)}(1-r)$.  It is intriguing to recall that when $p=q$,  the canonical diffusion limit has $(\Delta x)^2 \sim \Delta t$ as $\Delta t\to 0$; thus $\Delta x/\Delta t\to\infty$ which is implied in diffusive motion and continuous-time Markov chain with jumps. The above simultaneous $\Delta x=\eta\epsilon$, $\Delta t= \gamma^{-1}\epsilon$, and $\epsilon\to 0$ implies $\Delta x/\Delta t=\gamma\eta$ exists and $|\dot{x}|\le \gamma\eta$.  

Poisson counting in continuous time takes $\Delta t\to 0$ while keeps $\Delta x$ fixed; in the limit one first obtains a continuous-time Markov chain with master equation for $P_n(t)=$ $\mathbb{P}\{X(t)=n\Delta x\}$ with $t\in\mathbb{R}$, 
\[
  \frac{\rd P_n(t)}{\rd t} 
  = u P_{n-1} - (u+w)P_n + w P_{n+1},
\]
provided that one assumes $p=u\Delta t$ and $q=w\Delta t$. The time now becomes ``random''.  If one then takes the limit $n\to\infty$ and $\Delta x\to 0$ with $x=n\Delta x$ fixed, one has asymptotic $P_{x/\Delta x}(t)=\exp\big\{\hspace{-2pt}-\hspace{-1pt}\tilde{\varphi}(x,t)/\Delta x\big\}$ where $\tilde{\varphi}(x,t)$ satisfies
\begin{equation}
   \frac{\partial\tilde{\varphi}(x,t)}{\partial t} + \tilde{H}\left(x,\frac{\partial\tilde{\varphi}}{\partial x}\right)=0, \ 
   \tilde{H}(x,y) = \tilde{u}\big(e^{y} -1\big)
   +\tilde{w}\big(e^{-y}-1\big),
\end{equation}
in which $\tilde{u}=u\Delta x=p(\Delta x/\Delta t)$ and $\tilde{w}=w\Delta x=q(\Delta x/\Delta t)$.  Under the supposition of continuous probability in the limit of $\Delta t\to 0$ \cite{schilling-book}: $p=u\Delta t$, $q=w\Delta t$ and $r=1-(u+w)\Delta t$,
$\tilde{H}$ agrees with Eq. (\ref{HJEa5}), $H(x,y)=\gamma\Delta t\big\{u\big(e^{\eta y}-1\big)+w\big(e^{-\eta y}-1\big)\big\}$ is a particular case of the general Hamilton-Jacobi-Hu equation in (\ref{HJE-Hu}); the denominators in (\ref{xdot}) vanishes.

\section{List of symbols}

\begin{table}[!ht]
\begin{tabularx}{\textwidth}{p{0.18\textwidth}X}
\toprule
  Symbol & Description \\
  \midrule
  \underline{\textbf{\emph{Chemical Reactions}}}\\
    $V$    & volume of the reaction vessel\\
\addlinespace
  $\vec{n}(t)$    & molecular number of reactive species at time $t$\\  
\addlinespace
  $\vec{c}(t)$    & concentration of reactive species at time $t$\\  
\addlinespace
  $\nu_{ij}^{\pm}$    & stoichiometric coefficient for species $j$ on the left/right hand side of $i$'th reaction\\
  \addlinespace
  $\nu_{ij}$    & number change of species $j$ in the $i$'th reaction\\
  \addlinespace
  $r_i^{\pm}(\vec{n})$    & rate function for the $i$'th forward/backward reaction in the discrete version\\
  \addlinespace
  $R_i^{\pm}(\vec{c})$    & rate function for the $i$'th forward/backward
  reaction in the continuous version\\
\addlinespace
  $k_{i}^{\pm}$    & rate constant for the $i$'th forward/backward reaction\\  

\vspace{0.2cm}\underline{\textbf{\emph{Mathematical Models}}}\\
    $P_n(t)$    & probability for the system in state $\vec{n}$ at time $t$\\
\addlinespace
    $P_n^{ss},P_n^{eq}$    & steady-state/equilibrium probability for the system in state $\vec{n}$\\
\addlinespace
    $W_{n,n'}$    & transition rate from state $n'$ to state $n$\\
\addlinespace
    $J_{n,n'}$    & flux from state $n'$ to state $n$\\
\addlinespace
    $p(\vec{x},t)$    & probability density for the system reaching position $\vec{x}$ at time $t$\\
\addlinespace
    $p^{ss}(\vec{x}),p^{eq}(\vec{x})$    & stationary/equilibrium probability density for the system at position $\vec{x}$\\
\addlinespace
    $\vec{u}(\vec{x})$    & drift term at position $\vec{x}$\\
\addlinespace
    $\mathbf{D}(\vec{x})$    & diffusion matrix at position $\vec{x}$\\
\addlinespace
    $\vec{J}(\vec{x},t)$    & probability flux at position $\vec{x}$ and time $t$\\
\addlinespace

\vspace{0.2cm}\underline{\textbf{\emph{Large Deviation Theory}}}\\
    $\psi(\vec{c},t)$    & large deviation rate function w.r.t $\vec{c}$ at time $t$\\
\addlinespace
    $\psi^{ss}(\vec{c})$    & stationary large deviation rate function w.r.t $\vec{c}$\\
\addlinespace
    $H(\vec{c},\vec{\xi})$    & Hamiltonian function w.r.t $\vec{c}$ and $\vec{\xi}$\\
\addlinespace
    $L(\vec{c},\vec{\zeta})$    & Hamiltonian function w.r.t $\vec{c}$ and $\vec{\zeta}$\\
\addlinespace
    $\Psi^*(\vec{c},\vec{\xi})$    & dissipation potential function\\
\addlinespace
    $\Psi(\vec{c},\vec{\zeta})$    & Legendre-Fenchel duality of dissipation potential function\\
\addlinespace
  \bottomrule
\end{tabularx}
\captionsetup{labelformat=empty}
\caption{}
\end{table}

\begin{table}[ht]
\begin{tabularx}{\textwidth}{p{0.18\textwidth}X}
\toprule
  Symbol & Description \\
  \midrule
\vspace{0.2cm}\underline{\textbf{\emph{Nonequilibrium Thermodynamics}}}\\
    $T$    & temperature\\
\addlinespace
    $E(t)$    & internal energy at time $t$\\
\addlinespace    
    $S(t)$    & entropy function at time $t$\\
\addlinespace
    $F(t)$    & relative entropy/free energy at time $t$\\
\addlinespace
    $G(\vec{c})$    & Gibbs free energy\\
\addlinespace
    $A_i(\vec{c})$    & affinity for the $i$'th reaction\\
\addlinespace
    $\mu_j(c_j)$    & chemical potential for species $j$\\
\addlinespace
    $\mu_j^0(c_j)$    & standard chemical potential for species $j$\\
\addlinespace
    $\frac{\rd S}{\rd t}$    & instantaneous rate of entropy change\\
\addlinespace
    $\frac{\dbar Q}{\rd t}$    & heat exchange rate\\
\addlinespace
    $\frac{\dbar W}{\rd t}$    & work input rate\\
\addlinespace
    $\frac{\rd F}{\rd t}$    & free energy dissipation rate\\
\addlinespace
    $\frac{\dbar_i S}{\rd t}$    & entropy exchange rate\\
\addlinespace
    $\frac{\dbar_e S}{\rd t}$    & entropy production rate\\
\addlinespace
    $\dot{E}_{hk}$    & house-keeping energy input rate\\
\addlinespace
    $\frac{\rd S_{ex}}{\rd t}$    & excessive entropy production rate\\
\addlinespace
    $\frac{\rd S_{env}}{\rd t}$    & environment entropy change rate\\
\addlinespace

\vspace{0.2cm}\underline{\textbf{\emph{Stochastic Thermodynamics in Continuous Version}}}\\
    $s(\vec{x}_t,t)$    & stochastic entropy along trajectory $\vec{x}_t$ at time $t$\\
\addlinespace
    $\epsilon(\vec{x}_t,t)$    & stochastic energy along trajectory $\vec{x}_t$ at time $t$\\
\addlinespace
    $\frac{\dbar q_{tot}(\vec{x}_t,t)}{\rd t}$   & total heat absorption rate along trajectory $\vec{x}_t$ at time $t$\\
\addlinespace
    $\frac{\dbar q_{ex}(\vec{x}_t,t)}{\rd t}$   & excessive heat absorption rate along trajectory $\vec{x}_t$ at time $t$\\
\addlinespace
    $\frac{\dbar q_{hk}(\vec{x}_t,t)}{\rd t}$   & house-keeping heat absorption rate along trajectory $\vec{x}_t$ at time $t$\\
\addlinespace
    $\frac{\dbar w_{tot}(\vec{x}_t,t)}{\rd t}$   & instantaneous rate of work done along trajectory $\vec{x}_t$ at time $t$\\
\addlinespace
    $\frac{\dbar w_{ex}(\vec{x}_t,t)}{\rd t}$   & rate of excessive work done along trajectory $\vec{x}_t$ at time $t$\\
\addlinespace
    $\frac{\dbar w_{hk}(\vec{x}_t,t)}{\rd t}$   & rate of house-keeping work done along trajectory $\vec{x}_t$ at time $t$\\
\addlinespace
    $\frac{\rd s(\vec{x}_t,t)}{\rd t}$   & instantaneous entropy change rate along trajectory $\vec{x}_t$ at time $t$\\
\addlinespace
    $\frac{\dbar_e s(\vec{x}_t,t)}{\rd t}$   & entropy exchange rate along trajectory $\vec{x}_t$ at time $t$\\
\addlinespace
    $\frac{\dbar_i s(\vec{x}_t,t)}{\rd t}$   & entropy production rate along trajectory $\vec{x}_t$ at time $t$\\
\addlinespace
  \bottomrule
\end{tabularx}
\captionsetup{labelformat=empty}
\caption{}
\end{table}

\begin{table}[ht]
\begin{tabularx}{\textwidth}{p{0.18\textwidth}X}
\toprule
  Symbol & Description \\
  \midrule
\vspace{0.2cm}\underline{\textbf{\emph{Stochastic Thermodynamics in Discrete Version}}}\\
    $P[\gamma]$    & probability for the stochastic trajectory $\gamma$\\
\addlinespace
    $P[\tilde{\gamma}]$    & probability for the backward trajectory of $\gamma$\\
\addlinespace
    $P[\gamma^{\dagger}]$    & probability for the adjoint trajectory of $\gamma$\\
\addlinespace
    $D[\gamma]$    & Lebesgue measure of trajectory $\gamma$\\
\addlinespace
    $s(\vec{n},t)$    & stochastic entropy of state $\vec{n}$ at time $t$\\
\addlinespace
    $\lambda_t$    & external control variable\\
\addlinespace
    $\epsilon_n$    & energy of the system in state $\vec{n}$\\
\addlinespace
    $q_{n,n'}$    & heat absorbed by the system from the heat bath when jumping from state $n'$ to state $n$\\
\addlinespace
    $\Delta q_{tot}[\gamma]$   & total heat absorbed along trajectory $\gamma$\\
\addlinespace
    $\Delta q_{ex}[\gamma]$   & excessive heat absorbed along trajectory $\gamma$\\
\addlinespace
    $\Delta q_{hk}[\gamma]$   & house-keeping heat absorbed along trajectory $\gamma$\\
\addlinespace
    $\Delta w_{tot}[\gamma]$   & total work done along trajectory $\gamma$\\
\addlinespace
    $\Delta w_{ex}[\gamma]$   & excessive work done along trajectory $\gamma$\\
\addlinespace
    $\Delta w_{hk}[\gamma]$   & house-keeping work done along trajectory $\gamma$\\
\addlinespace
    $\frac{\dbar Q_{tot}}{\rd t}(t)$   & expected rate of total heat absorbed at time $t$\\
\addlinespace
    $\frac{\dbar Q_{ex}}{\rd t}(t)$   & expected rate of excessive heat absorbed at time $t$\\
\addlinespace
    $\frac{\dbar Q_{hk}}{\rd t}(t)$   & expected rate of excessive heat absorbed at time $t$\\
\addlinespace
    $\frac{\dbar W_{tot}}{\rd t}(t)$   & expected rate of total work done at time $t$\\
\addlinespace
    $\frac{\dbar W_{ex}}{\rd t}(t)$   & expected rate of excessive work done at time $t$\\
\addlinespace
    $\frac{\dbar W_{hk}}{\rd t}(t)$   & expected rate of house-keeping work done  at time $t$\\
\addlinespace
    $\Delta s[\gamma]$   & total entropy change along trajectory $\gamma$\\
\addlinespace
    $\Delta_e s[\gamma]$   & entropy exchange along trajectory $\gamma$\\
\addlinespace
    $\Delta_i s[\gamma]$   & total entropy production along trajectory $\gamma$\\
\addlinespace
    $\Delta s_{ex}[\gamma]$   & excessive entropy production along trajectory $\gamma$\\
\addlinespace
    $\Delta s_{hk}[\gamma]$   & house-keeping entropy production along trajectory $\gamma$\\
\addlinespace
  \bottomrule
\end{tabularx}
\captionsetup{labelformat=empty}
\caption{}
\end{table}

\section*{Acknowledgement}
This work was funded by the Guangdong Provincial Key Laboratory of Mathematical and Neural Dynamical Systems (DSNS2025002), Guangdong Basic and Applied Basic Research Foundation (2023A1515010157). L. H.  thanks the hospitality of Beijing Institute of Mathematical Sciences and Applications (BIMSA) during the preparation of this review. We thank Professors Jin Feng, Yuan Gao, Jian-Guo Liu, Liangrong Peng, and Xiangjun Xing  for their stimulating discussions.

\bibliographystyle{siamplain}
\bibliography{Multiscale}

\begin{thebibliography}{100}

\bibitem{alekseev_symb_dyn}
{\sc V.~Alekseev and M.~Yakobson}, {\em Symbolic dynamics and hyperbolic
  dynamic systems}, Physics Reports, 75 (1981), pp.~290--325.

\bibitem{berry1984_finite_time}
{\sc B.~Andresen, P.~Salamon, and R.~S. Berry}, {\em Thermodynamics in finite
  time}, Physics Today, 37 (1984), pp.~62--70.

\bibitem{aris1991}
{\sc R.~Aris, D.~G. Aronson, and H.~L. Swinney}, eds., {\em Patterns and
  Dynamics in Reactive Media}, Springer, New York, 1991.

\bibitem{beard_qian_book}
{\sc D.~A. Beard and H.~Qian}, {\em Chemical Biophysics: Quantitative Analysis
  of Cellular Systems}, Cambridge Univ. Press, London, 2008.

\bibitem{Chibbaro-book}
{\sc S.~Chibbaro, L.~Rondoni, and A.~Vulpiani}, {\em Reductionism, Emergence
  and Levels of Reality: {T}he Importance of Being Borderline}, Springer, New
  York, 2014.

\bibitem{chow2012fokker}
{\sc S.-N. Chow, W.~Huang, Y.~Li, and H.~Zhou}, {\em {F}okker-{P}lanck
  equations for a free energy functional or markov process on a graph}, Archive
  for Rational Mechanics and Analysis, 203 (2012), pp.~969--1008.

\bibitem{cox1954superposition}
{\sc D.~R. Cox and W.~L. Smith}, {\em On the superposition of renewal
  processes}, Biometrika, 41 (1954), pp.~91--99.

\bibitem{crooks1999entropy}
{\sc G.~E. Crooks}, {\em Entropy production fluctuation theorem and the
  nonequilibrium work relation for free energy differences}, Physical Review E,
  60 (1999), p.~2721.

\bibitem{cugliandolo2017rules}
{\sc L.~F. Cugliandolo and V.~Lecomte}, {\em Rules of calculus in the path
  integral representation of white noise langevin equations: the
  {O}nsager-{M}achlup approach}, Journal of Physics A: Mathematical and
  Theoretical, 50 (2017), p.~345001.

\bibitem{de1962}
{\sc S.~R. de~Groot and P.~Mazur}, {\em Non-Equilibrium Thermodynamics},
  North-Holland, Amsterdam, 1962.

\bibitem{qian_razo}
{\sc M.~J. del Razo and H.~Qian}, {\em A discrete stochastic formulation for
  reversible bimolecular reactions via diffusion encounter}, Communications in
  Mathematical Sciences, 14 (2016), pp.~1741--1772.

\bibitem{dembo2009large}
{\sc A.~Dembo and O.~Zeitouni}, {\em Large Deviations Techniques and
  Applications}, Springer Science, New York, 2009.

\bibitem{ding2022unified}
{\sc M.~Ding, F.~Liu, and X.~Xing}, {\em Unified theory of thermodynamics and
  stochastic thermodynamics for nonlinear langevin systems driven by
  non-conservative forces}, Physical Review Research, 4 (2022), p.~043125.

\bibitem{ding2022covariant}
{\sc M.~Ding and X.~Xing}, {\em Covariant nonequilibrium thermodynamics from
  ito-langevin dynamics}, Physical Review Research, 4 (2022), p.~033247.

\bibitem{eweinan}
{\sc W.~E}, {\em Principles of Multiscale Modeling}, Cambridge University
  Press, London, U.K., 2011.

\bibitem{einstein1905motion}
{\sc A.~Einstein et~al.}, {\em On the motion of small particles suspended in
  liquids at rest required by the molecular-kinetic theory of heat}, Annalen
  der physik, 17 (1905), p.~208.

\bibitem{elson}
{\sc E.~L. Elson and W.~W. Webb}, {\em Concentration correlation spectroscopy:
  a new biophysical probe based on occupation number fluctuations}, Annual
  Review of Biophysics and Bioengineering, 4 (1975), pp.~311--334.

\bibitem{epstein_book}
{\sc I.~R. Epstein and J.~A. Pojman}, {\em An Introduction to Nonlinear
  Chemical Dynamics: Oscillations, Waves, Patterns, and Chaos}, Oxford
  University Press, London, U.K., 1998.

\bibitem{erdi1989mathematical}
{\sc P.~{\'E}rdi and J.~T{\'o}th}, {\em Mathematical Models of Chemical
  Reactions: Theory and Applications of Deterministic and Stochastic Models},
  Manchester University Press, U.K., 1989.

\bibitem{Esposito2010Three}
{\sc M.~Esposito and C.~van~den Broeck}, {\em Three detailed fluctuation
  theorems.}, Physical Review Letters, 104 (2010), p.~090601.

\bibitem{esposito2010three-1}
{\sc M.~Esposito and C.~van~den Broeck}, {\em Three faces of the second law.
  {I}. master equation formulation}, Physical Review E, 82 (2010), p.~011143.

\bibitem{evans_book}
{\sc L.~C. Evans}, {\em Partial Differential Equations}, AMS Pub., Providence,
  R.I., 2nd~ed., 2010.

\bibitem{feinberg1972complex}
{\sc M.~Feinberg}, {\em Complex balancing in general kinetic systems}, Archive
  for Rational Mechanics and Analysis, 49 (1972), pp.~187--194.

\bibitem{feinberg}
{\sc M.~Feinberg}, {\em Foundations of Chemical Reaction Network Theory},
  Springer, New York, 2019.

\bibitem{feng2006large}
{\sc J.~Feng and T.~G. Kurtz}, {\em Large Deviations for Stochastic Processes},
  AMS Pub., Providence, R.I., 2006.

\bibitem{fisher-book}
{\sc R.~A. Fisher}, {\em The Genetical Theory of Natural Selection}, The
  Clarendon Press, Oxford, 1930.

\bibitem{freidlin1998random}
{\sc M.~I. Freidlin and A.~D. Wentzell}, {\em Random Perturbations of Dynamical
  Systems}, Springer, New York, 3rd~ed., 2012.

\bibitem{friedman1964}
{\sc A.~Friedman}, {\em Partial Differential Equations of Parabolic Type},
  Prentice-Hall, Englewood Cliffs, N.J., 1964.

\bibitem{gao2022revisit}
{\sc Y.~Gao and J.-G. Liu}, {\em Revisit of macroscopic dynamics for some
  non-equilibrium chemical reactions from a hamiltonian viewpoint}, Journal of
  Statistical Physics, 189 (2022), p.~22.

\bibitem{gao2023large}
{\sc Y.~Gao and J.-G. Liu}, {\em Large deviation principle and thermodynamic
  limit of chemical master equation via nonlinear semigroup}, Multiscale
  Modeling \& Simulation, 21 (2023), pp.~1534--1569.

\bibitem{ge2010physical}
{\sc H.~Ge and H.~Qian}, {\em Physical origins of entropy production, free
  energy dissipation, and their mathematical representations}, Physical Review
  E, 81 (2010), p.~051133.

\bibitem{Ge2016Mesoscopic}
{\sc H.~Ge and H.~Qian}, {\em Mesoscopic kinetic basis of macroscopic chemical
  thermodynamics: A mathematical theory}, Physical Review E, 94 (2016),
  p.~052150.

\bibitem{ge2016}
{\sc H.~Ge and H.~Qian}, {\em Nonequilibrium thermodynamic formalism of
  nonlinear chemical reaction systems with {W}aage--{G}uldberg's law of mass
  action}, Chemical Physics, 472 (2016), pp.~241--248.

\bibitem{gillespie1976general}
{\sc D.~T. Gillespie}, {\em A general method for numerically simulating the
  stochastic time evolution of coupled chemical reactions}, Journal of
  Computational Physics, 22 (1976), pp.~403--434.

\bibitem{gillespie2000chemical}
{\sc D.~T. Gillespie}, {\em The chemical {L}angevin equation}, The Journal of
  Chemical Physics, 113 (2000), pp.~297--306.

\bibitem{gnesotto2018broken}
{\sc F.~S. Gnesotto, F.~Mura, J.~Gladrow, and C.~P. Broedersz}, {\em Broken
  detailed balance and non-equilibrium dynamics in living systems: a review},
  Reports on Progress in Physics, 81 (2018), p.~066601.

\bibitem{goldstein1951}
{\sc H.~Goldstein}, {\em Classical Mechanics}, Addison-Wesley, New York, 1951.

\bibitem{exponential_nature}
{\sc P.~T. Greenland}, {\em Seeking non-exponential decay}, Nature, 335 (1988),
  pp.~298--298.

\bibitem{grzybowski2009chemistry}
{\sc B.~A. Grzybowski}, {\em Chemistry in Motion: Reaction-Diffusion Systems
  for Micro-and Nanotechnology}, John Wiley \& Sons, New York, 2009.

\bibitem{guggenheim1933}
{\sc E.~A. Guggenheim}, {\em Modern Thermodynamics by the Methods of {W}illard
  {G}ibbs}, Methuen \& Co., New York, 1933.

\bibitem{hill1977}
{\sc T.~L. Hill}, {\em Free Energy Transduction in Biology: The Steady-State
  Kinetic and Thermodynamic Formalism}, Academic Press, New York, 1977.

\bibitem{hopfield_jtb}
{\sc J.~J. Hopfield}, {\em Physics, computation, and why biology looks so
  different}, Journal of Theoretical Biology, 171 (1994), pp.~53--60.

\bibitem{horn1972necessary}
{\sc F.~Horn}, {\em Necessary and sufficient conditions for complex balancing
  in chemical kinetics}, Archive for Rational Mechanics and Analysis, 49
  (1972), pp.~172--186.

\bibitem{horn1972general}
{\sc F.~Horn and R.~Jackson}, {\em General mass action kinetics}, Archive for
  Rational Mechanics and Analysis, 47 (1972), pp.~81--116.

\bibitem{gang1986lyapounov}
{\sc G.~Hu}, {\em Lyapounov function and stationary probability distributions},
  Zeitschrift f{\"u}r Physik B Condensed Matter, 65 (1986), pp.~103--106.

\bibitem{jarzynski1997nonequilibrium}
{\sc C.~Jarzynski}, {\em Nonequilibrium equality for free energy differences},
  Physical Review Letters, 78 (1997), p.~2690.

\bibitem{jeans1907}
{\sc J.~H. Jeans}, {\em An Elementary Treatise on Theoretical Mechanics}, Ginn
  \& Co., Boston, 1907.

\bibitem{jiang2004mathematical}
{\sc D.-Q. Jiang, M.~Qian, and M.-P. Qian}, {\em Mathematical Theory of
  Nonequilibrium Steady States}, Springer-Verlag, Berlin, 2004.

\bibitem{jordan1998variational}
{\sc R.~Jordan, D.~Kinderlehrer, and F.~Otto}, {\em The variational formulation
  of the {F}okker-{P}lanck equation}, SIAM Journal on Mathematical Analysis, 29
  (1998), pp.~1--17.

\bibitem{kaiser2017acceleration}
{\sc M.~Kaiser, R.~L. Jack, and J.~Zimmer}, {\em Acceleration of convergence to
  equilibrium in {M}arkov chains by breaking detailed balance}, Journal of
  Statistical Physics, 168 (2017), pp.~259--287.

\bibitem{keizer1982nonequilibrium}
{\sc J.~Keizer}, {\em Nonequilibrium statistical thermodynamics and the effect
  of diffusion on chemical reaction rates}, The Journal of Physical Chemistry,
  86 (1982), pp.~5052--5067.

\bibitem{keizer1987statistical}
{\sc J.~Keizer}, {\em Statistical Thermodynamics of Nonequilibrium Processes},
  Springer-Verlag, New York, 1987.

\bibitem{khinchin1960mathematical}
{\sc A.~Khinchin}, {\em Mathematical Methods in the Theory of Queueing},
  Charles Griffin \& Co, London, 1960.

\bibitem{schilling-book}
{\sc D.~Khoshnevisan and R.~Schilling}, {\em From {L}\'{e}vy-Type Processes to
  Parabolic {SPDE}s}, Springer, New York, 2016.

\bibitem{kimmel2018inferring}
{\sc J.~C. Kimmel, A.~Y. Chang, A.~S. Brack, and W.~F. Marshall}, {\em
  Inferring cell state by quantitative motility analysis reveals a dynamic
  state system and broken detailed balance}, PLoS Computational Biology, 14
  (2018), p.~e1005927.

\bibitem{klebaner2012introduction}
{\sc F.~C. Klebaner}, {\em Introduction to Stochastic Calculus with
  Applications}, World Scientific, Singapore, 2012.

\bibitem{kondepudi2014modern}
{\sc D.~Kondepudi and I.~Prigogine}, {\em Modern Thermodynamics: From Heat
  Engines to Dissipative Structures}, John Wiley \& Sons, New York, 2014.

\bibitem{kramers1940brownian}
{\sc H.~A. Kramers}, {\em Brownian motion in a field of force and the diffusion
  model of chemical reactions}, Physica, 7 (1940), pp.~284--304.

\bibitem{kullback1951information}
{\sc S.~Kullback and R.~A. Leibler}, {\em On information and sufficiency}, The
  Annals of Mathematical Statistics, 22 (1951), pp.~79--86.

\bibitem{kurtz1972}
{\sc T.~G. Kurtz}, {\em The relationship between stochastic and deterministic
  models for chemical reactions}, The Journal of Chemical Physics, 57 (1972),
  pp.~2976--2978.

\bibitem{kurtz1978strong}
{\sc T.~G. Kurtz}, {\em Strong approximation theorems for density dependent
  {M}arkov chains}, Stochastic Processes and their Applications, 6 (1978),
  pp.~223--240.

\bibitem{kurtz1981clt}
{\sc T.~G. Kurtz}, {\em The central limit theorem for {M}arkov chains}, The
  Annals of Probability, 9 (1981), pp.~557--560.

\bibitem{langevin1908theorie}
{\sc P.~Langevin et~al.}, {\em Sur la th{\'e}orie du mouvement brownien}, CR
  Acad. Sci. Paris, 146 (1908), p.~530.

\bibitem{Li-Qian-Yi-jcp}
{\sc Y.~Li, H.~Qian, and Y.~Yi}, {\em Oscillations and multiscale dynamics in a
  closed chemical reaction system: {S}econd law of thermodynamics and temporal
  complexity}, Journal of Chemical Physics, 129 (2008), p.~154505.

\bibitem{Liujun-book}
{\sc J.~S. Liu}, {\em Monte Carlo Strategies in Scientific Computing},
  Springer-Verlag, New York, 2004.

\bibitem{Ma2015complete}
{\sc Y.-A. Ma, T.~Chen, and E.~Fox}, {\em A complete recipe for stochastic
  gradient mcmc}, in Advances in Neural Information Processing Systems, vol.~2,
  NeurIPS, 2015, pp.~2917--2925.

\bibitem{maas2011gradient}
{\sc J.~Maas}, {\em Gradient flows of the entropy for finite {M}arkov chains},
  Journal of Functional Analysis, 261 (2011), pp.~2250--2292.

\bibitem{machlup1953fluctuations}
{\sc S.~Machlup and L.~Onsager}, {\em Fluctuations and irreversible process.
  {II}. {S}ystems with kinetic energy}, Physical Review, 91 (1953), p.~1512.

\bibitem{martinez2019inferring}
{\sc I.~A. Mart{\'\i}nez, G.~Bisker, J.~M. Horowitz, and J.~M. Parrondo}, {\em
  Inferring broken detailed balance in the absence of observable currents},
  Nature Communications, 10 (2019), p.~3542.

\bibitem{mqw_paper_2}
{\sc B.~Miao, H.~Qian, and Y.-S. Wu}, {\em Emergence of {N}ewtonian
  deterministic causality from stochastic motions in continuous space and
  time}, arXiv:2406.02405,  (2024).

\bibitem{michel2020forward}
{\sc M.~Michel, A.~Durmus, and S.~S{\'e}n{\'e}cal}, {\em Forward event-chain
  monte carlo: Fast sampling by randomness control in irreversible markov
  chains}, Journal of Computational and Graphical Statistics, 29 (2020),
  pp.~689--702.

\bibitem{mielke2013geodesic}
{\sc A.~Mielke}, {\em Geodesic convexity of the relative entropy in reversible
  {M}arkov chains}, Calculus of Variations and Partial Differential Equations,
  48 (2013), pp.~1--31.

\bibitem{mielke2017non}
{\sc A.~Mielke, R.~I. Patterson, M.~A. Peletier, and D.~Michiel~Renger}, {\em
  Non-equilibrium thermodynamical principles for chemical reactions with
  mass-action kinetics}, SIAM Journal on Applied Mathematics, 77 (2017),
  pp.~1562--1585.

\bibitem{mielke2014relation}
{\sc A.~Mielke, M.~A. Peletier, and D.~M. Renger}, {\em On the relation between
  gradient flows and the large-deviation principle, with applications to markov
  chains and diffusion}, Potential Analysis, 41 (2014), pp.~1293--1327.

\bibitem{murray2003mathematical}
{\sc J.~D. Murray}, {\em Mathematical Biology, II: Spatial Models and
  Biomedical Applications}, Springer, New York, 2003.

\bibitem{exponential_prl}
{\sc E.~B. Norman, S.~B. Gazes, S.~G. Crane, and D.~A. Bennett}, {\em Tests of
  the exponential decay law at short and long times}, Phys. Rev. Lett., 60
  (1988), pp.~2246--2249.

\bibitem{onsager1953fluctuations}
{\sc L.~Onsager and S.~Machlup}, {\em Fluctuations and irreversible processes},
  Physical Review, 91 (1953), p.~1505.

\bibitem{otto2001geometry}
{\sc F.~Otto}, {\em The geometry of dissipative evolution equations: the porous
  medium equation},  (2001).

\bibitem{pavliotis2016stochastic}
{\sc G.~A. Pavliotis}, {\em Stochastic Processes and Applications}, Springer,
  New York, 2016.

\bibitem{peliti2021stochastic}
{\sc L.~Peliti and S.~Pigolotti}, {\em Stochastic Thermodynamics: An
  Introduction}, Princeton University Press, Princeton, New Jersey, 2021.

\bibitem{pence2011}
{\sc C.~H. Pence}, {\em ``describing our whole experience'': The statistical
  philosophies of {W}.{F}.{R}. {W}eldon and {K}arl {P}earson}, Stud. Hist.
  Philos. Biol. Biomed. Sci., 42 (2011), pp.~475--485.

\bibitem{peng2018generalized}
{\sc L.~Peng, Y.~Zhu, and L.~Hong}, {\em Generalized {O}nsager's reciprocal
  relations for the master and {F}okker-{P}lanck equations}, Physical Review E,
  97 (2018), p.~062123.

\bibitem{peng2018markov}
{\sc L.~Peng, Y.~Zhu, and L.~Hong}, {\em The {M}arkov process admits a
  consistent steady-state thermodynamic formalism}, Journal of Mathematical
  Physics, 59 (2018).

\bibitem{prigogine-book}
{\sc I.~Prigogine}, {\em Etude Thermodynamique des Ph\'{e}nom\`{e}nes
  Irr\'{e}versibles}, Dunod, Paris, 1947.

\bibitem{qianarpc07}
{\sc H.~Qian}, {\em Phosphorylation energy hypothesis: Open chemical systems
  and their biological functions}, Annual Review of Physical Chemistry, 58
  (2007), pp.~113--142.

\bibitem{qian_bj_08}
{\sc H.~Qian}, {\em Cooperativity and specificity in enzyme kinetics: {A}
  single-molecule time-based perspective}, Biophysical Journal, 95 (2008),
  pp.~10--17.

\bibitem{qian2022jctc}
{\sc H.~Qian}, {\em Statistical chemical thermodynamics and energetic behavior
  of counting: {G}ibbs’ theory revisited}, Journal of Chemical Theory and
  Computation, 18 (2022), pp.~6421--6436.

\bibitem{qianpnas04}
{\sc H.~Qian and E.~L. Elson}, {\em Fluorescence correlation spectroscopy with
  high-order and dual-color correlation to probe nonequilibrium steady states},
  Proceedings of the National Academy of Sciences, U.S.A., 101 (2004),
  pp.~2828--2833.

\bibitem{qian2021stochastic}
{\sc H.~Qian and H.~Ge}, {\em Stochastic Chemical Reaction Systems in Biology},
  Springer, Cham, Switzerland, 2021.

\bibitem{qian2016}
{\sc H.~Qian, S.~Kjelstrup, A.~B. Kolomeisky, and D.~Bedeaux}, {\em Entropy
  production in mesoscopic stochastic thermodynamics: nonequilibrium kinetic
  cycles driven by chemical potentials, temperatures, and mechanical forces},
  Journal of Physics: Condensed Matter, 28 (2016), p.~153004.

\bibitem{qian_qian_tang2002}
{\sc H.~Qian, M.~Qian, and X.~Tang}, {\em Thermodynamics of the general
  diffusion process: {T}ime-reversibility and entropy production}, Journal of
  Statistical Physics, 107 (2002), pp.~1129--1141.

\bibitem{rao2016nonequilibrium}
{\sc R.~Rao and M.~Esposito}, {\em Nonequilibrium thermodynamics of chemical
  reaction networks: Wisdom from stochastic thermodynamics}, Physical Review X,
  6 (2016), p.~041064.

\bibitem{reimann1999universal}
{\sc P.~Reimann, G.~J. Schmid, and P.~H\"{a}nggi}, {\em Universal equivalence
  of mean first-passage time and {K}ramers rate}, Physical Review E, 60 (1999),
  p.~R1.

\bibitem{risken1996fokker}
{\sc H.~Risken}, {\em Fokker-Planck Equation: Methods of Solutions and
  Applications}, Springer, New York, 1996.

\bibitem{schnakenberg1976}
{\sc J.~Schnakenberg}, {\em Network theory of microscopic and macroscopic
  behavior of master equation systems}, Reviews of Modern Physics, 48 (1976),
  p.~571.

\bibitem{schnoerr2014complex}
{\sc D.~Schnoerr, G.~Sanguinetti, and R.~Grima}, {\em The complex chemical
  {L}angevin equation}, The Journal of Chemical Physics, 141 (2014), p.~024103.

\bibitem{seifert2005entropy}
{\sc U.~Seifert}, {\em Entropy production along a stochastic trajectory and an
  integral fluctuation theorem}, Physical Review Letters, 95 (2005), p.~040602.

\bibitem{seifert2012stochastic}
{\sc U.~Seifert}, {\em Stochastic thermodynamics, fluctuation theorems and
  molecular machines}, Reports on Progress in Physics, 75 (2012), p.~126001.

\bibitem{seller_1984}
{\sc P.~H. Sellers}, {\em Combinatorial classification of chemical mechanisms},
  SIAM Journal on Applied Mathematics, 44 (1984), pp.~784--792.

\bibitem{shapiro_1969}
{\sc N.~Z. Shapiro}, {\em A generalized technique for eliminating species in
  complex chemical equilibrium calculations}, SIAM J. Appl. Math., 17 (1969),
  pp.~960--971.

\bibitem{sisan10}
{\sc D.~R. Sisan, D.~Yarar, C.~M. Waterman, and J.~S. Urbach}, {\em Event
  ordering in live-cell imaging determined from temporal cross-correlation
  asymmetry}, Biophysical Journal, 98 (2010), pp.~2432--2441.

\bibitem{sohl2014hamiltonian}
{\sc J.~Sohl-Dickstein, M.~Mudigonda, and M.~DeWeese}, {\em Hamiltonian monte
  carlo without detailed balance}, in International Conference on Machine
  Learning, PMLR, 2014, pp.~719--726.

\bibitem{stroock2008partial}
{\sc D.~W. Stroock}, {\em Partial Differential Equations for Probabalists},
  Cambridge University Press, London, U.K., 2008.

\bibitem{sung2018statistical}
{\sc W.~Sung}, {\em Statistical Physics for Biological Matter}, Springer, New
  York, 2018.

\bibitem{touchette2009large}
{\sc H.~Touchette}, {\em The large deviation approach to statistical
  mechanics}, Physics Reports, 478 (2009), pp.~1--69.

\bibitem{van2010three}
{\sc C.~van~den Broeck and M.~Esposito}, {\em Three faces of the second law.
  {II}. {F}okker-{P}lanck formulation}, Physical Review E, 82 (2010),
  p.~011144.

\bibitem{van1992stochastic}
{\sc N.~Van~Kampen}, {\em Stochastic Processes in Physics and Chemistry},
  North-Holland Publishing Co, 1992.

\bibitem{van1983}
{\sc N.~G. van Kampen}, {\em Stochastic processes in physics and chemistry},
  Elsevier, Amsterdam, 1983.

\bibitem{vol1972differential}
{\sc A.~I. Vol'pert}, {\em Differential equations on graphs}, Mathematics of
  the USSR-Sbornik, 17 (1972), p.~571.

\bibitem{volpert2014elliptic}
{\sc V.~Volpert}, {\em Elliptic Partial Differential Equations: Volume 2:
  Reaction-Diffusion Equations}, Springer, New York, 2014.

\bibitem{xie}
{\sc X.~S. Xie}, {\em Single-molecule spectroscopy and dynamics at room
  temperature}, Acc. Chem. Res., 29 (1969), pp.~598--606.

\bibitem{yu2018mathematical}
{\sc P.~Y. Yu and G.~Craciun}, {\em Mathematical analysis of chemical reaction
  systems}, Israel Journal of Chemistry, 58 (2018), pp.~733--741.

\bibitem{ZhuHongYangYong2015}
{\sc Y.~Zhu, L.~Hong, Z.~Yang, and W.-A. Yong}, {\em Conservation-dissipation
  formalism of irreversible thermodynamics}, Journal of Non-Equilibrium
  Thermodynamics, 40 (2015), pp.~67--74.

\end{thebibliography}
\end{document}